\documentclass[prd,nofootinbib,showpacs,superscriptaddress,preprint]{revtex4-1}
\pdfoutput=1
\usepackage{amsmath,amsfonts,amssymb,graphicx,natbib,bm}
\usepackage{color}
\usepackage{slashed}    
\usepackage[colorlinks,citecolor=blue]{hyperref}
\usepackage{color}
\usepackage{url}
\usepackage{cancel}
\usepackage{slashed}
\usepackage{multirow}
\usepackage{rotating}
\usepackage{braket}
\usepackage{float}
\usepackage{subfigure}
\usepackage{mathtools}
\usepackage{scalerel}
\usepackage{tgbonum}
\usepackage{placeins}

\usepackage[normalem]{ulem}

\usepackage[most]{tcolorbox}
\newtcbox{\mymath}[1][]{%
    nobeforeafter, math upper, tcbox raise base,
    enhanced, colframe=blue!30!black,
    colback=blue!30, boxrule=1pt,
    #1}

\def\vecsign#1{\rule[1.388\LMex]{\dimexpr#1-2.5pt}{.36\LMpt}%
  \kern-6.0\LMpt\mathchar"017E}

\def\theraysign#1{\rule{0pt}{17\LMpt}\rule[1.384\LMex]{\dimexpr#1-2.5pt}{.40\LMpt}%
  \kern-6.0\LMpt\mathchar"017E}

\usepackage{stackengine}
\stackMath

\newcommand{\be}{\begin{eqnarray*}}
\newcommand{\ee}{\end{eqnarray*}}

\newcommand{\bee}{\begin{eqnarray}}
\newcommand{\eee}{\end{eqnarray}}
\newcommand{\beeq}{\begin{equation}}
\newcommand{\eeq}{\end{equation}}

\renewcommand{\vec}{\bf}

\usepackage{mathtools}
\usepackage{xspace}

\newcommand{\ba}{\begin{array}}
\newcommand{\ea}{\end{array}}
\newcommand{\bd}{\begin{displaymath}}
\newcommand{\ed}{\end{displaymath}}
\newcommand{\besub}{\begin{subequations}}
\newcommand{\eesub}{\end{subequations}}
\newcommand{\bea}{\begin{eqnarray}}
\newcommand{\eea}{\end{eqnarray}}

\def\m{\mu}
\def\n{\nu}

\def\q2 {q^2}

\def\bt{\begin{table}}
\def\et{\end{table}}

\usepackage{pifont}
\newcommand{\cmark}{\ding{51}}%
\newcommand{\xmark}{\ding{55}}%

\begin{document}

\title{Effective Theory of Freeze-in Dark Matter}

\author{Basabendu Barman}
\email{bb1988@iitg.ac.in}
\author{Debasish Borah}
\email{dborah@iitg.ac.in}
\author{Rishav Roshan}
\email{rishav.roshan@iitg.ac.in}
\affiliation{Department of Physics, Indian Institute of Technology Guwahati, Assam 781039, India}


\begin{abstract}  

We perform a model independent study of freeze-in of massive particle dark matter (DM) by adopting an effective field theory framework. Considering the dark matter to be a gauge singlet Majorana fermion, odd under a stabilising symmetry $Z_2$ under which all standard model (SM) fields are even, we write down all possible DM-SM operators upto and including mass dimension eight. For simplicity of the numerical analysis we restrict ourselves only to the scalar operators in SM as well as in the dark sector. We calculate the DM abundance for each such dimension of operator considering both UV and IR freeze-in contributions which can arise before and after the electroweak symmetry breaking respectively. After constraining the cut-off scale and reheat temperature of the universe from the requirement of correct DM relic abundance, we also study the possibility of connecting the origin of neutrino mass to the same cut-off scale by virtue of lepton number violating Weinberg operators. We thus compare the bounds on such cut-off scale and corresponding reheat temperature required for UV freeze-in from the origin of light neutrino mass as well as from the requirement of correct DM relic abundance. We also briefly comment upon the possibilities of realising such DM-SM effective operators in a UV complete model.
\end{abstract}

\maketitle

{
  \hypersetup{linkcolor=black}
  \tableofcontents
}

\section{Introduction}
\label{sec:intro}

Cosmology based experiments like WMAP \cite{Hinshaw:2012aka} and PLANCK \cite{Aghanim:2018eyx}, through precise measurements of cosmic microwave background (CMB) anisotropies have suggested the presence of a mysterious, non-luminous and non-baryonic component of matter, known as dark matter (DM), giving rise to around $26\%$ of the present universe's energy density. In terms of density parameter $\Omega_{\rm DM}$ and $h = \text{Hubble Parameter}/(100 \;\text{km} ~\text{s}^{-1} 
\text{Mpc}^{-1})$, the present DM abundance is conventionally reported as \cite{Aghanim:2018eyx}: $\Omega_{\text{DM}} h^2 = 0.120\pm 0.001$ at 68\% CL. While cosmology based evidences are relatively more recent, astrophysical evidences for DM emerged long back starting with the galaxy cluster observations by Fritz Zwicky \cite{Zwicky:1933gu} back in 1933, observations of galaxy rotation curves in 1970's \cite{Rubin:1970zza} to the more recent observation of the bullet cluster \cite{Clowe:2006eq}. While all these evidences are purely based on gravitational interactions of DM, we do not have any knowledge about the particle aspects of DM. Since none of the Standard Model (SM) particles can satisfy the criteria for being a realistic DM candidate, several beyond standard model (BSM) proposals have been put forward out of which the weakly interacting massive particle (WIMP) \cite{Jungman:1995df,Bertone:2004pz,Feng:2010gw, Arcadi:2017kky} is the most popular one. WIMP paradigm considers thermal production of DM in the early universe from the SM bath \cite{Srednicki:1988ce, Gondolo:1990dk} with an interesting coincidence that a DM particle having mass and couplings around the electroweak scale can give rise to the correct DM abundance after thermal freeze-out. This is often referred to as the \textit{WIMP Miracle}~\cite{Kolb:1990vq}. The same interactions between DM and SM particles which lead to thermal production of DM, can also lead to DM-nucleon scattering with the possibility of leaving some signatures at direct detection experiments like LUX~\cite{Akerib:2016vxi}, PandaX-II~\cite{Tan:2016zwf, Cui:2017nnn}, XENON1T \cite{Aprile:2017iyp, Aprile:2018dbl}. However, the continuous absence of such signal in several direct detection experiments so far have already constrained DM-nucleon scattering rates very strictly, pushing it towards the region where coherent neutrino-nucleus scattering cross section may dominate, also dubbed as the neutrino floor \cite{Billard:2013qya}. Similar null results for WIMP type DM have also been reported at indirect DM detection experiments, and also the large hadron collider (LHC), all of which constrain the coupling strength of DM with SM particles.

While negative results in WIMP searches do not necessarily rule it out, it has motivated the particle physics community to look for beyond the thermal WIMP paradigm where the interaction scale of DM particle can be much lower than the scale of weak interaction i.e.\,\,DM may be more feebly interacting than the thermal WIMP paradigm. One such possibility is to consider the origin of DM to be purely non-thermal \cite{Hall:2009bx}. In such a scenario, DM interaction with the SM bath is so weak that it never attains thermal equilibrium at any epoch in the early universe. While the initial abundance of DM in such a scenario is negligible, it can be produced from out of equilibrium decays of heavy particles or annihilation of particles already present in the thermal plasma. Such a scenario where DM abundance freezes in from a negligible initial abundance to the observed abundance is known as {\it freeze-in}, and the candidates of such non-thermal DM produced via freeze-in are often classified into a group called FIMP (Feebly interacting) massive particle)(for a review on such a DM paradigm see, for example~\cite{Bernal:2017kxu}). If there exists renormalizable interactions between FIMP and the SM bath, then the non-thermal production of DM is effective at lowest possible temperature. If the mother particle is in thermal equilibrium with the bath then the maximum production of DM occurs when the temperature of the bath $T\simeq M_0$, the mass of mother particle. Therefore, the non-thermal criterion enforces the couplings to be extremely tiny via the following condition
$\left|\dfrac{\Gamma}{{\bf H}}\right|_{T\simeq M_0}<1$ \cite{Arcadi:2013aba}, where $\Gamma$
is the decay width. For the case of scattering, one has to replace $\Gamma$ by the interaction rate $n_{\rm eq}\, \langle \sigma {\rm v}\rangle$, $n_{eq}$ being the equilibrium number density of mother particle. These types of freeze-in scenarios are known as infra-red (IR)-freeze-in~\cite{Yaguna:2011qn,Chu:2011be,Blennow:2013jba,Merle:2015oja,Shakya:2015xnx,Hessler:2016kwm,Biswas:2016bfo,Konig:2016dzg,Biswas:2016iyh,Biswas:2016yjr,Bernal:2017kxu,Biswas:2018aib,Heeba:2018wtf,Zakeri:2018hhe,Becker:2018rve,Heeba:2019jho,Lebedev:2019ton,Barman:2019lvm,Bhattacharya:2019tqq,Koren:2019iuv} where DM production is dominated by the lowest possible temperature at which it can occur i.e. $T\sim M_0$, since for $T<M_0$, the number density of mother particle becomes Boltzmann suppressed. On the other hand, there exists another possibility where FIMP and SM sector are coupled via higher dimensional operators (dimension $d > 4$) only. In such a scenario, DM production is effective at high temperatures and very much sensitive to initial history like the reheat temperature of the universe. Due to the higher dimensional nature of such interactions, DM production happens via scattering only, specially at a temperature above the electroweak scale. This particular scenario is known as the ultra-violet (UV) freeze-in~\cite{Hall:2009bx,Elahi:2014fsa, McDonald:2015ljz,Chen:2017kvz,Biswas:2019iqm,Bernal:2019mhf, Bernal:2020bfj, Bernal:2020qyu}. It may also happen that a realistic FIMP scenario has a mixture of both IR as well as UV freeze-in where after a phase transition like the one at the electroweak scale, the DM can have renormalizable interactions with the SM bath. However, if DM mass is much higher than the scale of such phase transitions, then its production will be dominated by UV freeze-in only.

Motivated by these, in this work, we consider an effective field theory (EFT) approach for UV freeze-in of DM. Since scalar DM can have renormalizable interactions with the SM particles which no symmetries can prevent, we consider a singlet Majorana fermion, odd under a $Z_2$ stabilising symmetry, to be the DM candidate. Naturally, DM interactions with the SM particles can arise only at dimension (dim.) five or higher level, suppressed by appropriate powers of the cut-off scale. We first list out possible DM-SM operators upto dim.8. While calculating the DM relic abundance, we consider only scalar operators responsible for DM-SM interactions for simplicity. We then constrain the cut-off scale, DM mass, as well as reheat temperature from correct DM relic requirement by considering both UV as well as IR freeze-in contribution that may arise before and after electroweak symmetry breaking (EWSB) respectively. We find that the IR freeze-in contribution is sizeable only for dim.5 operators while it is negligible for higher dimensional operators unless we consider a very low reheat temperature $(\leq1 \; \text{TeV})$ of the universe. Also, as expected, such IR freeze-in contribution is insignificant if DM mass is above the electroweak scale. We then discuss the possibility of the same cut-off scale to be responsible or origin of light neutrino masses via Weinberg operators of dim.5 and 7 \cite{Weinberg:1979sa}. We also briefly comment on the scenario where DM-SM interactions can happen only via lepton number violating operators. We then discuss some possible UV completion of a few DM-SM operators. Finally, we briefly comment upon the possibility of inflaton decay into DM at radiative level by virtue of the effective DM-SM operators and show the differences in parameter space compared to the ones obtained by considering DM production purely from SM-DM operators.

This paper is organised as follows. In section \ref{sec:ops} we list out possible DM-SM operators upto and including dim. 8 followed by the details of possible interaction vertices that can arise before and after EWSB in section \ref{sec:process}. In section \ref{sec:dm-yld}, we compute DM relic abundance by considering only DM scalar operators as mentioned before. In section \ref{sec:nu-mass} we consider the possibility of DM production only through dim. 5 and dim. 7 operators and check the constraints from neutrino mass if it is assumed to be arising from Weinberg operators of the same dimensions. In section \ref{sec:uvcomplete} we briefly comment upon different possibilities of generating DM-SM effective operators within a UV complete framework and in section~\ref{sec:inf-dm} we show the production of DM via one loop decay of the inflaton. We also show the effects of non-instantaneous reheating on DM abundance in section \ref{sec:tmax} and finally we conclude in section \ref{sec:conc}.

\section{List of possible DM-SM operators up to and including dim.8}
\label{sec:ops}

In this section we list out possible operators upto dim.8 that can be formed by considering bilinears in DM fields. EFT analysis in the context of the WIMP type DM has been done extensively and can be found in~\cite{Beltran:2008xg,Cao:2009uw,Goodman:2010ku,Cheung:2012gi, DeSimone:2013gj,Matsumoto:2014rxa,Duch:2014xda,Liem:2016xpm,Brod:2017bsw,Arina:2020mxo} and the references therein. Here we perform a similar study for the FIMP DM, considering it to be a singlet Majorana fermion $(\chi)$.

\begin{table}[htb!]
\begin{center}
\begin{tabular}{|c|c|c|c|c|}
\hline
    Bilinear & Transformation under\\
    & $C$-operator\\ 
\hline        
$\overline{\chi}\chi$  & +   \\
\hline
$i\overline{\chi}\gamma^5\chi$  & +   \\
\hline
$\overline{\chi}\gamma^\mu\chi$  & $-$   \\
\hline
$\overline{\chi}\gamma^\mu\gamma^5\chi$  & +   \\
\hline
$\overline{\chi}\sigma^{\mu\nu}\chi$  & $-$   \\
\hline
\end{tabular}
\end{center}  
\caption{Possible DM bilinears and their transformation under charge conjugation operator.}
\label{tab:cc}
\end{table}

\begin{table}[htb!]
\begin{center}
\begin{tabular}{|c|c|c|c|c|c|c|c|c|c|c|}
\hline
DM & $1/\Lambda$ & $1/\Lambda^2$ & $1/\Lambda^3$ & $1/\Lambda^4$   \\
bilinear & & & &    \\
(dim.3) & & &    &    \\[0.5ex] 
\hline\hline
&    &  & $X_{\mu\nu}X^{\mu\nu},X_{\mu\nu}\widetilde{X^{\mu\nu}}$ &    \\
&  & & $\left|H^\dagger H\right|^2$  &  \\
$\overline{\chi^c}\chi,$ & $H^\dagger H$ &  & $\left|\mathcal{D}_\mu H\right|^2$ & $\left(\overline{\ell_L} \widetilde{H}\right)\left(\overline{\ell_L} \widetilde{H}\right)$    \\
$\overline{\chi^c}i\gamma^5\chi$&  &  & $i\overline{L}\slashed{\mathcal{D}}L,i\overline{R}\slashed{\mathcal{D}}R$ &    \\
&  &  & $\overline{L}HR,\overline{L}\widetilde{H}R$ &    \\
\hline\hline
$\overline{\chi^c}\gamma^\mu\gamma^5\chi$&&$\overline{L(R)}\gamma_\mu L(R)$&&$\overline{L(R)}\gamma_\mu L(R)\left(H^\dagger H\right)$\\
&&$iH^\dagger\mathcal{D}_\mu H$&&$iH^\dagger\mathcal{D}_\mu H\left(H^\dagger H\right)$\\
\hline\hline
\end{tabular}
\end{center}
\caption{Possible operators up to and including dim.8 with scalar, pseudoscalar and axial vector bilinears in the DM fields, invariant under SM gauge symmetry. Here $X_{\mu\nu}\in B_{\mu\nu},W^a_{\mu\nu},G^a_{\mu\nu}$, $L\in Q_L,l_L$ are the SM left-handed doublet fermions, $R\in Q_R,e_R$ are the SM right-handed singlet fermions, and $\slashed{\mathcal{D}}=\gamma_\mu\mathcal{D}^\mu$ is the covariant derivative for the SM fields.}
\label{tab:scalar-ops}
\end{table}

Since we are imposing a $Z_2$ symmetry for ensuring the stability of the DM, hence any interaction term involving the DM fields has to be at least bilinear in $\chi$. Due to the fact that the Majorana fermion is its own anti-particle, those bilinears which are odd under charge conjugation vanish identically. We first chalk out the bilinears that can be formed out of the DM fields with Majorana nature in Table~\ref{tab:cc}. As we see, it is only possible to construct those operators which have scalar, axial vector and pseudoscalar interactions in the DM fields, while the vector current and dipole moments vanish. Since we are interested in UV freeze-in that requires the new physics at a scale $\Lambda\geq\rm TeV$, hence we can choose our EFT basis at the scale of $\Lambda$, and write all the operators below $\Lambda$. The generalised DM-SM non-renormalizable interaction in such case can be written as:

\bea
\mathcal{L}\supset \frac{c_{ij}\mathcal{O}^{(d)j}_{\rm SM}\mathcal{O}^{(d^{'})i}_{\rm DM}}{\Lambda_{ij}^{d+d^{'}-4}},
\label{eq:nonren-op}
\eea

where $\mathcal{O}^{(d^{'})}_{\rm DM}$ is a dark sector operator of mass dimension $d^{'}$, $\mathcal{O}^{(d)}_{SM}$ is the operator in the visible sector of mass dimension $d$. The parameter $\Lambda_{ij}$ is a dimensionfull scale and $c_{ij}$ is the dimensionless Wilson coefficient. If $d+d^{'}>4$, then the interaction is associated with an effective non-renormalizable operator of the form presented in Eq.~\eqref{eq:nonren-op}. Since a DM bilinear itself makes up dim. 3, hence we need to construct gauge invariant Lorentz contracted SM operators upto dim.5. Now, there are 13 dimension 4, 1 dimension 5, 63 dimension 6 and 20 dimension 7 operators invariant under the Standard Model gauge group~\cite{Lehman:2014jma}. Four of the dimension 6 operators violate baryon number conservation, leaving 59 that conserve baryon number~\cite{Lehman:2014jma,Grzadkowski:2010es}. As mentioned earlier, we consider SM operators upto dim.5 only which take part in DM-SM interactions, which amounts to a maximum dimension of eight for DM-SM operators. This not only limits the number of operators but also ensures the kinematics involved in scattering processes to be simple. All DM-SM interactions are encoded by higher-dimensional operators, with a cut-off scale $\Lambda$, which is the mass scale of the heavy fields integrated out to obtain the low-energy Lagrangian: 

\bea
\mathcal{L}=\mathcal{L}_{\text{SM}}+\mathcal{L}_{\text{DM}}+\mathcal{L}_5+\mathcal{L}_7+\mathcal{L}_8,
\eea

where $\mathcal{L}_{\text{SM(DM)}}$ is the renormalizable SM (DM) Lagrangian and $\mathcal{L}_d$ corresponds to the operators of dimension $d>4$.

In the renormalizable level the Lagrangian for the DM field has the  form:

\bea
\mathcal{L}_{\text{DM}} = i\overline{\chi^c}\slashed{\partial}\chi-M_\chi\overline{\chi^c}\chi, 
\eea

as the DM is electroweak singlet with zero hypercharge. In  Table~\ref{tab:scalar-ops} we list the possible operators that can be built out of SM and DM fields upto dimension 8. Generically, the EFT description is valid as long as $\Lambda\gtrsim\frac{M_\chi}{2\pi}$~\cite{Busoni:2013lha,Busoni:2014gta} and hence it is justified to integrate out the heavy fields with masses roughly of the order of the cut-off scale. But in case of UV freeze-in scenario, as we shall see, the reheat temperature of the universe is also involved which we consider to be $\gtrsim\text{TeV}$. Hence in our prescription the EFT framework is valid as long as $\Lambda\gtrsim\frac{M_\chi}{2\pi},T_{\text{RH}}$. The formulation of the EFT at scale $\mu$ depends on which degrees of freedom are relevant at that particular scale. The operators we have listed in Table~\ref{tab:scalar-ops} are in the basis of unbroken electroweak phase, valid at or above the electroweak scale $\mu=\mu_{\text{EW}}\sim m_Z$.

\section{Decay and annihilation processes for freeze-in}
\label{sec:process}

In this section we would like to specify all the decay and annihilation processes that can lead to the freeze-in production of the DM. As mentioned earlier, before EWSB, DM interacts with SM bath only via $n\to m$ scattering processes (with $n,m\geq 2$) arising out of higher dimensional operators (UV freeze-in) whereas in the post-EWSB phase there exists the possibility of IR freeze-in as well via decays. Therefore, we discuss DM interactions during pre- and post-EWSB phases separately in the following subsections.

\subsection{Before EWSB}
\label{sec:before-ewsb}

Here we are going to consider all processes that can arise before EWSB {\it i.e.,} at temperature $T>T_{EW}\simeq 160~\rm GeV$. We know, all SM particles are massless above $T_{EW}$ and the Goldstone bosons (GB) are physical fields. Hence, we define the $SU(2)_L$ scalar as:

\bea H=
\begin{pmatrix}
\phi^+ \\ \phi^0 
\end{pmatrix},
\label{eq:gb}
\eea

where we have both the charged and the neutral GBs. Now let us compute the possible processes one by one, according to the total mass dimension of DM-SM operator listed in Tab.~\ref{tab:scalar-ops}.

\subsubsection{Dimension 5 operator}
\label{sec:dim5}

The lowest dimension gauge-invariant operators that can be written down involving the interaction of a Majorana fermion DM and the SM sector are of dim.5 and involve the Higgs doublet bilinear:

\bea
\mathcal{O}_{S}^5 = \frac{1}{\Lambda}\overline{\chi^c}\chi \left(H^\dagger H\right), 
\label{eq:dim-5}
\eea

where the subscript indicates the nature of the dark operator, while the superscript denotes the total mass dimension. We shall use this convention throughout. Now, substituting Eq.~\eqref{eq:gb} in Eq.~\eqref{eq:dim-5} we obtain

\bea
\mathcal{O}_{S}^5 = \frac{1}{\Lambda}\overline{\chi^c}\chi\left(\phi^+\phi^-+\phi^0\overline{\phi^0}\right),
\label{eq:dim-5a}
\eea

which shows there are only 4-point interaction processes for the DM production from GB annihilation for dimension 5 operator. 

\subsubsection{Dimension 7 operator}
\label{sec:dim7}

The dim.7 DM-Sm operators involve dim.4 SM operators. As a result, in dimension 7 several different interactions emerge:

\bea\begin{split}
\mathcal{O}_S^7 &= \frac{1}{\Lambda^3}\overline{\chi^c}\chi\Bigg\{B_{\mu\nu}B^{\mu\nu}+W^a_{\mu\nu}W^{a\mu\nu}+G^a_{\mu\nu}G^{a\mu\nu}+B_{\mu\nu}\widetilde{B^{\mu\nu}}\\&+W^a_{\mu\nu}\widetilde{W^{a\mu\nu}}+G^a_{\mu\nu}\widetilde{G^{a\mu\nu}}+\left|H^\dagger H\right|^2+\left(\mathcal{D}_\mu H\right)^\dagger\left(\mathcal{D}^\mu H\right)\\&+\overline{l_L}He_R+\overline{Q_L}Hq_R+\overline{Q_L}\widetilde{H}q_R+i\overline{l_L}\slashed{\mathcal{D}}l_L+i\overline{Q_L}\slashed{\mathcal{D}}Q_L+i\overline{e_R}\slashed{\mathcal{D}}e_R+i\overline{q_R}\slashed{\mathcal{D}}q_R+h.c.\Bigg\},     
    \end{split}\label{eq:dim-7}
\eea
where the dual field strength tensor is defined as: $\widetilde{X_{\mu\nu}}=\epsilon_{\mu\nu\alpha\beta}X^{\alpha\beta}$, $\widetilde{H}=i\sigma^2 H^{\star}$ where $\sigma^a$ are the Pauli spin matrices. We define the covariant derivative for SM field: $\mathcal{D}_\mu=\partial_\mu-i g_2\tau^a W_\mu^a-i\frac{g_1}{2}Y B_\mu$, with $Q=T_{3L}+Y/2$ as the electromagnetic charge and $\tau^a=\sigma^a/2$ $(a=1,2,3)$. All gauge bosons have two degrees of freedom or in other words they are massless. Now, the non-abelian field strength tensors are defined as:

\bea
X_{\m\n}^a = \partial_\m X^a_\n-\partial_\n X^a_\m + g_X \epsilon^{abc} X_\m^b X_\n^c,
\eea

where $g_X$ is the appropriate coupling constant for the SM non-abelian gauge sector. The last term gives rise to self-interaction vertices involving three and four gauge bosons. Therefore, the gauge kinetic terms give rise to $2\to 2$, $3\to 2, 2 \to 3, 2 \to 4, 4 \to 2$ and $3\to 3$ scattering processes for DM production. 

Next is the term involving the Higgs doublet: $\left|H^\dagger H\right|^2$ that gives rise to the following interaction vertices upon expansion:

\bea
\frac{1}{\Lambda^3}\overline{\chi^c}\chi\left(\phi^+\phi^-\phi^+\phi^-
+\phi^0\overline{\phi^0}\phi^0\overline{\phi^0}+2\phi^+\phi^-\phi^0\overline{\phi^0}\right).
\eea

All of the above interactions, as one can see, give rise to $4\to 2, 2 \to 4, 3 \to 3$ scattering processes for DM production. The expansion of the scalar kinetic term $\left|\mathcal{D}_\mu H\right|^2$ before EWSB is given by Eq.~\eqref{eq:app-sca-kin-bewsb} in Appendix. The terms within the first parenthesis give rise to $2\to 2$ scattering for DM production involving GBs. Then we also have $4\to 2, 2 \to 4, 2\to3, 3\to2$ and $ 3 \to 3$ scattering for DM production involving gauge bosons and GBs. Note that the gauge bosons in this regime are massless, and hence have two degrees of freedom.

We then have the interactions involving SM Yukawa terms:

\bea\begin{split}
&\frac{1}{\Lambda^3}\overline{\chi^c}\chi\Bigg\{\overline{\nu_L}\phi^+e_R+\overline{e_L}\phi^0e_R+\overline{u_L}\phi^+ d_R+\overline{d_L}\phi^0 d_R+\overline{u_L}\overline{\phi^0}u_R-\overline{d_L}\phi^- u_R+h.c.\Bigg\}.    
    \end{split}
\eea

All these processes are $3\to 2, 2\to 3$ scattering processes for DM production involving both leptons and quarks. 

For the operators involving SM fermion kinetic terms we have different hypercharge for left and right-handed fermions. We use the following notation for generic fermion doublet and fermion singlet:

\bea
{SU(2)_L \; \text{doublet}}: \left(\psi_L,\xi_L\right),~{SU(2)_L  \;\text{singlet}}:\xi_R. 
\eea

With this we can now expand the corresponding interaction operator as:


\bea\begin{split}
&\frac{1}{\Lambda^3}\overline{\chi^c}\chi\Big\{...\Big\}_{\text{kin}}+\frac{1}{\Lambda^3}\overline{\chi^c}\chi\frac{1}{2}\Bigg\{\overline{\psi_L}\gamma^\m\left(g_2 W_\m^3+g_1 \frac{Y_L}{2} B_\m\right)\psi_L+g_2\overline{\psi_L}\gamma^\m\left(W_{1\mu}+i W_{2\m}\right)\xi_L\\&+\frac{g_1}{2}\overline{\xi}\gamma^\mu\frac{1}{2}\biggl[\underbrace{\left(Y_L+Y_R\right)}_{Y_T}-\gamma^5\underbrace{\left(Y_L-Y_R\right)}_{Y_D}\biggr]\xi B_\mu-g_2\overline{\xi_L}\gamma^\mu\xi_LW_\mu^3\Bigg\},     
    \end{split}\label{eq:ferm-kin-dim7}
\eea

where $\Big\{...\Big\}_{\text{kin}}=\left(\overline{\psi_L}\gamma^\m\partial_\m\psi_L+\overline{\xi_L}\gamma^\m\partial_\m\xi_L+\overline{\xi_R}\gamma^\m\partial_\m\xi_R\right)$ and $\xi=\xi_L+\xi_R$. $Y_{L,R}$ is the hypercharge corresponding to left and right-handed fermions. Note that, we have to consider both leptons and quarks in this case. Here the pure kinetic terms with ordinary derivatives give rise to $2\to 2$ processes, while others are $3\to 2, 2\to 3, 3\to3$ processes for DM production. This completes the  interactions involving dim.7 DM-SM operators. 

\subsubsection{Dimension 8 operator}
\label{sec:dim8}

Let us now examine the term involving dim.5 Weinberg operator, which leads to dim.8 DM-SM operator:

\bea
\mathcal{O}_S^8 = \frac{1}{\Lambda^4}\overline{\chi^c}\chi\left(\overline{\ell_L} \widetilde{H}\right)\left(\overline{\ell_L} \widetilde{H}\right),
\label{eq:dim-8}
\eea

where $\ell_L$ stands for SM leptons only, as operators with quarks are not invariant under SM colour symmetry. This is the only possible dimension 8 operator leading to interactions between DM and SM. Expansion of eq.~\eqref{eq:dim-8} gives rise to $4\to 2, 2\to 4, 3 \to 3$ processes for DM production before EWSB. 


\begin{figure}[htb!]
$$
\includegraphics[scale=0.45]{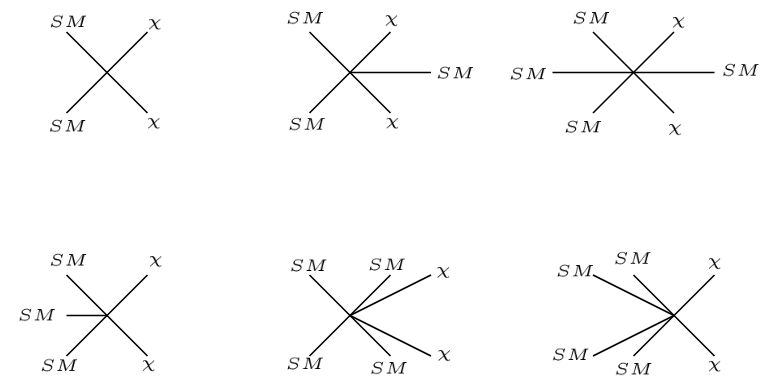}
$$
\caption{Possible $n\to2$ (with $n\geq 2$) annihilation channels for freeze-in production of the DM before EWSB corresponding to dim.5, dim.7 and dim.8 operators.} 
\label{fig:ann-bewsb}
\end{figure}

\begin{table}[htb!]
\begin{center}
\begin{tabular}{|c|c|c|c|c|c|c|c|c|c|c|}
\hline
Operator & $1\to n, n \geq 2$ & $2\to 2$ & $3\to 2, 2 \to 3$ & $4\to 2, 2 \to 4$ & $5\to 2, 2 \to 5$ & $6\to 2, 2 \to 6$ \\
type & & &  & $3\to 3$ & $3 \to 4, 4 \to 3$ & $3 \to 5, 5 \to 3$ \\
& & &  &  &  & $4 \to 4$ \\ [0.5ex] 
\hline\hline
$\mathcal{O}_S^5$ & \xmark & \cmark & \xmark & \xmark & \xmark & \xmark\\
$\mathcal{O}_S^7$ & \xmark & \cmark & \cmark & \cmark & \xmark & \xmark \\
$\mathcal{O}_S^8$ & \xmark & \xmark & \xmark & \cmark & \xmark & \xmark\\
\hline\hline
\end{tabular}
\end{center}
\caption{Annihilation/decay channels for DM-SM operators upto dimension 8 before EWSB.}
\label{tab:ann-bewsb}
\end{table}
%

Table \ref{tab:ann-bewsb} summarises possible $n\to m$ scattering processes leading to DM production arising from DM-SM operators of dim.5, 7, 8 before EWSB. The corresponding Feynman diagrams for all such processes are shown in figure \ref{fig:ann-bewsb}.

\subsection{After EWSB}
\label{sec:after-ewsb}

In this subsection we will investigate possible DM-SM interactions that can arise after EWSB {\it i.e.,} for $T<T_{EW}$. In this regime the Higgs doublet can be expanded around its vacuum expectation value (VEV) denoted by $v_h$ and can be expressed in unitary gauge as:

\bea H =
\begin{pmatrix}
0 \\ \frac{h+v_h}{\sqrt{2}}
\end{pmatrix},\label{eq:higgs}
\eea

where the Goldstone modes are being eaten up by the gauge bosons and they become massive with three degrees of freedom. 

Once again, we now proceed as in Sec.~\ref{sec:before-ewsb} to find possible decay/annihilation channels for DM production that can arise after EWSB. The dimension 5 operator in Eq.~\eqref{eq:dim-5} now can be rewritten as:

\bea
\frac{1}{\Lambda}\overline{\chi^c}\chi\left(h^2+2 h v_h+v_h^2\right),
\eea

where the first term is the usual 4-point interaction that is present before EWSB as well, while last term serves as another mass term for the DM, which leads to the resulting mass: $\overline{\chi^c}\left(M_\chi+\frac{2 v_h^2}{\Lambda}\right)\chi$, however the correction term is suppressed for large $\Lambda$. Notice that there is also a decay term which gives rise to IR freeze-in, proportional to a dimensionless effective coupling. 

\begin{figure}[htb!]
$$
\includegraphics[scale=0.45]{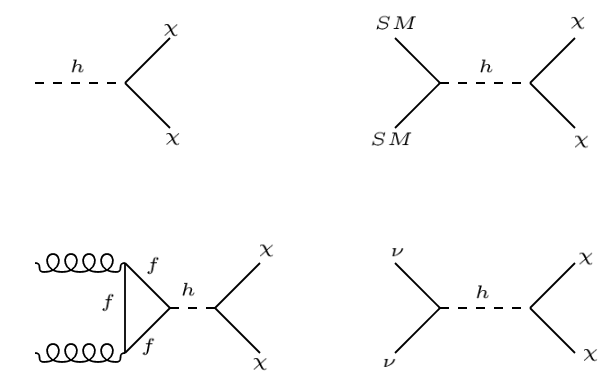}
$$
\caption{Possible $2\to2$ annihilation and decay channels for freeze-in production of the DM {\it after} EWSB corresponding to dim.5, dim.7 and dim.8 operators. The top left, top right and bottom left diagrams are present both in dim.5 and dim.7 cases, while the bottom right diagram is for dim.8. $SM$ in the top right diagram stands for all SM particles including fermions, gauge bosons and Higgs itself. In the bottom left diagram $f$ stands for SM fermions, however we only consider the contribution of top quark.}\label{fig:ann-aewsb}
\end{figure}

\begin{table}[htb!]
\begin{center}
\begin{tabular}{|c|c|c|c|c|c|c|c|c|}
\hline
Operator  & $1\to 2$ & $2\to 2$ & $3\to 2, 2 \to 3$ & $4\to 2, 2\to 4$  \\
type &  &  &  & $3\to 3$  \\[0.5ex] 
\hline\hline
$\mathcal{O}_S^5$ & \cmark & \cmark & \xmark & \xmark \\
$\mathcal{O}_S^7$ & \cmark & \cmark & \cmark & \cmark  \\
$\mathcal{O}_S^8$ & \xmark & \cmark & \cmark & \cmark  \\
\hline\hline
\end{tabular}
\end{center}
\caption{Annihilation/decay channels for DM-SM operators upto dimension 8 after EWSB.}
\label{tab:ann-aewsb}
\end{table}

For dimension 7 operators in eq.~\eqref{eq:dim-7} the gauge kinetic terms provide $2\to 2$, $3\to 2, 2\to 3$, $4\to 2, 2 \to 4$ and $3 \to 3$ processes as was the case before EWSB. The term involving $\left|H^\dagger H\right|^2$ can now be expanded as:

\bea\begin{split}
&\frac{1}{2\Lambda^3}\overline{\chi^c}\chi\left(h+v_h\right)^4=\frac{1}{2\Lambda^3}\overline{\chi^c}\chi\left(h^4+6 v_h^2 h^2+v_h^4+4 v_h^3 h+4 v_h h^3\right),     
    \end{split}
\eea

where the second term is a 4-point vertex proportional to the square of the VEV. The first term can give rise to $4\to 2$ process, whilst the last term is a $3\to 2$ process again proportional to the VEV. Depending upon the mass of DM, these same operators can also give rise to $3 \to 3, 2 \to 3$ scattering for DM production as well, if kinematically allowed at temperatures below EWSB. Lastly, there is again a decay process that gives rise to IR freeze-in, proportional to a dimensionless effective coupling. Now, after EWSB, the SM gauge bosons mix and give rise to physical fields as:

\bea
\begin{pmatrix}
B_\m \\ W_{3\m} 
\end{pmatrix}=\begin{pmatrix}
c_w & -s_w\\ s_w & c_w
\end{pmatrix}\begin{pmatrix}
A_\m \\ Z_\m
\end{pmatrix},
\eea

where $c(s)_w$ is the (co)sine of the Weinberg angle. As evident from Eq.~\eqref{eq:app-sca-kin-aewsb} in Appendix, the scalar kinetic term, after EWSB, gives rise to $h-Z(W)-Z(W)$, $h-h-Z(W)-Z(W)$, pure kinetic term for $h$, along with the mass term for the heavy gauge bosons as in Eq.~\eqref{eq:app-sca-kin-aewsb}. Therefore, the dimension 7 operator in Eq.~\eqref{eq:dim-7} shall consist of $2\to 2$, $3\to 2$ and $4\to 2$ vertices. All of such vertices have explicit gauge boson mass dependence and hence exist only after EWSB. The post-EWSB fermion kinetic term shall give rise to several charge and neutral current interactions involving SM leptons and quarks. All these are 3-point interaction vertices. Therefore, interactions like $\frac{1}{\Lambda^3}\overline{\chi^c}\chi\left(\overline{f}\slashed{\mathcal{D}}f\right)$ will produce $2\to2, 3\to 2,3\to3$ annihilation channels for DM production. The SM Yukawa interactions {\it viz.,} $H\bar{f}f$ result in 3-body vertices involving the Higgs and SM fermions. So, the dim.7 terms including SM Yukawa interactions shall produce $3\to 2,3\to3$ interactions for freeze-in. 

At dim.8 level, we have the Weinberg operator, which on expansion, after EWSB, gives rise to: $\frac{1}{2} \overline{\nu_L}~\overline{\nu_L} (h+v_h)^2$. Such an operator, therefore, gives a $2\to 2$ process proportional to $v_h^2$, a $3\to 2$ process proportional to the VEV and a $4\to 2$ suppressed by $\frac{1}{\Lambda^4}$. The post-EWSB production, being dominantly IR freeze-in, are sizeable only if DM mass is below the EWSB scale and the cut-off scale is not too high so that the dimensionless couplings proportional to $v_h/\Lambda$ remain sizeable enough. Since for IR freeze-in the abundance increases with increase in such couplings \cite{Hall:2009bx}, lowering the cut-off for a particular dimension of DM-SM operators will lead to increase in IR freeze-in contribution. All these annihilation and decay processes that arise after EWSB, are listed in Table \ref{tab:ann-aewsb} while the corresponding Feynman diagrams are shown in figure \ref{fig:ann-aewsb}.

\section{Dark matter yield from annihilation and decay}
\label{sec:dm-yld}

In this section we would like to compute the freeze-in yield of the DM $\chi$ before and after EWSB. Since the dark sector and visible sector in our case communicate via operators of different dimension suppressed by powers of some high scale $\Lambda$ as shown in Eq.~\eqref{eq:nonren-op}, it leads to the UV freeze-in scenario before EWSB. In this case the yield is completely determined by the cut-off scale $\Lambda$ and the reheat temperature $T_{\text{RH}}$ of the universe\footnote{Once inflation ends, the thermalisation of the universe occurs, leading to a radiation dominated phase. This is the reheating epoch~\cite{Allahverdi:2010xz}, which takes the universe to a radiation-dominated phase after the end of inflation.  Success of the big bang nucleosynthesis (BBN) puts a lower bound on the reheating temperature {\it i.e.} $T_{\text{RH}}\gtrsim \mathcal{O}(1)$ MeV \cite{deSalas:2015glj}.}. Before EWSB, all SM fields have zero mass, but the DM is still massive because of its bare mass, while after EWSB all the SM particles acquire masses. Since one can then expand the scalar field around its minima, the decay channels also appear along with the $n\to2$ annihilation processes. Also, before EWSB, as we have seen in the last section, there are no decay channels that lead to DM production. In order to determine the DM abundance at present temperature, we need to solve the Boltzmann equation (BEQ) to obtain the number density of $\chi$. The BEQs involved in this case are elaborated in Appendix.~\ref{sec:app-beq2to2}, ~\ref{sec:app-beq3to2} and~\ref{sec:app-beq4to2}.

\begin{figure}[htb!]
$$
\includegraphics[scale=0.42]{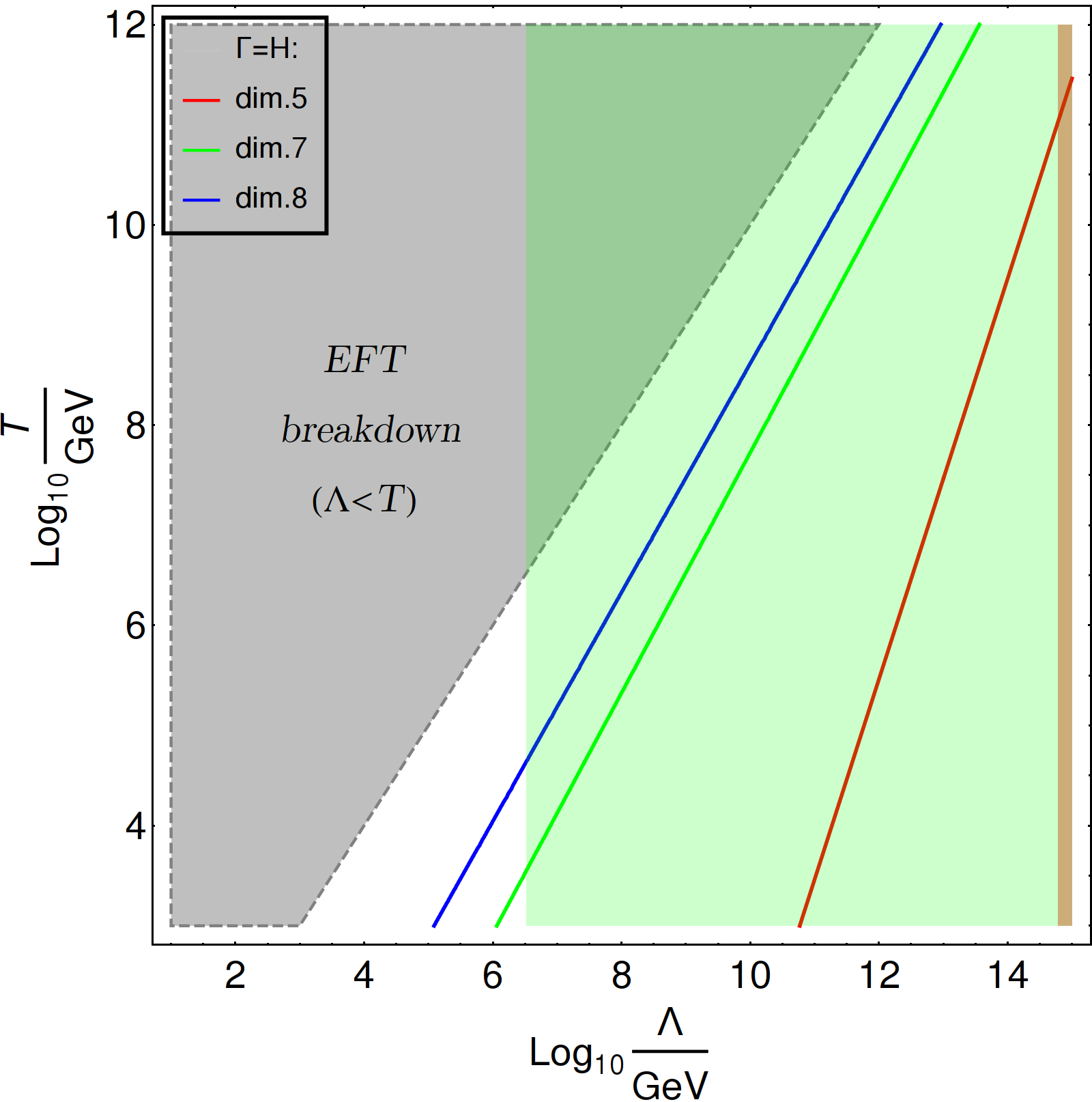}
$$
\caption{Along the contours with different colours the thermalisation condition $\Gamma_{\text{int}}=H$ is satisfied (see Appendix.~\ref{sec:app-therm-cond}), where all interaction rates are calculated for processes before EWSB. The grey shaded region is where EFT is no longer valid as $T>\Lambda$. The green shaded region satisfies light neutrino mass generated from L-violating dimension 7 SM operator, while the red shaded region satisfies light neutrino mass generated from L-violating dimension 5 SM operator.}\label{fig:rate}
\end{figure}

The decay processes after EWSB are dominantly $1\to 2$, where the Higgs decays to produce a pair of DM particles, if kinematically allowed. As one can understand, since the decay happens in the rest frame of the mother particle, hence the mass of the decaying particle is involved in this case ({\it i.e.,} the Higgs mass). As a consequence, decay always gives rise to IR freeze-in, where the yield does not depend on the reheat temperature, in contrast to standard UV freeze-in set-up. Also, since only Higgs decay is involved in our case, hence the DM mass $M_\chi$ is necessarily below $m_h/2\sim 62~\rm GeV$ in order to have non-zero contribution from decay. After EWSB there is also one gluon initiated process that can produce DM in the final state via Higgs mediation through a triangle loop (as shown in Fig.~\ref{fig:ann-aewsb}). The effective $ggH$ coupling has the form: $\frac{-ig_s^2}{32\pi^2}\frac{m_h^2}{v_h}\mathcal{F}\left(x\right)$~\cite{Buschmann:2014sia}, where $\mathcal{F}\left(x\right)=x\biggl[1+\left(1-x\right)f\left(x\right)\biggr]$ with
 \[
    f(x)=\left\{
                \begin{array}{ll}
                  \Bigg(\sin^{-1}\sqrt{\frac{1}{x}}\Bigg)^2,~~x>1\\
                  -\frac{1}{4}\Bigg({\text{ln}}\Biggl[\frac{1+\sqrt{1-x}}{1-\sqrt{1-x}}\Biggr]-i\pi\Bigg)^2~~x<1,\\
                  \end{array}
              \right.
  \]
and $x=\frac{4m_t^2}{m_h^2}$, where we are considering only top quark contribution in the triangle loop. Total yield due to annihilation and decay after EWSB is therefore coming from both IR and UV processes. 

Before going into the details of the Boltzmann equation (BEQ) for determining the DM relic abundance, we would first like to put a constraint on the cut-off scale $\Lambda$ such that the DM is out of equilibrium, ensuring its non-thermal production. In order to determine that, we need to calculate the scattering rate and compare it with the corresponding Hubble rate, the details of which can be found in Appendix.~\ref{sec:app-therm-cond}. In Fig.~\ref{fig:rate} we have shown the constraint on $\Lambda$ in the bi-dimensional plane of $T-\Lambda$ such that the DM-SM interaction is always out of equilibrium. In the plot, the straight line contours with different colours correspond to the condition $\mathcal{R}=\frac{\Gamma_{n\to2}}{H}=1$ for dimension 5 (red), 7 (green) and 8 (blue) operators. Here $H (T) = \sqrt{\frac{\pi^2}{90}g_{\star\rho}(T)} T^2/M_{pl}$ is the Hubble rate with $g_{\star\rho}(T)$ being the relativistic energy degrees of freedom at temperature T, and $M_{pl}$ is the Planck mass. The region to the left of each contour is where the DM thermalises with the SM bath. As the rate goes roughly as $\mathcal{R}\propto\frac{T M_{pl}}{\Lambda^n}$, hence the condition for thermalisation: $T.M_{pl}\gtrsim\Lambda^n$, also overlaps with the condition where the effective formalism breaks down: $T>\Lambda^n$. For operators with higher dimension the DM can be kept out of equilibrium for a smaller $\Lambda$ as $\mathcal{R}\sim\frac{1}{\Lambda^n}$, while lower dimensional operators need a larger $\Lambda$ to ensure non-thermal DM production. This is exactly reflected in Fig.~\ref{fig:rate}, where we see for dim.8 interaction $\Lambda\gtrsim 10^4~\rm GeV$ is the minimum cut-off scale that guarantees that the DM remains out of thermal equilibrium, while dim.5 demands $\Lambda\gtrsim 10^{10}~\rm GeV$.

The Boltzmann equation (BEQ) for the DM yield consists of all the processes that appear before and after EWSB, which includes decay and scattering diagrams. The yield is defined as the ratio of DM number density $n_\chi$ to the entropy density $s$: $Y_\chi=n_\chi/s$, where $s=\frac{2\pi^2}{45} g_{\star s}(T)T^3$ and $g_{\star s}$ is the relativistic entropy of degrees of freedom. BEQ corresponding to decay is derived in Appendix.~\ref{sec:app-beq-decay}, while those due to the scatterings are elaborated in Appendix.~\ref{sec:app-beq2to2}, ~\ref{sec:app-beq3to2} and~\ref{sec:app-beq4to2}. Now, the total yield at present epoch (at temperature $T_0$) is a sum of the contribution from yield before EWSB and yield after EWSB, which can be written as:

\bea\begin{split}
Y_\chi^\text{total} \left(T_0\right)_{T_{\text{RH}}>T_{\text{EW}}}&\simeq\Biggl[\Bigg\{\int_{T_\text{EW}}^{T_\text{RH}} dT\frac{1}{512\pi^6}\int_{T_\text{EW}}^{T_\text{RH}}\frac{dT}{s(T).H(T)}\int_{0}^\infty ds d\Omega \left(\frac{\sqrt{s}}{2}\right)^2 \overline{\left|\mathcal{M}\right|}^2_{12\to34}\\& \frac{1}{\sqrt{s}} K_1\left(\frac{\sqrt{s}}{T}\right)\Bigg\}+\Bigg\{\int_{T_\text{EW}}^{T_\text{RH}}\frac{dT}{s(T).H(T)}\frac{1}{64\left(2\pi\right)^7}\\&\int_0^\infty dss^{3/2}\overline{\left|\mathcal{M}\right|}^2_{123\to 45}K_1\left(\frac{\sqrt{s}}{T}\right)\int_0^1 dx_1 \int_{1-x_1}^1 dx_2\Bigg\}+\\&\Bigg\{\int_{T_\text{EW}}^{T_\text{RH}}\frac{dT}{s(T).H(T)}\frac{1}{64\left(2\pi\right)^9}\int_0^\infty ds \sqrt{s} \overline{\left|\mathcal{M}\right|}^2_{1234\to 56} K_1\left(\frac{\sqrt{s}}{T}\right)\\&\int_0^{\sqrt{s}} ds_{12}\int_0^{(\sqrt{s}-\sqrt{s_{12}})^2} ds_{34} \sqrt{1+\frac{s_{12}^2}{s^2}-\frac{2 s_{12} s_{34}}{s^2}+\frac{s_{34}^2}{s^2}-\frac{2 s_{12}}{s}-\frac{2 s_{34}}{s}}\\&\int\frac{d\cos\theta_{12}}{2} \int \frac{d\cos\theta_{34}}{2}\Bigg\}\Biggr]+\Biggl[\int_{T_\text{0}}^{T_\text{EW}} dT\frac{m_h^2 \Gamma_{h\to\chi\chi}}{2\pi^2}\frac{K_1\left(m_h/T\right)}{s(T).H(T)}+\\&\frac{1}{512\pi^6}\int_{T_\text{0}}^{T_\text{EW}}\frac{dT}{s(T).H(T)}\int_{\text{max}\left(4M_\chi^2,4m_{\text{SM}}^2\right)}^\infty ds d\Omega\frac{1}{4}\sqrt{\left(s-4m_{\text{SM}}^2\right)\left(s-4M_\chi^2\right)}\\&\overline{\left|\mathcal{M}\right|}^2_{12\to34} \frac{1}{\sqrt{s}} K_1\left(\frac{\sqrt{s}}{T}\right)\Biggr] 
    \end{split}
    \label{eq:totyld}
\eea

\begin{figure}[htb!]
$$
\includegraphics[scale=0.4]{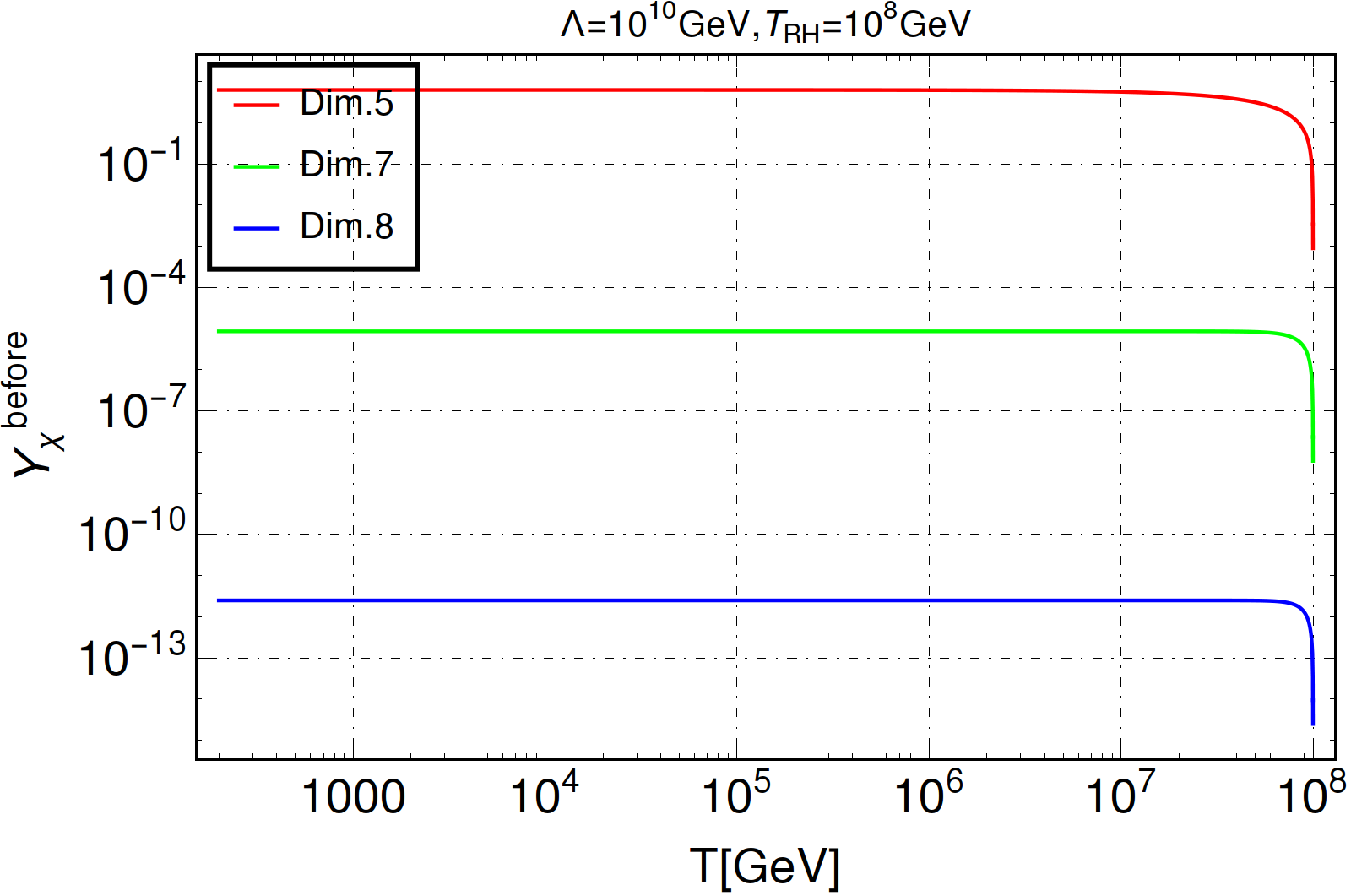}~~
\includegraphics[scale=0.4]{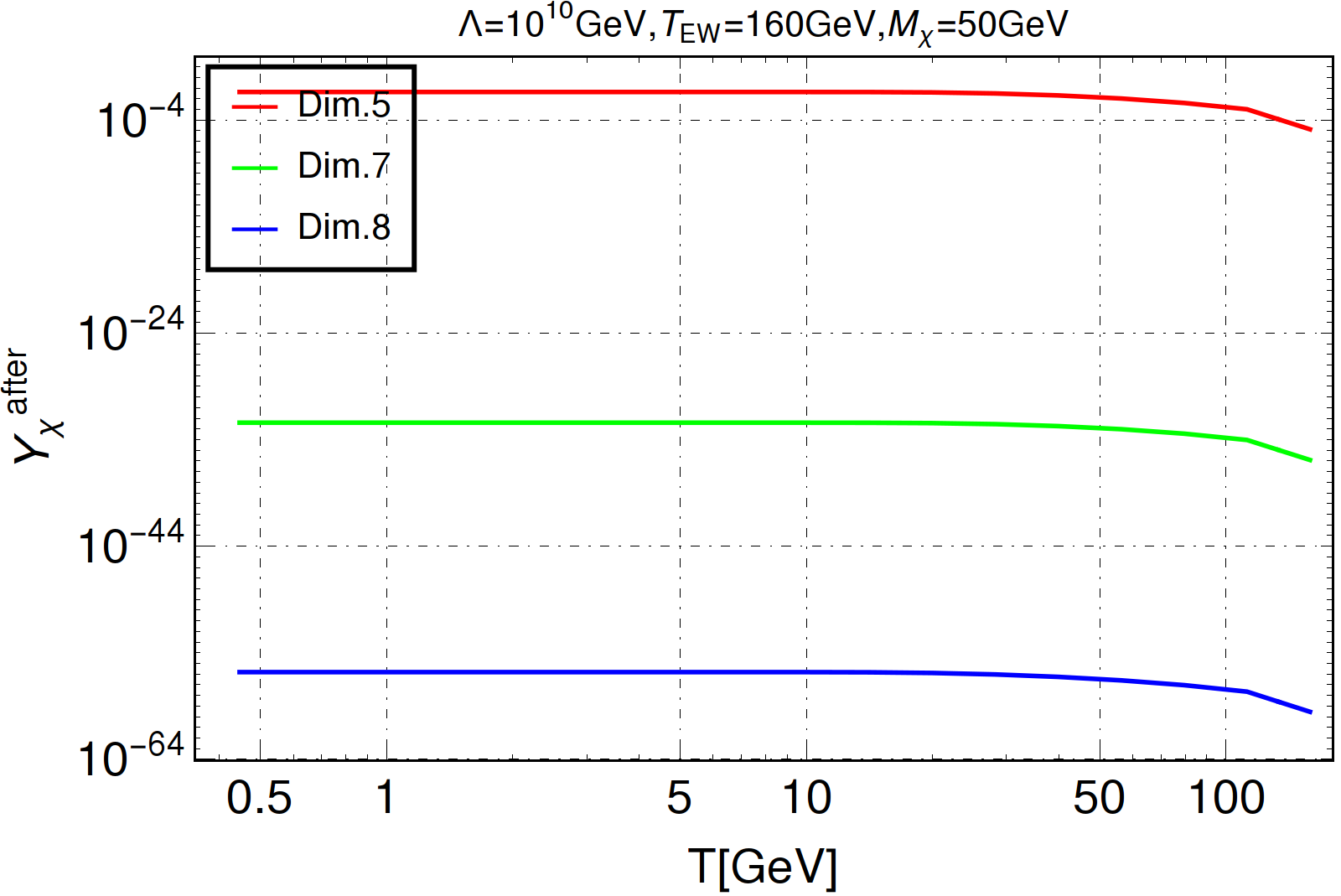}
$$
\caption{Contribution of operators with different dimensions to the DM yield before (left) and after (right) EWSB. In both cases we have chosen $\Lambda=10^{10}~\text{GeV}, T_{\text{RH}}=10^8~\rm GeV$.}\label{fig:yld-dim}
\end{figure}

\begin{figure}[htb!]
$$
\includegraphics[scale=0.4]{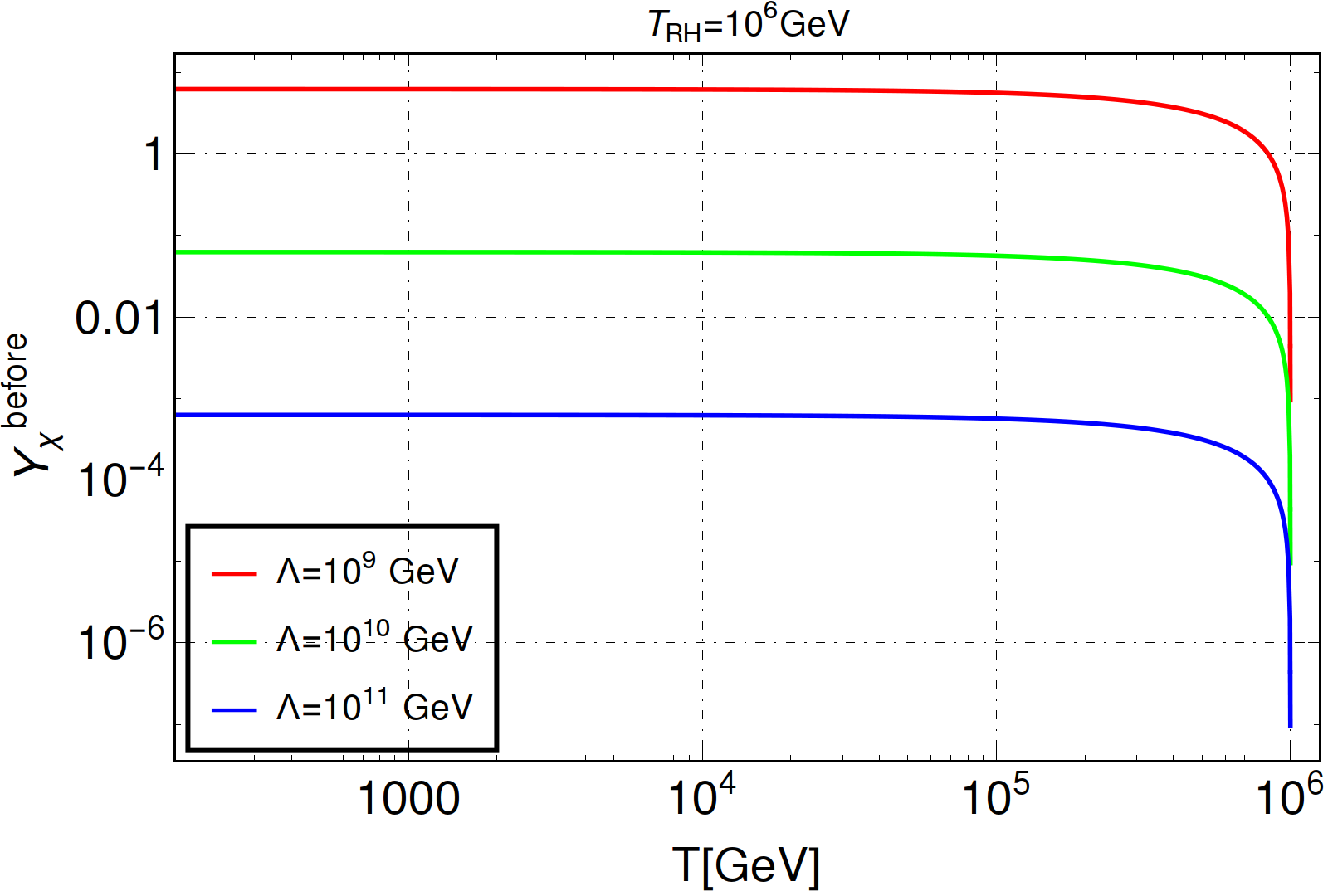}~~
\includegraphics[scale=0.4]{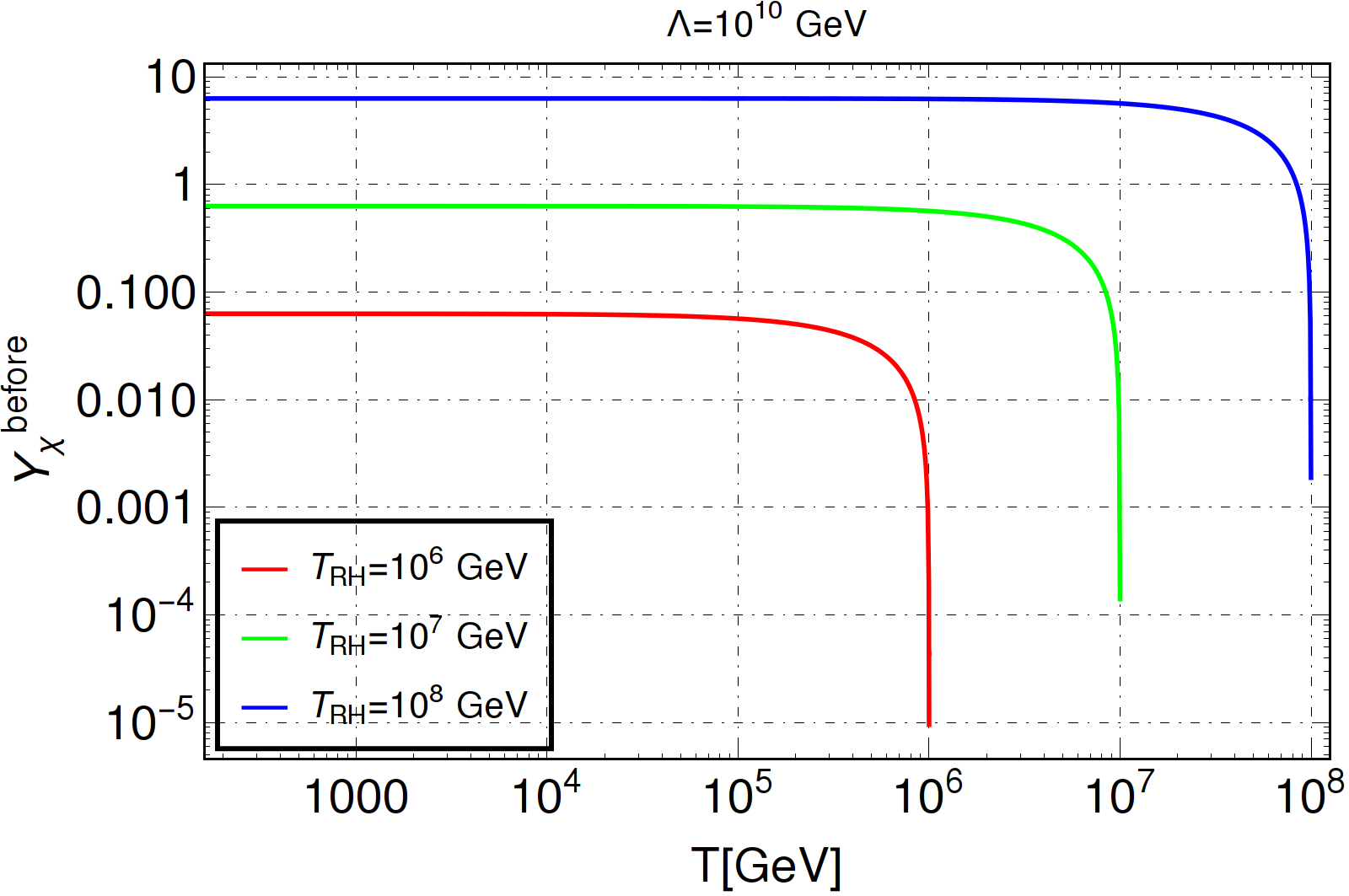}
$$
$$
\includegraphics[scale=0.32]{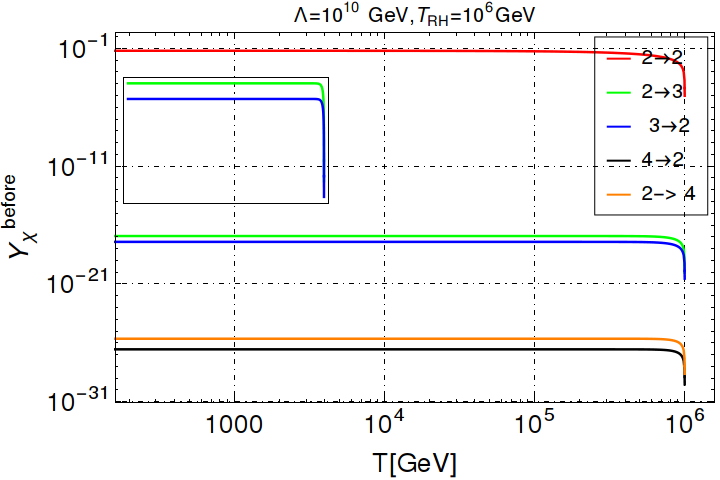}~~
\includegraphics[scale=0.4]{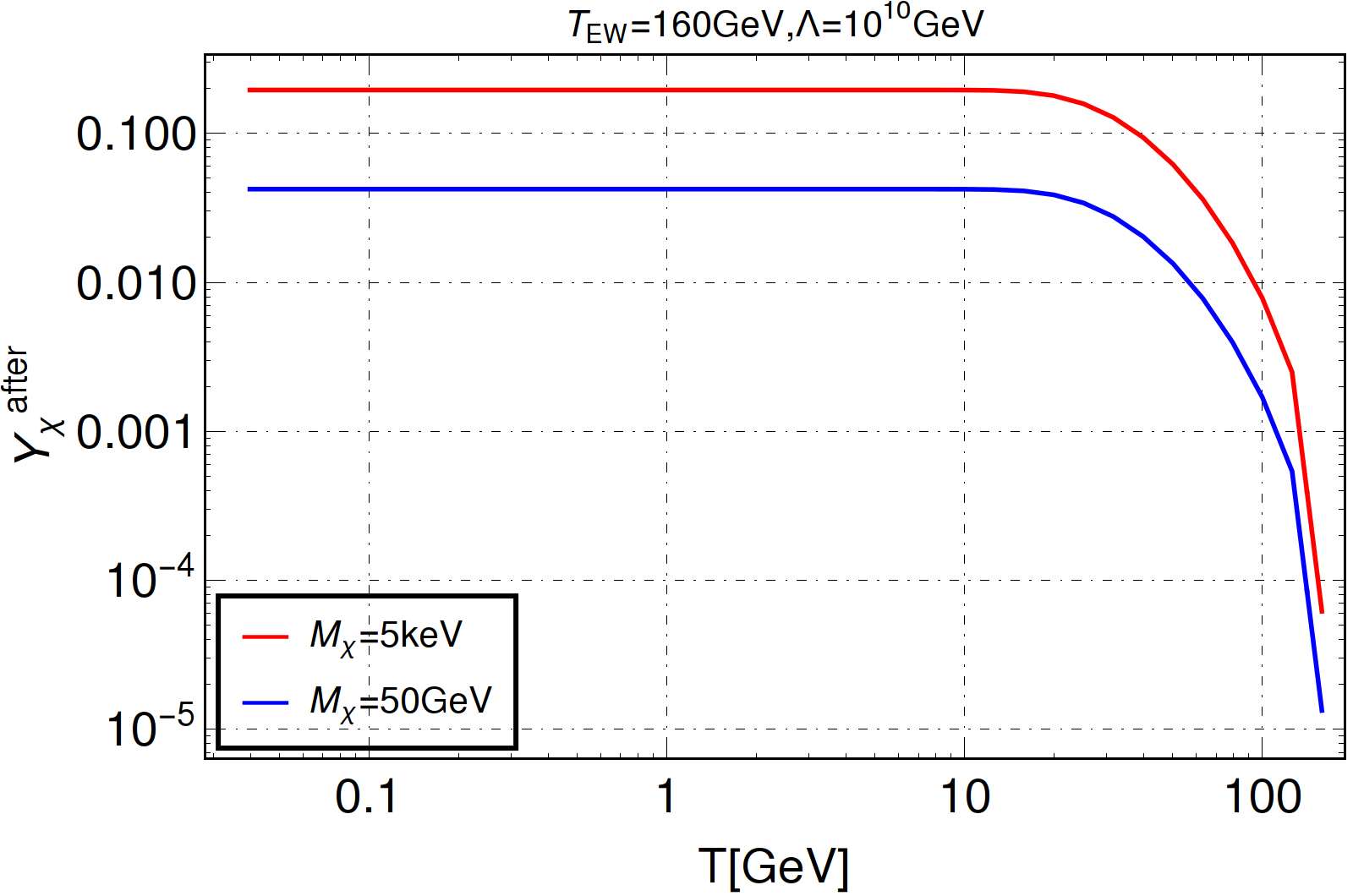}
$$
\caption{Top Left: DM yield before EWSB considering all $2\to2$, $2\to3$, $3\to2$ and $4\to2$ channels for a fixed reheat temperature $T_{\text{RH}}=10^6~\rm GeV$, where different coloured curves correspond to different choices of $\Lambda:\{10^9,10^{10},10^{11}\}~\rm GeV$ shown in red, green anbd blue respectively. Top Right: DM yield before EWSB for a fixed $\Lambda=10^{10}~\rm GeV$ for different choices of the reheat temperature $T_{\text{RH}}:\{10^6,10^{7},10^{8}\}~\rm GeV$ shown in red, green and blue respectively. Bottom Left: Individual contribution of $2\to2$, $2\to3$, $3\to2$ and $4\to2$ processes in DM yield for a fixed $\Lambda=10^{10}~\rm GeV$ and reheat temperature $T_{\text{RH}}=10^6~\rm GeV$. Bottom Right: Variation of DM yield with temperature considering contribution from $1\to2$ decay and $2\to2$ annihilations only after EWSB. All states are considered to be massive with DM mass $M_\chi=5~\rm keV$ (red) and $M_\chi=50~\rm GeV$ (blue).}\label{fig:yld-bewsb}\label{fig:yldplots}
\end{figure}

where the first big parenthesis $\biggl[...\biggr]$ takes care of the yield before EWSB that includes contributions from $n\to m$ processes. The second big parenthesis $\biggl[...\biggr]$ includes contribution from processes after EWSB due to $1\to2$ decay and $2\to2$ annihilations. Also note that in case of yield after EWSB the lower limit of the integration depends on the mass of the particles involved in the process. One can then obtain the relic abundance of the DM at present epoch using:

\bea
\Omega_\chi h^2 = 2.75\times 10^8\frac{M_\chi}{\text{GeV}} Y^{\text{total}}_\chi\left(T_0\right),\label{eq:dm-relic}
\eea

which need to satisfy the PLANCK~\cite{Aghanim:2018eyx} observed limit: $\Omega_{\text{DM}} h^2=0.120\pm 0.001$. 

The contributions of operators with different dimensions to the DM yield (before and after EWSB) are shown in Fig.~\ref{fig:yld-dim}, where in both the plots the red, green and blue curves correspond to dim.5, dim.7 and dim.8 operators respectively. Although the final yield, as evident from Eq.~\eqref{eq:totyld}, is a sum of the yield before and after EWSB, this exercise helps us to understand the dynamics of the DM yield with the bath temperature before and after EW symmetry breaking occurs. This also indicates which processes dominate over the others. Here we see, dim.5 interactions always have dominant contribution over the others both before and after the EWSB. This leads us to the fact that dim.5 interactions play the deciding role in determining the total yield as well as the DM relic abundance. In Fig.~\ref{fig:yldplots} we have illustrated how the DM yield varies with the bath temperature before and after EWSB separately when all the operators with different dimensions are considered together. In the top left panel of Fig.~\ref{fig:yldplots} we have shown the variation of DM yield $Y_\chi$ with temperature $T$ considering all $n\to2$ channels before EWSB for a fixed reheat temperature $T_{\text{RH}}=10^6~\rm GeV$ for illustration. With the change in the effective scale $\Lambda$ the yield also changes as shown by the red, green and blue curves corresponding to $\Lambda=\{10^9,10^{10},10^{11}\}~\rm GeV$ respectively, and as expected, for larger $\Lambda$ the yield is small. As the reheat temperature is fixed, hence all the curves originate from the same point at high temperature and the yield becomes maximum at $T\sim T_{\text{RH}}$. The yield freezes-in immediately $T\sim T_{\text{RH}}$, which is a typical feature of UV freeze-in. Since we are considering the era before EWSB, all SM particles are massless, but the DM, because of its bare Majorana mass, is still massive. However, the yield is very loosely dependent on the DM mass because of the involvement of two large scales in the theory, namely the cut-off scale and the reheat temperature. As a result ignoring the DM mass does not change the outcome. In the top right panel of Fig.~\ref{fig:yldplots} we have again shown how the DM yield before EWSB varies with the temperature for a fixed choice of the effective scale $\Lambda=10^{10}~\rm GeV$. In this case we choose three different reheat temperature: $\{10^6,10^7,10^8\}~\rm GeV$ to illustrate the effects on $Y_\chi$. As we can notice, with the change in $T_{\text{RH}}$ the upper limit of the integration in the first parenthesis of  Eq.~\eqref{eq:totyld} changes, resulting in the change in corresponding yield. For larger $T_{\text{RH}}$ we achieve a larger yield following Eq.~\eqref{eq:totyld}. Also, all the curves originate from different $T$ with the change in $T_{\text{RH}}$, but the flavour of UV freeze-in prevails as the yield in each case is maximum at $T\sim T_{\text{RH}}$. Note that, all these curves end before the electroweak phase transition temperature $T_{EW}\simeq 160~\rm GeV$ ensuring DM production only before EWSB era and hence dominance of UV freeze-in. In the bottom left panel of Fig.~\ref{fig:yldplots} we illustrate contribution from different $n\to2$ processes, where the red curve is due to $2\to2$, the blue and green curves are for $3\to2$ and $2\to3$ and the black curve is due to $4\to2$ processes. As expected, the $2\to2$ processes dominate over all the others. Although the $2\to3$ and $3\to2$ processes almost overlap on each other, but a close scrutiny (see inset) shows that $2\to3$ processes are more relevant than $3\to2$, while $4\to2$ processes are the most suppressed ones. The $3\to3$ processes are also sub-dominant in the presence of $2\to3$ and $3\to2$ processes, hence we do not show them here. Finally, in the bottom right panel of Fig.~\ref{fig:yldplots} we have shown the yield after EWSB when all SM particles are considered to be massive along with the DM. We consider the IR dominated $1\to2$ decay and all $2\to2$ annihilation channels that lead to DM pair production. Since all the states are massive, the $2\to2$ channels dominate over other $n\to2$ channels for $n>2$. As a result, we only consider the decay and $2\to2$ annihilation processes in this regime. Here we show the variation of yield for DM mass $M_\chi:\{5~\text{keV},50~\text{GeV}\}$. For both the cases the $h\to\chi\chi$ channel also contributes, while for $M_\chi>m_h/2$ only $2\to2$ annihilation channels contribute. Due to the absence of decay modes the yield after EWSB is negligibly small for DM masses larger than the EWSB scale{\footnote{For exmple, the yield after EWSB corresponding to a DM of mass 500 GeV is $\sim 10^{-13}$ at $T\simeq 0.1~\rm GeV$.}}. For a DM with mass $\sim\text{keV}$ the yield saturates at $T\sim 10~\text{GeV}$, much earlier than BBN. This implies that there is no damping in the matter power spectrum due to late DM formation, which otherwise puts a very strong bound on the DM mass~\cite{Irsic:2017ixq,Murgia:2018now}. Note that, in the presence of decay, the yield after EWSB is comparable with that before EWSB. Note here, due to the effect of IR freeze-in, the DM freezes in when $x=m/T\sim\mathcal{O}(1)$. Also note that these yields do not correspond to the right relic abundance, rather these are just to illustrate how the DM yield builds up with the temperature.

\begin{figure}[htb!]
$$
\includegraphics[scale=0.35]{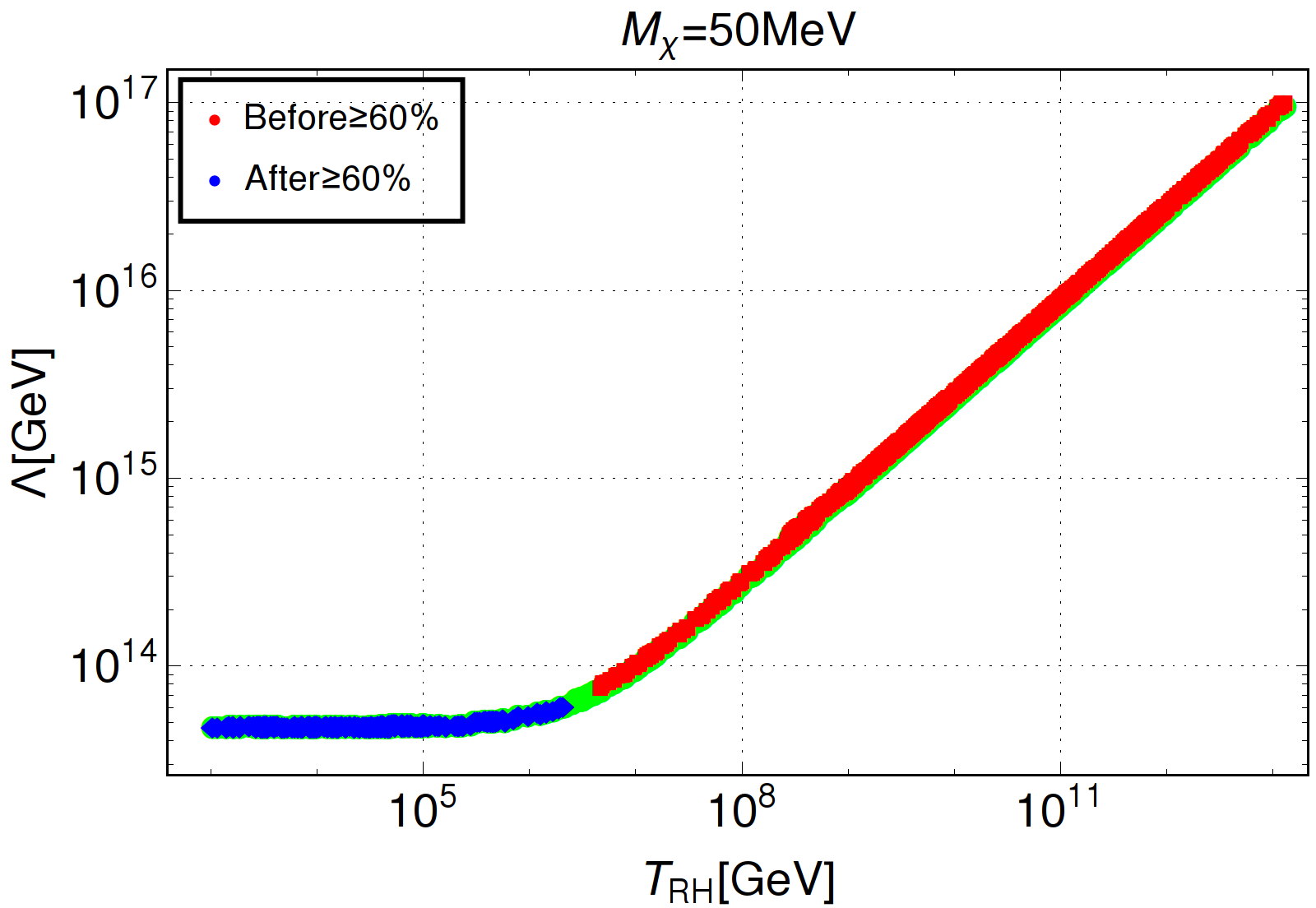}~~
\includegraphics[scale=0.35]{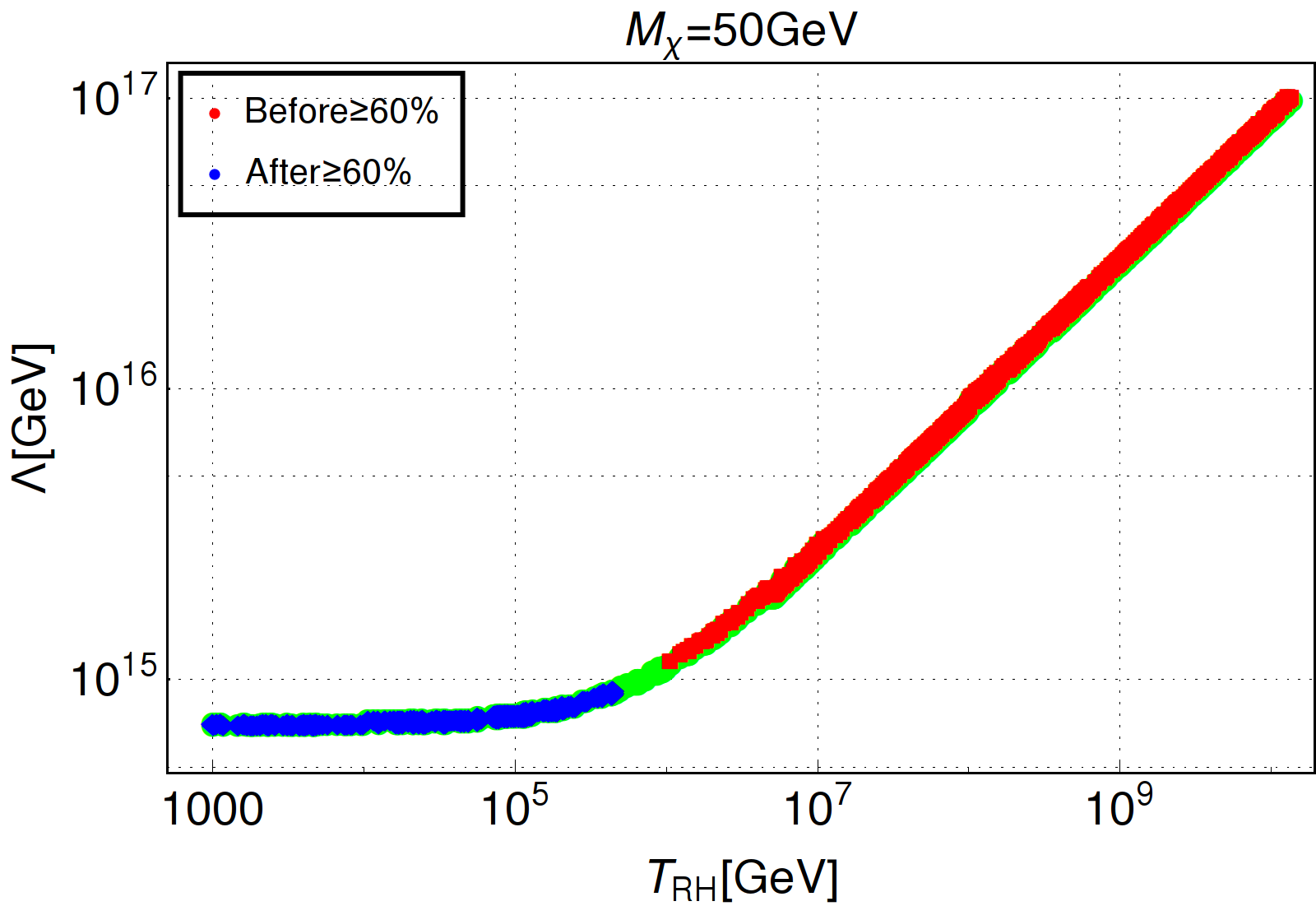}
$$
$$
\includegraphics[scale=0.35]{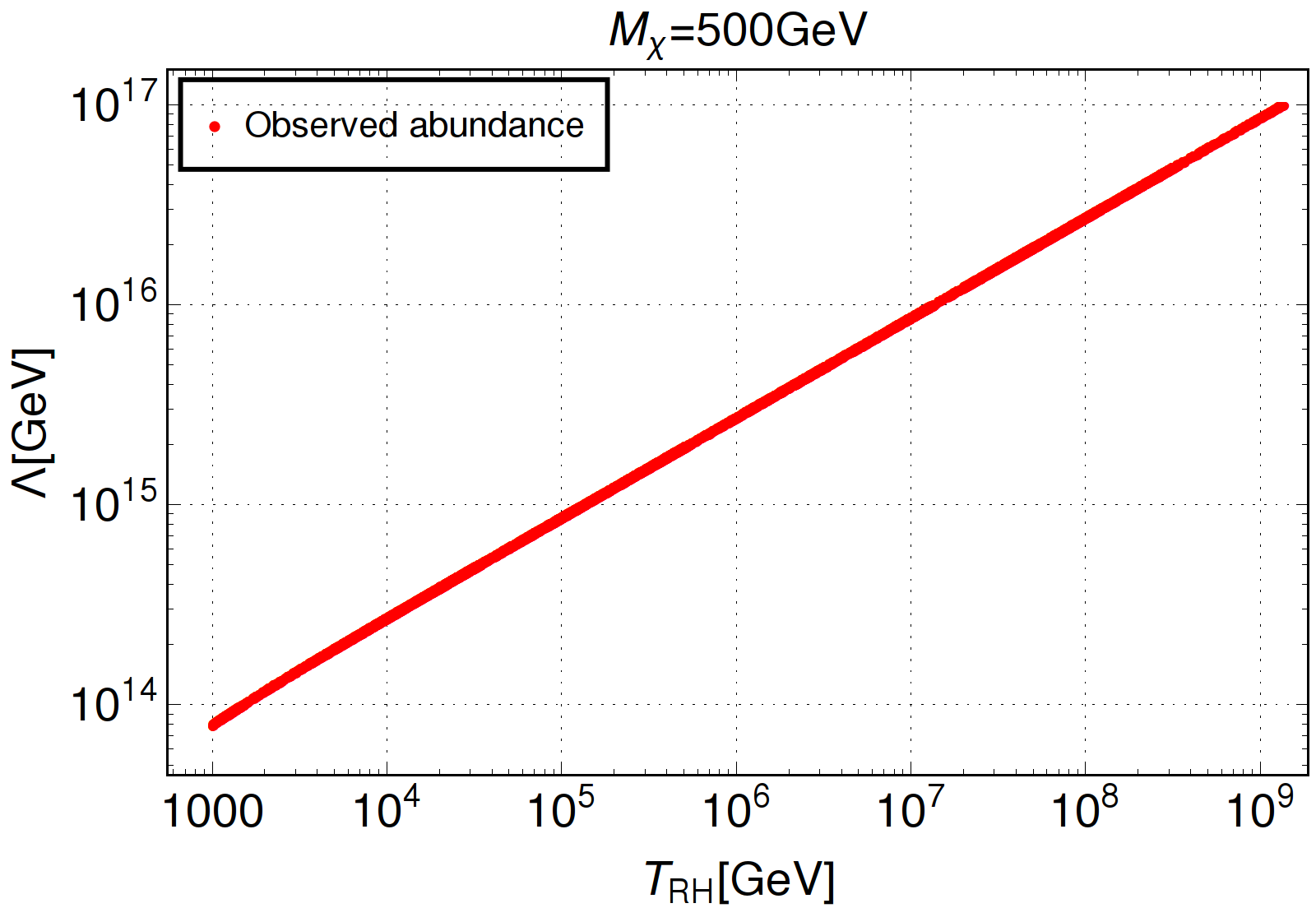}~~
\includegraphics[scale=0.35]{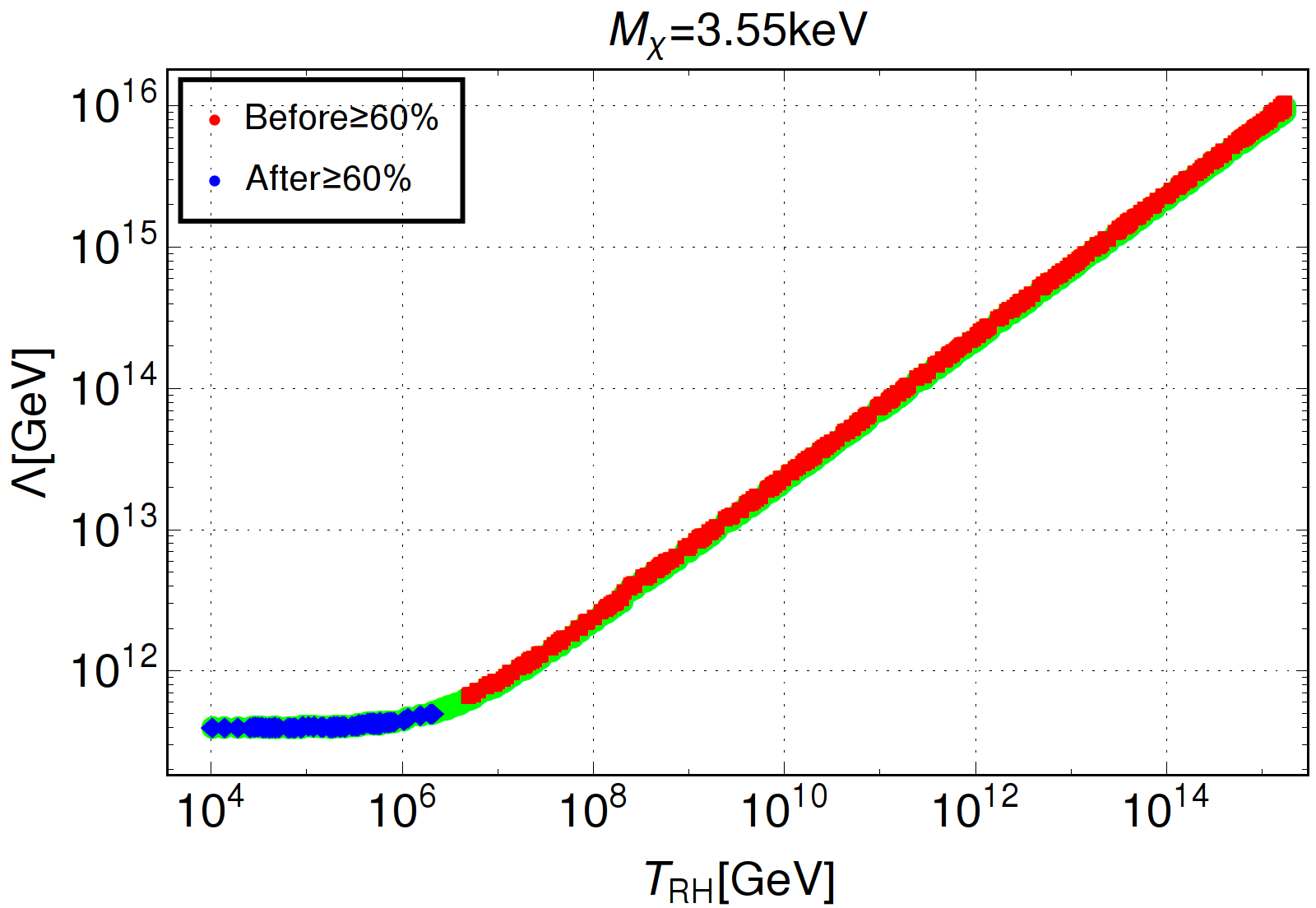}
$$
\caption{Top Left: Constraints on $\Lambda$ and $T_{\text{RH}}$ from the observed relic abundance for DM mass of 50 MeV. Here all the green colored points satisfy the PLANCK observed relic density. The red points have dominant contribution from before EWSB (UV) while the blue points have dominant contribution from after EWSB (IR) processes. Top Right: Same as top left but for DM mass of 50 GeV. Bottom Left: For 500 GeV DM mass the red points satisfy observed relic abundance. Here all contribution comes from before EWSB channels. Bottom Right: Observed relic abundance for 3.55 keV DM in the bi-dimensional plane of $\Lambda-T_{\text{RH}}$, where the colour codes are same as the top panel plots.}\label{fig:relic}
\end{figure}

\begin{figure}[htb!]
$$
\includegraphics[scale=0.45]{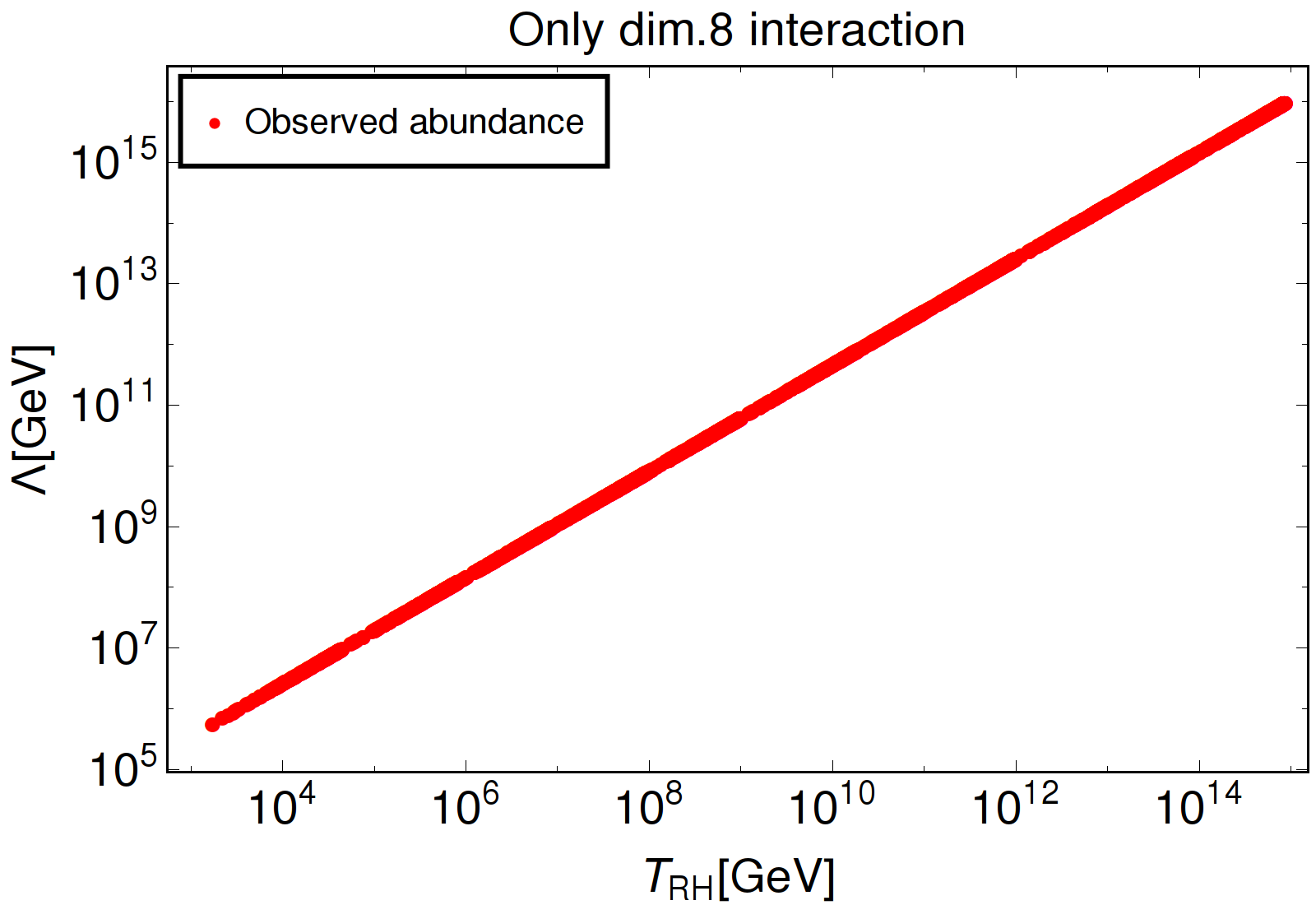}
$$
\caption{Parameter space satisfying relic abundance with only dim.8 interactions taken into account. DM mass is assumed to be $M_\chi=50~\rm GeV$.}\label{fig:dimRelic0}
\end{figure}

In Fig.~\ref{fig:relic} we have shown the parameter space satisfying PLANCK observed relic abundance in $\Lambda-T_{\text{RH}}$ plane for different choices of the DM mass. The relic abundance is computed using Eq.~\eqref{eq:dm-relic} with all the dimensions taken into account together. We choose DM masses  $M_\chi:\{50~\text{MeV},50~\text{GeV},500~\text{GeV}\}$ (clockwise from top left) for illustrating the resulting parameter space. We also have shown that a DM mass of 3.55 keV satisfies the desired relic abundance in the bottom right panel of Fig.~\ref{fig:relic}. DM with mass in that ballpark has received lots of attention in the context of the 3.55 keV $X$-ray observation~\cite{Bulbul:2014sua,Boyarsky:2014jta,Boyarsky:2014ska,Mambrini:2015nza,Cappelluti:2017ywp,Cappelluti:2017ywp}. In the top left panel of Fig.~\ref{fig:relic} we show the region satisfying relic abundance for a DM mass of 50 MeV. Here all points satisfy the observed relic abundance. For DM mass $M_\chi<m_h/2$, the $h\to\chi\chi$ channel plays the important role as it enhances the DM yield after EWSB. Since after EWSB IR freeze-in dominates (as all the states are massive), hence we see a part of the parameter space independent of the reheat temperature. This corresponds to $\Lambda\sim 10^{13}~\rm GeV$ for $M_\chi=50~\rm MeV$ and $\Lambda\sim 10^{15}~\rm GeV$ for $M_\chi=50~\rm GeV$ (top right panel). Beyond $T_{\text{RH}}\sim 10^7~\rm GeV$, $\Lambda$ rises linearly with $T_{\text{RH}}$ for $M_\chi=50~\rm MeV$ as UV freeze-in starts contributing. For $M_\chi=50~\rm GeV$ this linear rise starts at  $T_{\text{RH}}\sim 10^6~\rm GeV$ (top right panel). We have also shown the percentage contribution to the observed relic abundance coming from processes before and after EWSB in the top panel plots and also in the bottom right panel. Here we see more than 60\% contribution to observed relic abundance from processes before EWSB comes at a larger reheat temperature (red points). This means UV freeze-in is more effective for larger $T_{\text{RH}}$. This is understandable as from Eq.~\eqref{eq:totyld} we see that $Y_\chi^\text{UV}\propto T_{\text{RH}}$ as we are considering $T_{\text{EW}}<<T_{\text{RH}}$. Now, a larger $T_{\text{RH}}$ also calls for a larger $\Lambda$ to satisfy the relic abundance as from Eq.~\eqref{eq:dm-relic} we see that for a fixed DM mass $\Omega_\chi\propto\frac{T_{\text{RH}}}{\Lambda^n}$. Contribution from processes after EWSB is more profound at a lower reheat temperature where IR freeze-in dominates (blue points). As explained before, in that region the relic abundance is almost independent of the reheat temperature, which is again attributed to Eq.~\eqref{eq:totyld}, where we see $Y_\chi^\text{IR}\propto T_{\text{EW}}$ as $T_0<<T_{\text{EW}}$. For DM mass $\leq T_{\text{EW}}$ this pattern remains the same as evident from the top panel and bottom right panel plots. On the other hand, for DM mass of 500 GeV (left bottom panel of Fig.~\ref{fig:relic}) all of the contribution comes from processes before EWSB (as $M_\chi>T_{\text{EW}}$) where only UV freeze-in is in action. As a result, there is a linear rise of $\Lambda$ with the increase in reheat temperature throughout. Note that, in this case, the reheat temperature can also be $\sim\text{TeV}$ as for a massive DM one has to dial down the reheat temperature in order to satisfy the relic abundance since $\Omega_\chi\propto\frac{M_\chi T_{\text{RH}}}{\Lambda^n}$. This is also reflected in the other plots where we see for a lighter DM one requires a larger $T_{\text{RH}}$ to obtain the right abundance. One important point to note here is the fact that in presence of all the operators, dim.5 interactions dominate over the others which is understandable from the $\frac{1}{\Lambda}$ suppression compared to other dimensions where the suppression is even stronger. 

In Fig.~\ref{fig:relic} we have considered possible operators upto dim.8 where naturally dim.5 operators dominate. We have taken all operators at the same time to analyse the parameter space allowed by relic abundance. The required cut-off scale in the vertical axes of the plots shown in Fig.~\ref{fig:relic} clearly depicts the dominance of dim.5 operator. However, depending on the UV completion, only one such dimension may be allowed for DM-SM operators. For a comparison, we show the corresponding scanned plots for a scenario where DM-SM interaction occurs only through dim.8 scalar operators in Fig.~\ref{fig:dimRelic0}. While the correlation of the reheat temperature and the cut-off scale remains the same as in Fig.~\ref{fig:relic}, but the required $\Lambda$ becomes substantially small as one can expect. Another interesting observation is that, as we go to higher dimensional operators for DM-SM interactions beyond dim.5, the IR freeze-in contribution to DM relic becomes more and more negligible, specially when reheat temperature of the universe is kept well above 1 TeV.

\section{Connection to neutrino mass}
\label{sec:nu-mass}

Within the SM field content it is possible to generate light neutrino Majorana mass via operators of different mass dimension that violate lepton number by two units $(\Delta L=2)$~\cite{Babu:2001ex,deGouvea:2007qla,Angel:2012ug,Cepedello:2017eqf,Gargalionis:2019drk,Herrero-Garcia:2019czj} and are suppressed by some scale $\Lambda_\nu$. Thus, a more natural explanation for the smallness of $m_\nu$ is that they are generated (via some underlying new physics) at a scale $\Lambda_\nu$ (higher than the electroweak scale), and manifest
themselves at low energies through effective higher dimensional operators. As we know, $d=5$ seesaw operators are the lowest dimensional effective neutrino mass operators. Now, for such a dim.5 operator one can express the light neutrino mass in terms of the Higgs VEV and the effective scale $\Lambda_\nu$ at which the lepton number is broken:

\bea
m_\nu\simeq\frac{v_h^2}{\Lambda_\nu}\label{eq:dim5-nu},
\eea

which indicates that in order to generate light neutrino mass in the right ballpark the scale $\Lambda_\nu\gtrsim 10^{11}~\rm TeV$. Similarly, for neutrino mass generated from lepton number violating SM operators in dim.7, one can write: 

\bea
m_\nu\simeq\frac{v_h^4}{\Lambda_\nu^3}\label{eq:dim7-nu},
\eea

which gives rise to $\Lambda_\nu\gtrsim 10^3~\rm TeV$ in order to get neutrino mass in the desired ballpark. Now, a non-zero neutrino mass can be generated only after EWSB, while DM relic abundance has contribution both from processes before after EWSB. In order to make a connection between the freeze-in scale and the scale at which light neutrino mass can be generated, we compare the freeze-in scale considering only dim.5 and dim.7 interactions where it is also possible to generate neutrino mass via SM operators. 

\begin{figure}[htb!]
$$
\includegraphics[scale=0.42]{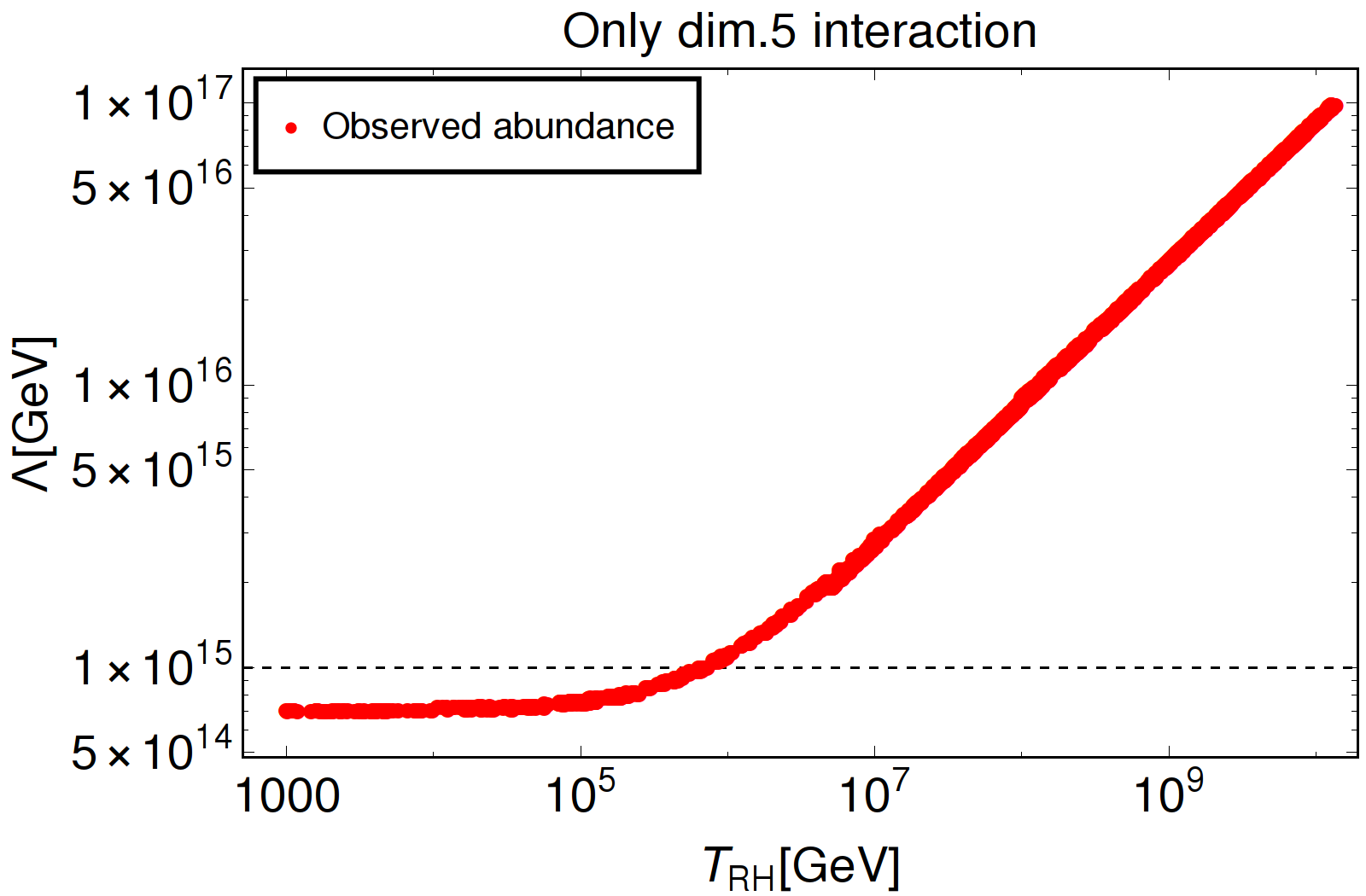}~~
\includegraphics[scale=0.4]{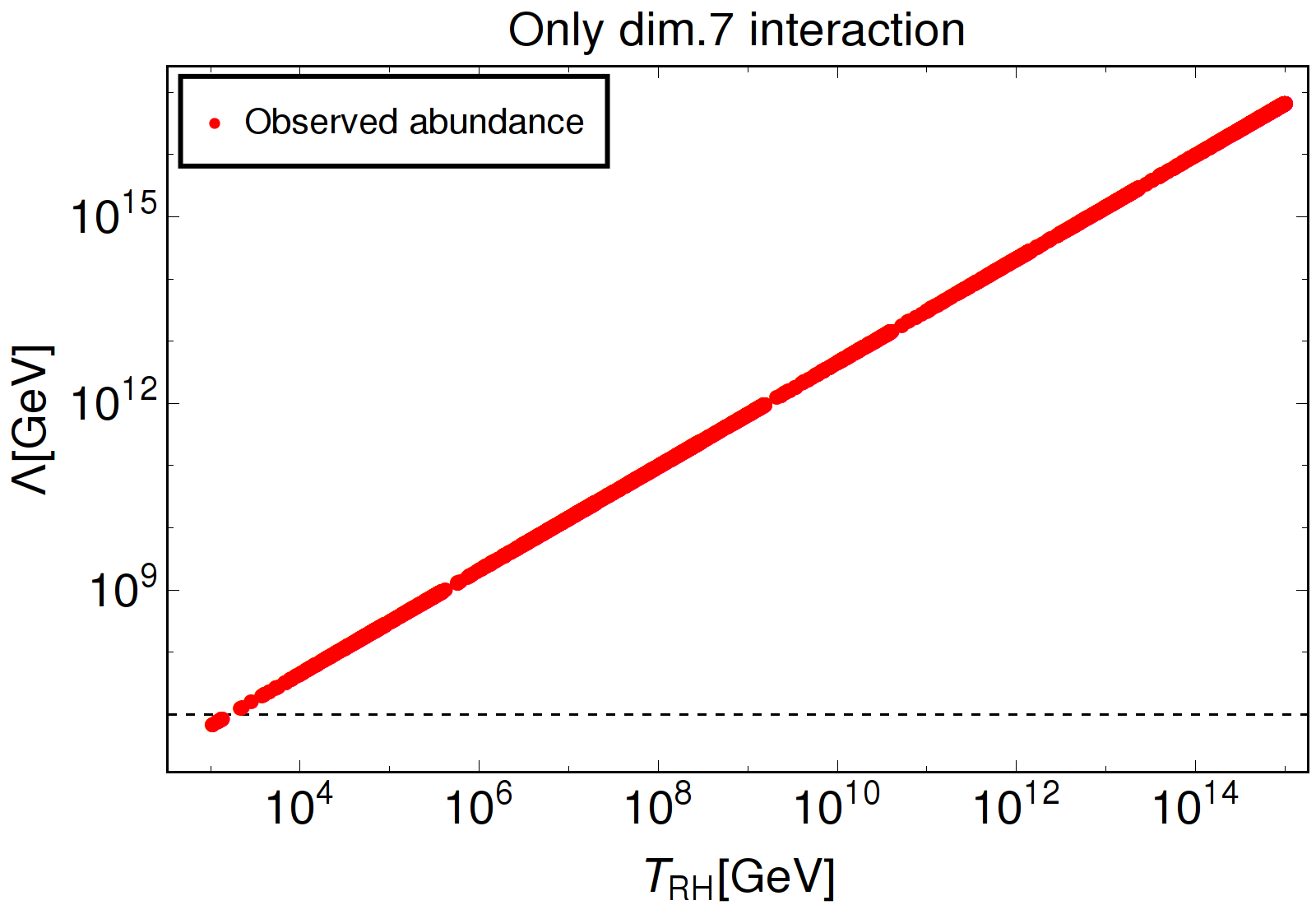}
$$
\caption{Top: Parameter space satisfying the PLANCK observed relic abundance taking into account only dim.5 (left) and only dim.7 (right) interactions for $M_\chi=50~\rm GeV$. In both the plots the black dashed line is roughly the minimum required $\Lambda_\nu$ to generate light neutrino mass following Eq.~\eqref{eq:dim5-nu} and Eq.~\eqref{eq:dim7-nu}.}\label{fig:dimRelic}
\end{figure}

In Fig.~\ref{fig:dimRelic} we show regions satisfying the PLANCK observed relic abundance in $T_{\text{RH}}-\Lambda$ bi-dimensional plane for a fixed DM mass of 50 GeV. In the left panel we depict the parameter space for only dim.5 DM-SM operator and in the right panel we show the same where the contribution comes only from dim.7 DM-SM interactions. While the relative contribution of UV and IR freeze-in contribution for dim.5 operator will be similar to the ones shown in Fig.~\ref{fig:relic}, for dim.7 the effect of IR freeze-in is negligible as noted before for dim.8 operator. This is because for dim.7, due to the cut-off scale suppression, in order to make the post-EWSB processes effective, the reheat temperature has to go below $\sim 1~\text{TeV}$. We are not considering reheat temperature below a TeV as mentioned earlier. In both the plots we also show minimum $\Lambda_\nu$ required to get light neutrino mass (Eq.~\eqref{eq:dim5-nu}, Eq.~\eqref{eq:dim7-nu}) via the black dashed lines. Hence, in both cases all of the region above the black dashed line is compatible with the neutrino mass. From both the plots it is evident that $\Lambda$ required for right relic abundance both in the case of dim.5 and dim.7 coincide with $\Lambda_\nu$ obtained from Eq.~\eqref{eq:dim5-nu} and Eq.~\eqref{eq:dim7-nu}. This indicates, the scale of UV freeze-in where only dim.5 and dim.7 operators contribute, can simultaneously explain light neutrino mass. Such a possibility of simultaneous origin of neutrino mass and DM-SM operator through effective operators that violate lepton number, in turn, constrains the reheat temperature of the universe as seen from the plots. 


\section{Possible UV Completion}
\label{sec:uvcomplete}

Several possibilities have been discussed in the literature which naturally give rise to feeble DM-SM couplings. For example, the authors of \cite{Biswas:2018aib} considered loop suppressions as origin of FIMP interactions, in \cite{Mambrini:2013iaa, Bhattacharyya:2018evo} the possibility of DM-SM interactions via superheavy neutral gauge bosons was discussed. On the other hand, the authors of \cite{Kim:2017mtc, Kim:2018xsp} considered clockwork origin of FIMP couplings. In this section, we briefly comment upon the possibility of generating some of the DM-SM effective operators within a complete theory. This is similar to the UV completion of the Weinberg operator of light neutrino masses \cite{Weinberg:1979sa} via seesaw mechanism \cite{Minkowski:1977sc, GellMann:1980vs, Mohapatra:1979ia, Schechter:1980gr, Mohapatra:1980yp, Lazarides:1980nt, Wetterich:1981bx, Schechter:1981cv}.

Let us start with the dim.5 operator between DM and SM. The only possible scalar operator is $ (\overline{\chi^c}\chi) (H^{\dagger} H)/\Lambda$. If we consider a UV complete theory, where in addition to the $Z_2$ odd DM, there exists a pair of vector like lepton doublet $\psi_{L,R}$ odd under the same $Z_2$ symmetry, the additional relevant terms in the Lagrangian are 
\begin{equation}
-\mathcal{L} \supset M_{\psi} \overline{\psi} \psi + Y_1 \overline{\psi_L} \tilde{H} \chi + Y_2 \overline{\psi^c_R} H \chi + {\rm h.c.}
\end{equation}
At a scale $\mu \ll M_{\psi}$, the heavy vector like leptons can be integrated out, resulting in an operator like 
$ (\overline{\chi^c}\chi) (H^{\dagger} H) Y_1 Y_2/M_{\psi}$. If reheat temperature of the universe is smaller than $M_{\psi}$, these additional vector like leptons are not present in the thermal bath and hence DM-SM interactions mimic as a dimension five operator with a cut-off $\Lambda = M_{\psi}$.

Similarly, one can generate dimension seven operator of the type  $ (\overline{\chi^c}\chi) (H^{\dagger} H)^2/\Lambda^3$. Consider the presence of $Z_2$ odd singlet fermion $\psi_L$ and a $Z_2$ even singlet scalar $\phi$. The relevant new terms in the Lagrangian are 
\begin{equation}
-\mathcal{L} \supset M_{\psi} \overline{\psi^c_L} \psi_L + \left(Y_1 \overline{\psi_L} \phi \chi + {\rm h.c.} \right) + \mu_{\phi} \phi H^{\dagger} H + M^2_{\phi} \phi \phi
\end{equation}
At a scale $\mu \ll M_{\psi}, M_{\phi}, \mu_{\phi}$, the heavy fields $\psi_L, \phi$ can be integrated out, resulting in DM-SM operator of the type $ (\overline{\chi^c}\chi) (H^{\dagger} H)^2 Y^2_1 \mu^2_{\phi}/(M^4_{\phi}M_{\psi})$. Considering $\mu_{\phi} \approx M_{\psi} \approx M_{\phi} \approx \Lambda$ and order one Yukawa couplings, this leads to the expected dimension seven operator $ (\overline{\chi^c}\chi) (H^{\dagger} H)^2/\Lambda^3$. In the same way, one can also generate other operators discussed in the above analysis.

\section{Dark matter production via radiative inflaton decay}
\label{sec:inf-dm}

\begin{figure}[htb!]
$$
\includegraphics[scale=0.45]{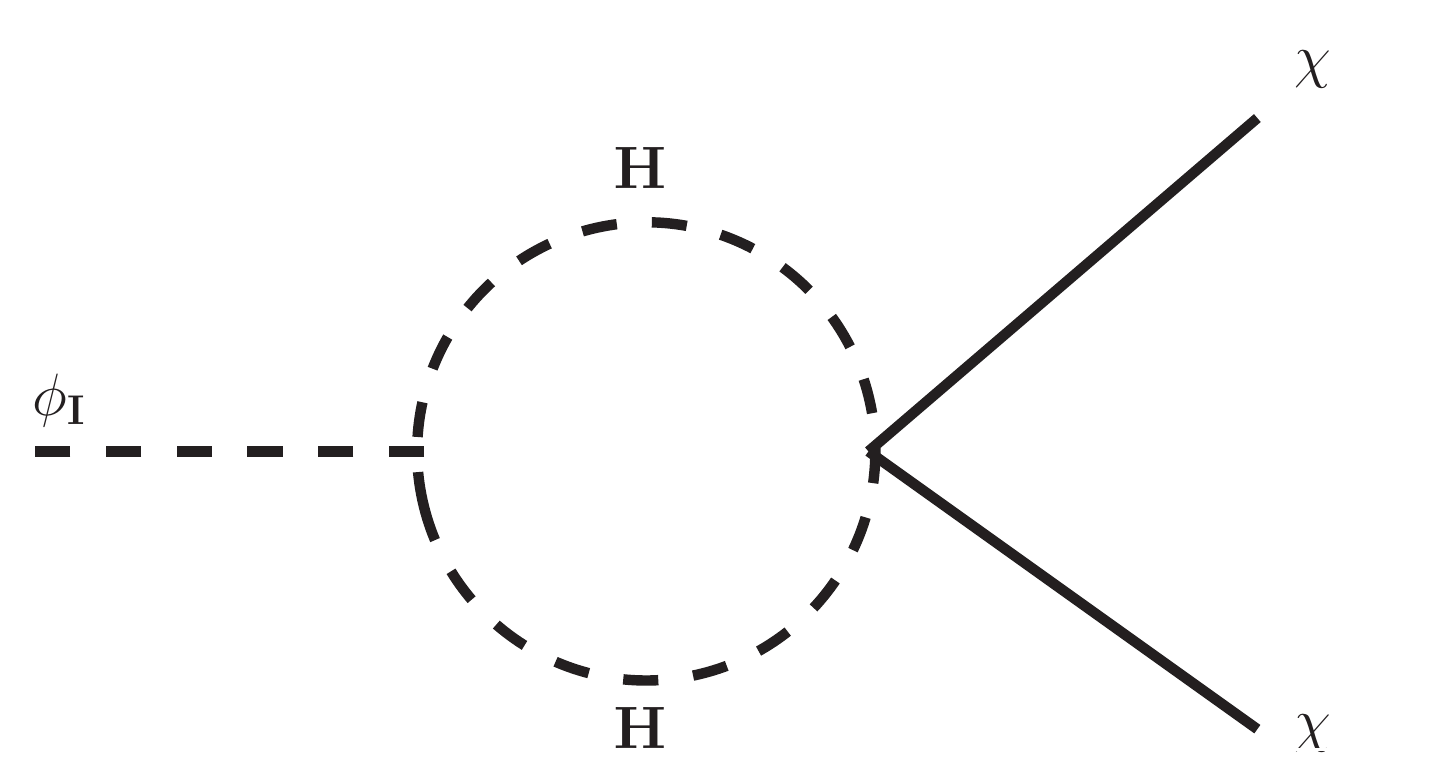}
$$
\caption{Decay of the inflaton to the DM pair via Higgs loop at 1-loop level.}\label{fig:inflt-dec}
\end{figure}

In this section we look into the possibility of DM production from the decay of inflaton. Since we are interested in freeze-in production of DM via effective DM-SM operators, we do not consider any direct coupling of inflaton to DM. However, inflaton has direct couplings to the SM particles, required for reheating~\cite{Allahverdi:2010xz}.
In such a scenario, even though inflaton does not have direct coupling with DM at tree level, one can not ignore the  decay of the inflaton to the DM via SM loops~\cite{Kaneta:2019zgw}. As we extensively have shown earlier the dominance of dim.5 operator over the others for DM production via freeze-in, hence we only show the results for such dim.5 interaction where the DM is produced via one-loop decay of the inflaton as shown in Fig.~\ref{fig:inflt-dec}. We neglect direct coupling of the DM to the inflaton and consider the following effective Lagrangian between the inflaton, SM Higgs, SM fermions and the DM $\chi$:
\bea
\mathcal{L}\supset \mu_{h} \phi_I\left(H^\dagger H\right)+y_f\phi_I ff + \frac{1}{\Lambda} \overline{\chi^c}\chi \left(H^\dagger H\right),
\eea
where $\mu_h$ is the Higgs-inflaton coupling (having mass dimension one) which we consider to be as large as the inflaton mass, and $y_f$ is the SM fermion-inflaton Yukawa coupling, which is considered to be of order one. The one-loop decay of the inflaton via Higgs exchange is given by:
\bea
\Gamma_{\phi_I}^{\text{loop}} =  \frac{M_I}{2\pi}L^2,
\eea
with the loop contribution
\bea
L =  \frac{\mu_h}{16\pi^2\Lambda}\Biggl(1+\ln\Bigl[\frac{\Lambda^2}{M_I^2}\Bigr]\Biggr)
\eea
the details of which can be found in Appendix~\ref{sec:app-loop}. Here we have ignored the masses of the SM particles as well as that of the DM since the inflaton itself is very heavy and the decay happens at a very high temperature. Now, the tree-level inflaton decay to the Higgs and to the SM fermions is given by:
\bea\begin{split}
\Gamma_{\phi_I\to hh}&=\frac{\mu_h^2}{16\pi M_I};~~\Gamma_{\phi_I\to ff}&=\frac{N_c y_f^2}{8\pi} M_I.     
    \end{split}
\eea
We define the branching ratio of inflaton decay into DM at one-loop as
\bea\begin{split}
{\rm Br}&=\frac{\Gamma_{\phi_I}^{\text{loop}}}{\Gamma_{\phi_I\to hh}+\Gamma_{\phi_I\to ff}+\Gamma_{\phi_I}^{\text{loop}}}.    
    \end{split}\label{eq:brnch}
\eea
Following the prescriptions in~\cite{Chung:1998rq,Giudice:2000ex,Garcia:2017tuj,Kaneta:2019zgw} for non-instantaneous thermalization (finite decay width of the inflaton) one can write the relic abundance of the DM due to inflaton decay as:
\bea\begin{split}
\Omega_{\chi}^{\text{inf}} h^2&= M_{\chi} {\rm Br} \Biggl(\frac{T_{\text{RH}}}{M_I}\Biggr).    
    \end{split}\label{eq:rel-inf}
\eea
where the reheat temperature is defined via:
\bea
\frac{3}{2} c H\left(T_\text{RH}\right)=\Gamma_{\phi_I},
\label{eq:inf-trh}
\eea
$c$ being a numerical constant taken to be $c\approx 1.2$~\cite{Pradler:2006hh,Ellis:2015jpg,Kaneta:2019zgw} and $H\left(T_\text{RH}\right)$ is the Hubble parameter at $T=T_{\rm RH}$. Next, we compute the bound on the reheat temperature, branching fraction and cut-off scale with the condition that the inflaton decay gives rise to all of the observed DM density. We consider only inflaton decay as the source of DM production, for large cut-off scales which we consider, production from thermal bath is negligible at this stage as was shown by the authors of~\cite{Kaneta:2019zgw}.
\begin{figure}[htb!]
$$
\includegraphics[scale=0.36]{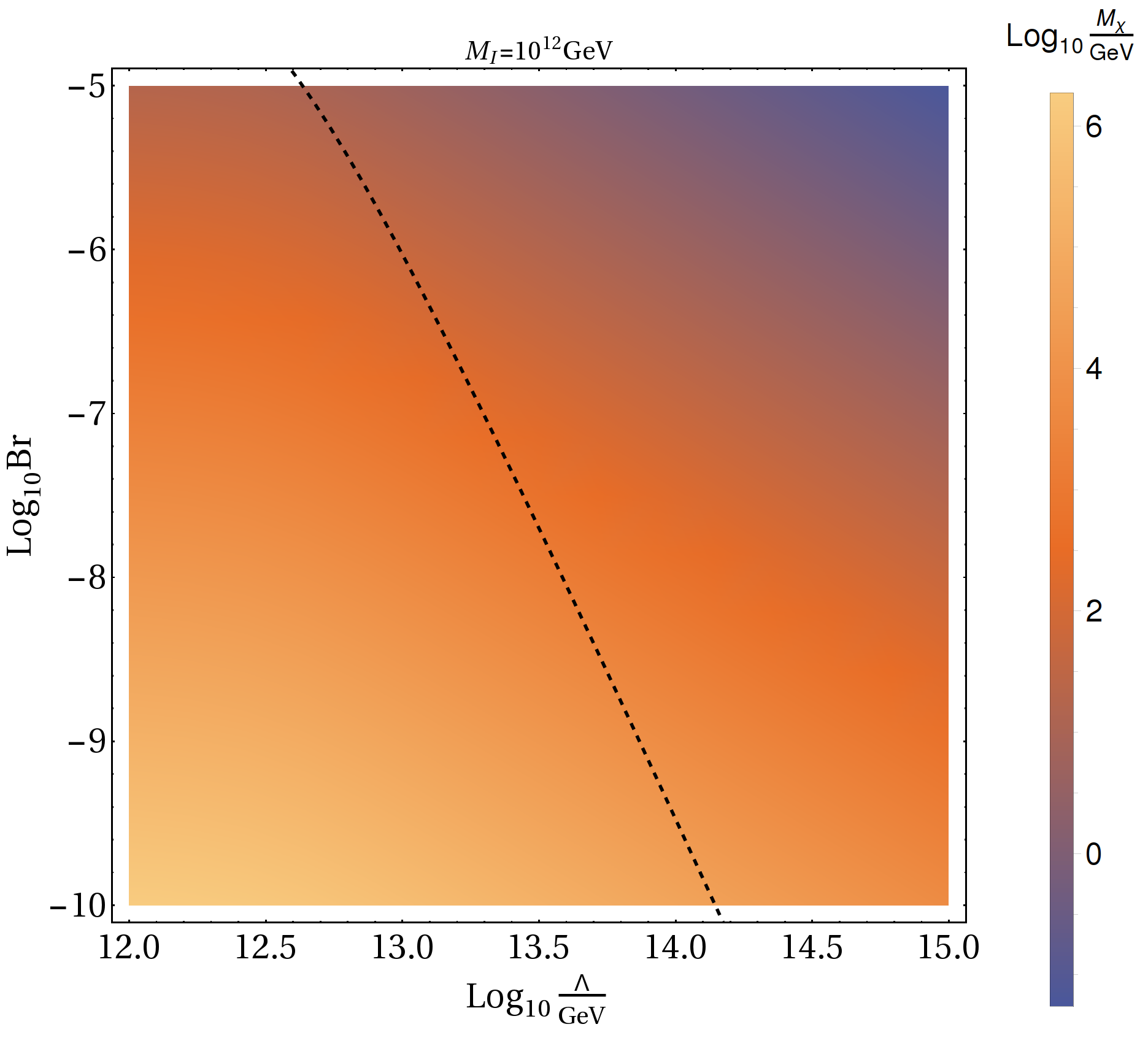}
\includegraphics[scale=0.36]{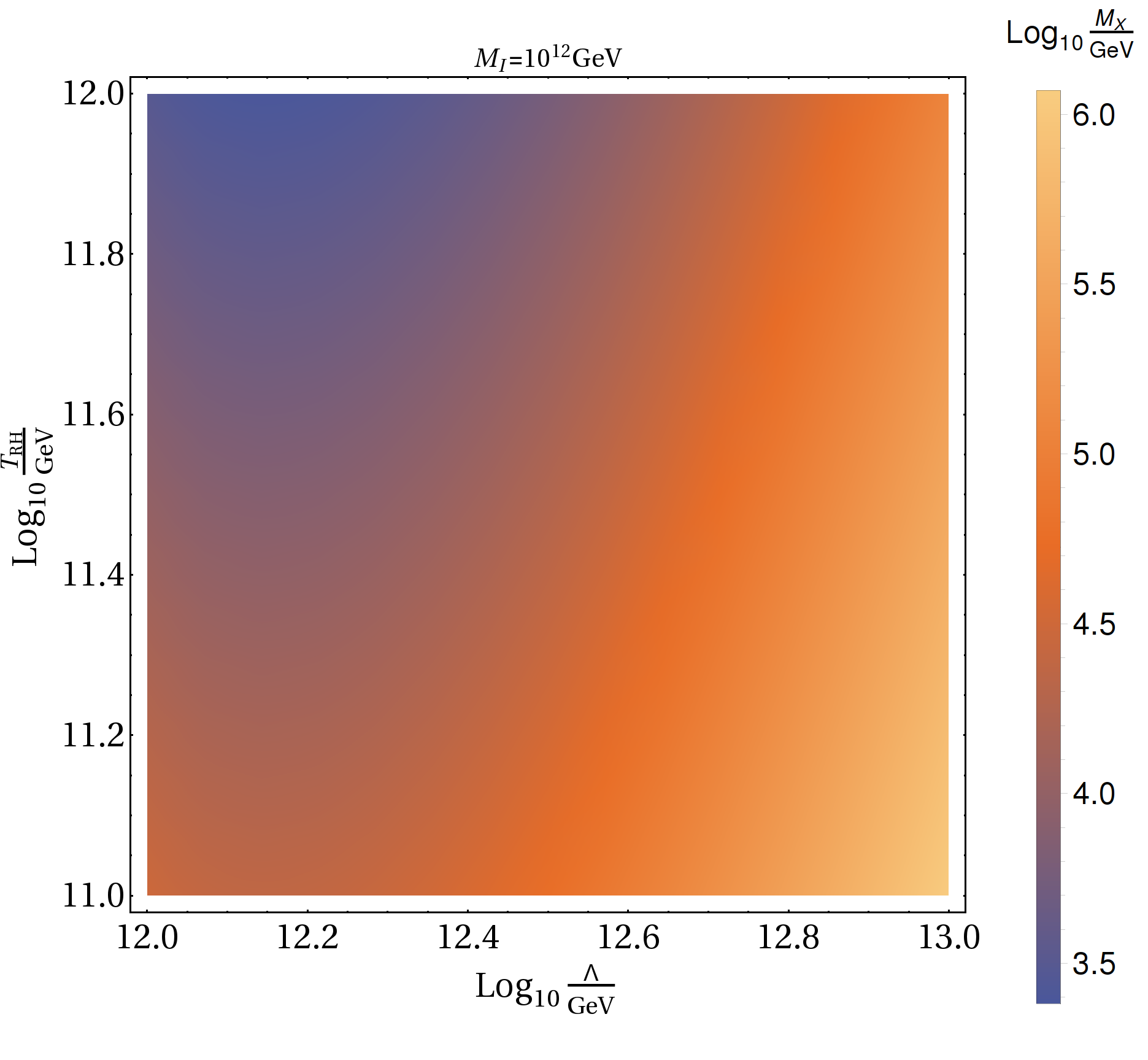}
$$
\caption{Left: Parameter space in the bi-dimensional plane of ${\rm B}-\Lambda$ obeying relic abundance. The black dashed line corresponds to the branching ratio following Eq.~\eqref{eq:brnch}. Right: Relic density allowed region in $T_\text{RH}-\Lambda$ plane for different DM masses shown with the colour code. For both the plots we choose $M_I=10^{12}~\rm GeV$ and $y_f=1$.}\label{fig:relic-infl}
\end{figure}

In the left panel of Fig.~\ref{fig:relic-infl} we show the parameter space for PLANCK observed relic density for inflaton mass of $M_I=10^{12}~\rm GeV$. For each value of $\{\Lambda,{\rm Br}\}$ the DM mass $M_\chi$ needed to obtain the correct relic abundance is colour coded by the scale at the right of the plot. Note that, for lighter DM mass one naturally requires a larger branching ratio, as expected from Eq.~\eqref{eq:rel-inf}. The black dashed line in this figure corresponds to the branching ratio following Eq.~\eqref{eq:brnch}, applicable to the particular dim.5 DM-SM effective operator we consider here. In other regions of this plot, the branching ratio is independently varied so that correct DM relic is obtained for chosen DM mass and cutoff scale. In the right panel of Fig.~\ref{fig:relic-infl} we show the PLANCK allowed parameter space in $T_\text{RH}-\Lambda$ plane where for each value of $\{\Lambda,T_\text{RH}\}$ the required $M_\chi$ to obtain right relic abundance is shown by the colour code. Since the reheat temperature has an inverse relation with the branching ratio according to Eq.~\eqref{eq:inf-trh}, hence in this plot we see a complementary dependence of DM mass for different $T_\text{RH}$ on the cut-off scale. It is interesting to note that, correct DM abundance for such dim.5 scenario can be produced from radiative inflaton decay for a relatively smaller value of cut-off scale $\Lambda$ compared to what we found in our previous analysis considering only DM-SM operators to be responsible for DM production. This can be realised by comparing the parameter space shown in Fig.~\ref{fig:relic-infl} with the ones in Fig. \ref{fig:relic} and \ref{fig:dimRelic}.


\section{Effect of non-instantaneous inflaton decay}
\label{sec:tmax}

In calculating the DM relic abundance in Sec.~\ref{sec:dm-yld} we made a crucial assumption: $Y_\chi\left(T=T_\text{RH}\right)=0$, which implies that the DM abundance is zero at the end of reheating when the the thermal bath reaches an equilibrium temperature $T=T_\text{RH}$. Such an assumption is very commonplace in freeze-in analysis which takes into account the fact that the reheat temperature is the largest temperature that the thermal bath can achieve, which is true if reheating is an instantaneous process. The reheat temperature $(T_\text{RH})$ is usually calculated by assuming an instantaneous conversion of the energy density in the inflaton field into radiation when the decay width of the inflaton is equal to the Hubble expansion rate. In reality, however, reheating is not an instantaneous process~\cite{Chung:1998rq,Giudice:2000ex,Kaneta:2019zgw}. For example, if the inflaton is described by a simple model with quadratic potential, the radiation dominated phase follows a prolonged stage of matter domination during which the energy density of the universe is dominated by the coherent oscillations of the inflaton field. For different choices of inflaton potential, the equation of state during this pre-reheating or preheating phase can be different from the one in matter dominated phase. The temperature of the thermal bath during this phase can reach a value much larger than $T_\text{RH}$ and is typically denoted by: $T_\text{max}\sim\left(H_I M_\text{pl}\right)^{1/4}\sqrt{T_\text{RH}}$, where for a simple model of inflation $H_I$ can be identical with the inflaton mass~\cite{Chung:1998rq}. The temperature subsequently comes down from $T_\text{max}$ to  $T_\text{RH}$ when the inflaton decay is completed. The DM, thus can be produced at a temperature $T=T_\text{max}$ ({\it i.e.,} prior to the inflation ends), which is higher than $T_\text{RH}$. This practically indicates that the DM can have a non-zero abundance at $T=T_\text{RH}$ when the inflaton decay is completed and the radiation dominated era begins.

We consider that the inflaton decays dominantly into the radiation, and the DM is not produced in thermal equilibrium during reheating. With this assumption, to determine the DM abundance at $T=T_\text{RH}$, one has to first solve a set of coupled BEQ involving the inflaton density $\rho_\phi$ and the radiation density $ \rho_R$~\cite{Kaneta:2019zgw} in order to find the evolution of temperature with expansion of the universe:

\bea\begin{split}
&\dot{\rho_\phi}+ 3H\rho_\phi = -\Gamma_\phi\rho_\phi \\&   \dot{\rho_R}+ 4H\rho_R = \Gamma_\phi\rho_\phi \\& H^2 = \frac{\rho_\phi}{3M_\text{pl}^2} + \frac{\rho_R}{3M_\text{pl}^2}, 
    \end{split}\label{eq:beq-cpl}
\eea

\noindent where $\Gamma_\phi$ is the decay width of the inflaton. The temperature of the thermal bath is found to be scaled as $T\propto a^{-3/8}$ (contrary to $T\propto a^{-1}$ as in radiation domination) as $T$ decreases from $T_\text{max}$ to  $T_\text{RH}$. Therefore, before reheating is completed, for a given temperature, the universe expands faster than in the radiation-dominated phase. In order to capture this effects, we can parametrise the DM production rate via annihilation \footnote{We can do the similar exercise for DM production from inflaton decay before EWSB era.} following the prescription given in~\cite{Kaneta:2019zgw}: 

\bea
R_\text{ann} = \frac{T^{\xi+6}}{\Lambda^{\xi+2}}. 
\label{eq:ann-rate}
\eea

with the BEQ for the DM number density evolution being:

\bea
\dot{n_\chi}+3Hn_\chi=R_\text{ann}.\label{eq:beq-chi}
\eea

Note that, the annihilation rate in Eq.~\eqref{eq:ann-rate} corresponds to a cross section~\cite{Bernal:2020qyu}:

\bea
\langle\sigma v\rangle = \frac{T^\xi}{\Lambda^{\xi+2}},
\eea

\noindent with $\xi\geq 0$ is an even integer and the reaction rate has the form $R_\text{ann}=n^2_{\rm SM}\langle\sigma v\rangle$ where $n_{\rm SM} \propto T^3$ is the equilibrium number density of the radiation bath consisting of SM particles. This cross section is generated by a non-renormalisable operator with mass dimension: $d=5+\xi/2$. Such an interaction rate may correspond to pre-EWSB or UV freeze-in scenario in our framework where the DM is produced from the annihilation of the bath particles.  It is then possible to solve Eq.~\eqref{eq:beq-chi} analytically~\cite{Kaneta:2019zgw} to find the DM number density (yield) at $T=T_\text{RH}$ with inputs from Eq.~\eqref{eq:beq-cpl} and Eq.~\eqref{eq:ann-rate}:

\bea
  n_\chi\left(T_\text{RH}\right)\simeq\left\{
                \begin{array}{ll}
                  \frac{24\sqrt{30}}{5g{_{\star \rho}}(T)\pi}\frac{T_\text{RH}^{\xi+4}M_\text{pl}}{\left(6-\xi\right)\Lambda^{\xi+2}},~~\xi<6\\
                  \frac{24\sqrt{30}}{5g{_{\star \rho}}(T)\pi}\frac{M_\text{pl}T_\text{RH}^{10}}{\Lambda^8}\log\Bigl(\frac{T_\text{max}}{T_\text{RH}}\Bigr)~~,\xi=6\\
                  \frac{24\sqrt{30}}{5g{_{\star \rho}}(T)\pi}\frac{T_\text{max}^{\xi-6}T_\text{RH}^{10}}{\left(\xi-6\right)\Lambda^{\xi+2}}~~,\xi>6.
                  \end{array}
              \right.\label{eq:ndm-tmax}
  \eea

The DM relic abundance at present epoch for a mass $M_\chi$ can be deduced from Eq.~\eqref{eq:ndm-tmax}:

\bea
\Omega_\chi h^2\simeq 6\times 10^6\Biggl[\frac{n\left(T_\text{RH}\right)}{T^3_\text{RH}}\Biggr]\Biggl(\frac{M_\chi}{\text{GeV}}\Biggr).
\label{eq:rel-tmax}
\eea

\begin{figure}[htb!]
$$
\includegraphics[scale=0.4]{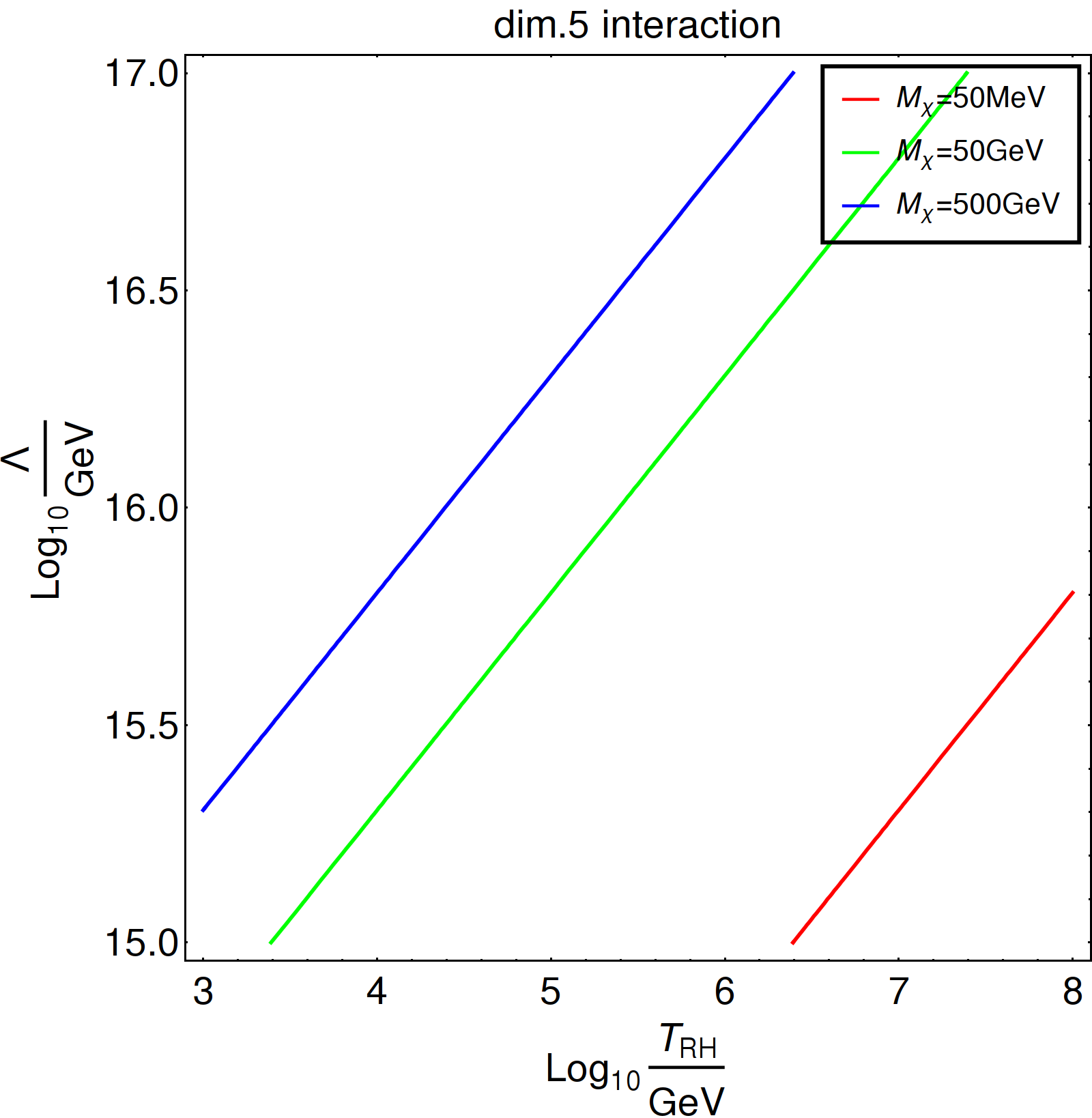}~~~
\includegraphics[scale=0.4]{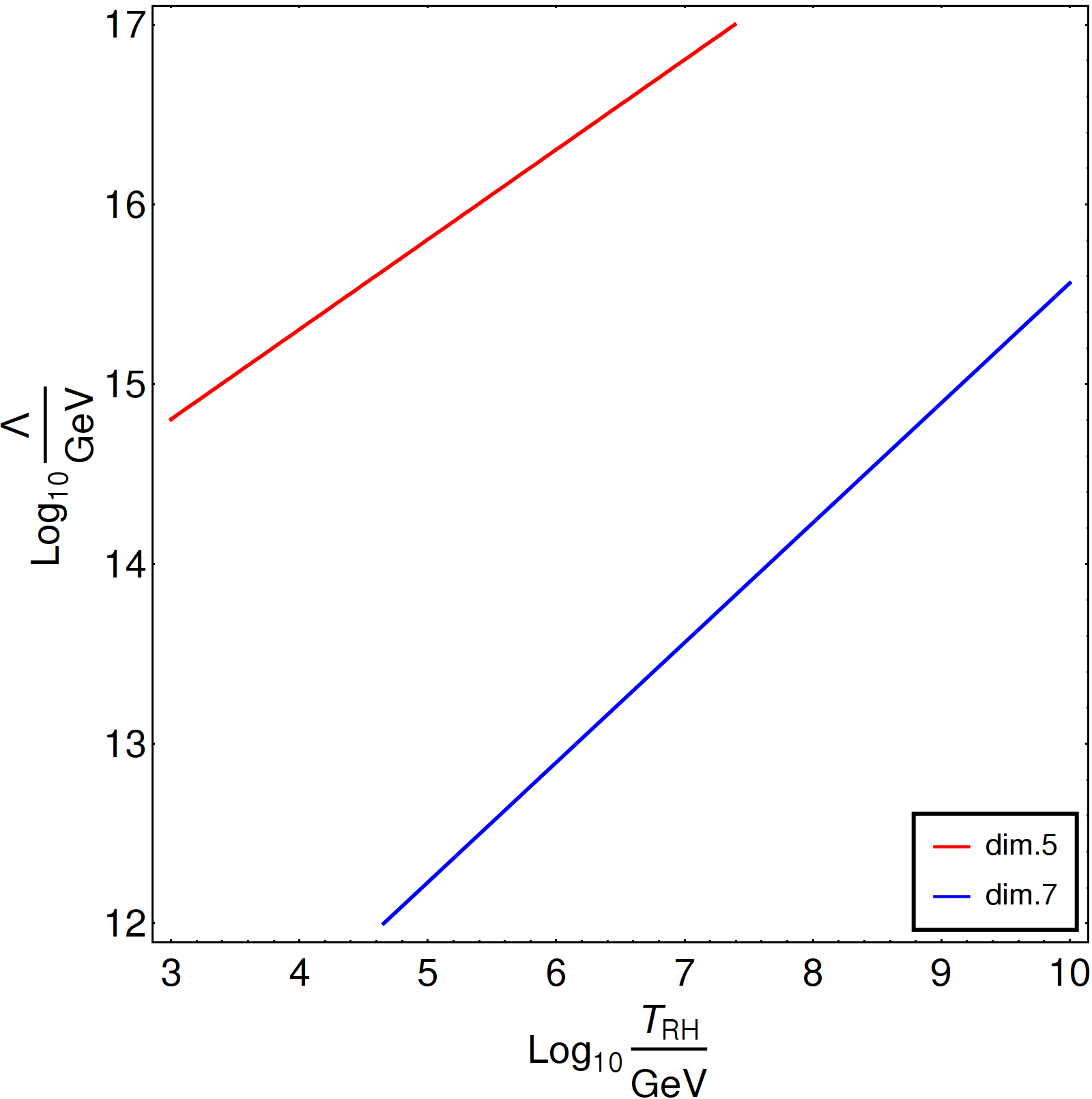}
$$
\caption{Left panel: Contours satisfying PLANCK observed relic abundance following Eq.~\eqref{eq:rel-tmax} in $\Lambda$-$T_\text{RH}$ plane for different choices of the DM mass and $\xi=0~(d=5)$ case. Right panel: Comparison of relic density allowed parameter space for $\xi=0~(d=5)$ (in red) and $\xi=4~(d=7)$ (in blue) for $M_\chi=50~\rm GeV$.}\label{fig:cntr-tmax}
\end{figure}

As it is evident from Eq.~\eqref{eq:ndm-tmax}, for $\xi\leq 6$ (corresponding to interactions of dimension $d\leq 8$), there is either no direct dependence of DM relic abundance on $T_\text{max}$ or the dependence on $T_\text{max}$ being logarithmic, is mild. For $\xi>6$ ({\it i.e.,} $d>8$) the DM number density at the reheat temperature is directly proportional to appropriate powers of $T_\text{max}$ and hence the effect of non-instantaneous inflaton decay on DM abundance at reheat temperature becomes important. We can thus infer, since in the present analysis we are dealing with operators $d\leq 8$, including the DM production before $T=T_\text{RH}$ will not lead to significant change in our results obtained earlier, assuming instantaneous reheating.

In order to verify this, we choose particular examples of dim.5 and dim.7 interactions corresponding to $\xi=\{0,4\}$. The DM number density in these cases are given by (Eq.~\eqref{eq:ndm-tmax}):

\bea\begin{split}
n_\chi^{d=5}\left(T_\text{RH}\right) = \frac{24\sqrt{30}}{5g{_{\star \rho}}(T)\pi}\frac{T_\text{RH}^4 M_\text{pl}}{6\Lambda^{2}},~~n_\chi^{d=7}\left(T_\text{RH}\right) = \frac{24\sqrt{30}}{5g{_{\star \rho}}(T)\pi}\frac{T_\text{RH}^8 M_\text{pl}}{2\Lambda^6}.     
    \end{split}
\eea

The resulting relic density allowed parameter space is shown in Fig.~\ref{fig:cntr-tmax}, where we stick to the UV freeze-in in the pre-EWSB regime and hence there is no effect of decay. One can compare Fig.~\ref{fig:cntr-tmax} with Fig.~\ref{fig:dimRelic} where we have illustrated the relic density allowed region for dim.5 and dim.7 interactions without considering effects from non-instantaneous reheating. Although we have taken into account contributions from all the processes appearing after EWSB in Fig.~\ref{fig:dimRelic}, but still one can notice, in order to satisfy the observed abundance, the constraint on the cut-off scale $\Lambda$ remains nearly the same. This is true for the region with linear dependence of the cut-off scale on the reheat temperature (where effects of IR freeze-in is sub-dominant). This implies, there is no serious departure in the outcome of the analysis for operators with dimension $d\leq 8$ due to non-instantaneous reheating and hence our assumption of DM yield to be zero at $T=T_\text{RH}$ in calculations shown in previous sections remains valid.

\section{Conclusion}
\label{sec:conc}

We have classified the simplest possible operators connecting dark matter (DM) and the standard model (SM) particles relevant for UV freeze-in scenario up to and including dim.8. Considering the DM to be a singlet Majorana fermion odd under an unbroken $Z_2$ symmetry we first list out possible DM-SM operators. Since UV freeze-in is a high scale phenomena, we write down all these operators at a scale above the electroweak (EW) symmetry breaking so that the SM operators appearing in the interactions are invariant under SM gauge symmetry. The DM being a fermion we only have operators of dim.5, 7, 8 that are invariant under SM gauge symmetry for DM-SM interactions.

After enlisting the possible operators upto dim.8 we consider the simplest possibility for relic abundance calculation where the DM operators emerge as scalar bilinears. While including other Lorentz structures is not going to change our conclusions significantly, but choice of scalar DM operators keep the analysis very simple. For each possible dimension of these operators we first check the required cut-off scale $\Lambda$ to ensure the non-thermal production of the DM, thus in turn constraining it. We note, dim.5 operator requires the cut-off scale to be at least $\gtrsim 10^{11}$ GeV in order to keep the DM out of equilibrium, whereas dim. 8 interactions can significantly reduce this scale, allowing $\Lambda$ to be as low as $\sim 10^5$ GeV. By keeping the effective scale $\Lambda$ in the range required to satisfy the non-thermal DM criteria, we then move on to calculate the DM relic by considering both UV and IR freeze-in contributions with the latter arising after the electroweak symmetry breaking, and more relevant for DM mass below electroweak scale. We thus constrain the cut-off scale and reheat temperature $T_{\text{RH}}$ from the requirement of observed DM relic abundance, in agreement with PLANCK 2018 data. After taking all possible operators upto dim.8 at the same time, we also constrain the relevant parameters from the requirement of relic abundance by taking each dimensional operator one at a time. We check the relative contribution of UV and IR freeze-in to the total DM relic abundance. While for DM mass above the electroweak scale, yield due to IR freeze-in is negligible as expected, for lower DM mass, IR freeze-in is sizeable only for dim.5 DM-SM operators. This is found to be true especially when the reheat temperature of the universe is kept well above $1~\text{TeV}$.

Finally, we explore the possibility of constraining the relevant parameters $\{\Lambda,T_{\text{RH}}\}$ simultaneously from DM relic and neutrino mass criteria, assuming the DM-SM interaction and neutrino mass generation from operators at the same dimension. If neutrino mass arises from Weinberg type operators, then such a scenario is restricted to dim.5 and 7 only. We find, DM and neutrino mass originating from dim.7 operators demand the reheat temperature to be more than $10^3$ GeV while dim.5 operator requires $T_{\text{RH}}\gtrsim 10^6$ GeV. We briefly comment on the possibility of realising some of these effective DM-SM operators within a UV complete theory. For the sake of completeness, we briefly comment upon the possibility of DM production from inflaton decay. Even if inflaton does not have any direct coupling with the DM, the required inflaton coupling with the SM particles which eventually reheats the universe also lead to inflaton-DM coupling at one-loop level by virtue of DM-SM effective operators. We consider the dim.5 DM-SM operator and the corresponding one-loop decay of inflaton into DM to show that correct DM abundance can be satisfied even with smaller values of cut-off scale $\Lambda$ compared to what we found in case of DM production purely from DM-SM operators of dim.5. We also check the effects of non-instantaneous reheating on DM production and show that for DM-SM operators with dimension $d \leq 8$, such effects are negligible and our assumption of vanishing or negligible DM abundance at $T=T_{\rm RH}$ remains valid.

The UV freeze-in scenario does not have much prospects for direct detection, but it can have some indirect detection prospects, for example, generation of monochromatic photon lines \cite{Biswas:2019iqm}. Also, since the DM yield in UV freeze-in is very much sensitive to the reheat temperature of the universe, it is worth exploring the consequences within specific inflationary models~\cite{Bernal:2020qyu} that may leave some footprints in other cosmological observations. We leave  studies of detection prospects for such DM models and their connection to specific inflationary scenarios to future works.

\acknowledgments

DB acknowledges the support from Early Career Research Award from DST-SERB, Government of India (reference number: ECR/2017/001873). BB would like to thank Shakeel Ur Rahaman and Joydeep Chakrabortty for helping out with the {\tt Mathematica} based package {\tt GrIP}~\cite{Banerjee:2020bym}. BB and RR would like to acknowledge email communications with Fatemeh Elahi and Anirban Biswas. RR would also like to thank Arunansu Sil for fruitful discussions. The authors thank Yann Mambrini for useful comments on the first preprint version of this work.


\appendix
\section{Scalar kinetic term}
\label{sec:app-scal-kin}
With the covariant derivative defined in Sec.~\ref{sec:dim7} we derive here the expressions for scalar kinetic term before and after EWSB.
\subsection{Before EWSB}

Before EWSB the $SU(2)_L$ scalar has the form given in Eq.~\eqref{eq:gb} where the Goldstone bosons are physical. The scalar kinetic term the reads:

\bea\begin{split}
\left|\mathcal{D}_\m H\right|^2&\supset\left(\partial_\mu\phi^+\partial^\mu\phi^-+\partial_\mu\phi^0\partial^\mu\overline{\phi^0}\right)+\frac{g_1^2}{4}B_\mu B^\mu\left(\phi^+\phi^-+\phi^0\overline{\phi^0}\right)\\&
+\frac{g_2^2}{4}\sum_{i=1,2,3} W_{i\m}W^{i\m}\left(\phi^+\phi^-+\phi^0\overline{\phi^0}\right).
    \end{split}\label{eq:app-sca-kin-bewsb}
\eea

\subsection{After EWSB}

After EWSB the $SU(2)_L$ scalar doublet is written as in Eq.~\eqref{eq:higgs}, by expanding around its minima. As a result the scalar kinetic term turns out to be:

\bea\begin{split}
\left|\mathcal{D}_\m H\right|^2 &\supset\frac{1}{2}\partial_\m h\partial^\m h+\left(h+v_h\right)^2\Bigg\{\frac{g_2^2}{4} W_\m^+ W^{-\m}+\frac{g_2^2}{8c_w^2} Z_\m Z^\m\Bigg\}.
    \end{split}\label{eq:app-sca-kin-aewsb}
\eea

\section{Boltzmann Equation for decay and annihilations}
\label{sec:app-beq}

Here we would like to derive the Boltzmann equation (BEQ) for processes corresponding to decay and annihilation. As we are considering only $1\to2$ and $2\to2$ processes after EWSB, where all the states involved in the subsequent processes are massive, while before EWSB we are considering all $n\to m$ processes with $n,m\geq 2$ where the SM particles are massless but the DM is massive. However, considering zero DM mass before EWSB does not affect our reults. 


\subsection{BEQ for $1\to 2$ process}
\label{sec:app-beq-decay}

For a decay process $h(p)\to \chi(p_1),\chi(p_2)$ the evolution of number density of $\chi$ is given by:

\bea
\dot{n_\chi}+3 H n_\chi = \int d\Pi_1 d\Pi_2 d\Pi \left(2\pi\right)^4 \delta^4\left(p_1+p_2-p\right) \overline{\left|\mathcal{M}\right|}_{\text{decay}}^2 f_h,
\label{eq:beq-decay}
\eea

where $d\Pi_j=g_j\frac{d^3 p_j}{2E_j\left(2\pi\right)^3}$ are Lorentz invariant phase space elements, and $f_i$ is the phase space density of the particle $i$:

\bea
n_i = \frac{g_i}{\left(2\pi\right)^3}\int d^3p f_i, 
\label{eq:num-density}
\eea

is the particle density of species $i$ possessing $g_i$ internal degrees of freedom (DOF). In writing Eq.~\eqref{eq:beq-decay} we make two important assumptions:

\begin{itemize}
                                                                                                                                                  \item The initial $\chi$ abundance is negligible
so that we may set $f_\chi=0$.
\item  Neglect Pauli-blocking/stimulated
emission effects, i.e. approximating $1\pm f_i\approx 1$.  

\end{itemize}

With these we can then write the BEQ as:

\bea
\begin{split}
\dot{n_\chi}+3 H n_\chi = g_h\int\Gamma_{h\to\chi\chi}^{'} f_h \frac{d^3p_h}{\left(2\pi\right)^3}=n_h^{\text{eq}}\langle\Gamma_{h\to \chi\chi}\rangle, 
\end{split}
\label{eq:beq-decay-2}
\eea

where the decay width is defined as~\cite{Hall:2009bx}:

\bea
\Gamma_{h\to \chi\chi}=\int \frac{1}{2m_h}\frac{\overline{\left|\mathcal{M}\right|}_{\text{decay}}^2}{g_h} \left(2\pi\right)^4\delta^4\left(p_1+p_2-p\right)d\Pi_1 d\Pi_2,
\eea

with $\Gamma_{h\to \chi\chi}^{'}=\frac{m_h}{E_h}\Gamma_{h\to \chi\chi}\equiv\frac{\Gamma_{h\to \chi\chi}}{\gamma}$ and $g_h=1$. The quantity $\langle\Gamma_{h\to \chi\chi}\rangle$ is the thermal averaged decay width, defined as:

\bea
\langle\Gamma_{h\to \chi\chi}\rangle = m_h\frac{\int d\Pi_h\Gamma_{h\to \chi\chi} f_h}{\int d\Pi_h E_h f_h}.
\label{eq:beq-decay-3}
\eea

As Higgs is in thermal equilibrium, if we consider the Maxwell-Boltzmann distribution: $f_h^{\text{eq}}=exp\left(-E_h/T\right)$, then one can write Eq.~\eqref{eq:beq-decay-3} as~\cite{Hall:2009bx,Ahmed:2017dbb}:

\bea
\begin{split}
\langle\Gamma_{h\to \chi\chi}\rangle = m_h \Gamma_{h\to \chi\chi} \frac{\int_{m_h}^\infty dE_h\sqrt{E_h^2-m_h^2} exp\left(-E_h/T\right)}{\int_{m_h}^\infty dE_h E_h\sqrt{E_h^2-m_h^2} exp\left(-E_h/T\right)}=\frac{K_1\left(m_h/T\right)}{K_2\left(m_h/T\right)}\Gamma_{h\to \chi\chi},
\end{split}
\label{eq:beq-decay-4}
\eea

where we have used the relation: $E_h^2=p_h^2+m_h^2$. Similarly, one can show (following Eq.~\eqref{eq:num-density}): $n_h^{\text{eq}}=\frac{T}{2\pi^2} m_h^2 K_2\left(m_h/T\right)$ for equilibrium distribution of $h$. Substituting these two in Eq.~\eqref{eq:beq-decay-2} we get~\cite{Hall:2009bx}:

\bea
\dot{n_\chi}+3 H n_\chi = \frac{ m_h^2\Gamma_{h\to \chi\chi}T}{2\pi^2}K_1\left(m_h/T\right).
\label{eq:beq-decay-5}
\eea

In terms of yield $Y_\chi=n_\chi/s$, this can be recasted as:

\bea
\begin{split}
Y_\chi^\text{decay} = -\int_{T_{max}}^{T_{min}} \frac{ m_h^2\Gamma_{h\to \chi\chi}}{2\pi^2}\frac{K_1\left(m_h/T\right)}{s(T).H(T)} dT
\end{split}
\label{eq:yld-temp-decay}
\eea

It is possible to express Eq.~\eqref{eq:yld-temp-decay} in terms of $x=m_h/T$:

\bea
Y_\chi^\text{decay} = \frac{45}{1.66~4\pi^4}\frac{ M_{pl}\Gamma_{h\to \chi\chi}}{m_h^2 g_{*s} \sqrt{g_{*\rho}}}\int_{x_{min}}^{x_{max}} dx~x^3 K_1(x).
\label{eq:yld-x-decay}
\eea

Now, $x_{min}=0$ corresponds to $T\approx\infty$, while $x_{max}=\infty$ corresponds to $T\approx 0$. Therefore, on integration, we obtain:

\bea
Y_\chi^\text{decay} = \frac{135 }{8\pi^3(1.66)g_{*s}\sqrt{g_{*\rho}}}\left(\frac{M_{pl}\Gamma_{h\to \chi\chi}}{m_h^2}\right),
\label{eq:yld-x-decay1}
\eea

which gives an analytical expression for yield from decay. 

\subsection{BEQ for $2\to 2$ process}
\label{sec:app-beq2to2}

Now for $2\to 2$ processes: $12\to34$ the evolution of number density of $\chi$ can be written as:

\bea
\dot{n_\chi}+3 H n_\chi = \int d\Pi_1 d\Pi_2 d\Pi_3d\Pi_4\left(2\pi\right)^4 \delta^4\left(p_3+p_4-p_1-p_2\right) \overline{\left|\mathcal{M}\right|}_{12\to34}^2 f_1 f_2,
\label{eq:beq-ann-22}
\eea

$\overline{\left|\mathcal{M}\right|}_{12\to34}^2$ is the amplitude squared for the $2\to2$ process. Now, let us write the averaging over initial and sum over final states as a general form~\cite{Gondolo:1990dk,Edsjo:1997bg}:

\bea\begin{split}
W_{ij}^{\text{n-body}}&=\frac{1}{S_f}\int\overline{\left|\mathcal{M}\right|}^2 \left(2\pi\right)^4\delta^4\left(p_i+p_j-\sum_f p_f\right)\prod_f \frac{d^3\vec{p_f}}{\left(2\pi\right)^3 2E_f},    
    \end{split}
\eea

where $S_f$ is the symmetry factor accounting for identical final state particles. For two-body final state (which is our case) this can be written as:

\bea\begin{split}
W^{\text{2-body}}_{ij\to xy} = \frac{\left|\vec{p_{xy}}\right|}{16\pi^2 S_{xy}\sqrt{s}} \int\overline{\left|\mathcal{M}\right|}_{ij\to xy}^2 d\Omega,      
    \end{split}\label{eq:w-factor}
\eea

where $p_{xy}$ is the final center of mass (CM) momentum, $S_{xy}=2$ for identical final states. Average over initial internal degrees of freedom is implied. With this we can then recast Eq.~\eqref{eq:beq-ann-22} as:

\bea\begin{split}
\dot{n_\chi}+3 H n_\chi &= \sum_{ij} \int d\Pi_i d\Pi_j W_{ij} e^{-E_i/T} e^{-E_j/T}.  
    \end{split}
\eea

This, on changing integration variable to~\cite{Gondolo:1990dk,Edsjo:1997bg}: $E_+=E_i+E_j,E_-=E_i-E_j$ and $s=m_i^2+m_j^2-2 \left|\vec{p_i}\right|\left|\vec{p_j}\right|\cos\theta$ gives rise to:

\bea
\dot{n_\chi}+3 H n_\chi = \frac{T}{32\pi^4}\sum_{ij}\int ds \left|\vec{p_{ij}}\right| W_{ij} K_1\left(\frac{\sqrt{s}}{T}\right),
\label{eq:beq-ann-22a}
\eea

where $\theta$ is the angle between $\vec{p_i},\vec{p_j}$ and $\left|\vec{p_{ij}}\right|$ is the initial CM momentum. Now, substituting Eq.~\eqref{eq:w-factor} in Eq.~\eqref{eq:beq-ann-22a} we obtain the final expression for the number density evolution of $\chi$:

\bea\begin{split}
\dot{n_\chi}+3 H n_\chi &=  \frac{T}{512\pi^6}\int_{s=4M_\chi^2}^\infty ds d\Omega \left|\vec{p_{12}}\right| \left|\vec{p_{34}}\right|\overline{\left|\mathcal{M}\right|}_{12\to 34}^2\frac{1}{\sqrt{s}} K_1\left(\frac{\sqrt{s}}{T}\right),    
    \end{split}
\eea

where $K_1(...)$ is the modified Bessel function of the second kind of order 1 and $\left|\vec{p_{a,b}}\right|=\frac{1}{2\sqrt{s}}\sqrt{s-(m_a+m_b)^2}\sqrt{s-(m_a-m_b)^2}\to \frac{\sqrt{s}}{2}$ in the limit $m_{a,b}\to 0$. Note that, the lower limit of the integration is zero if we consider all states to be massless.

Let us now recast this equation in terms of DM yield: $Y_\chi=n_\chi/s$, where $s$ is the entropy per comoving volume. On changing variable one can write:

\bea
\begin{split}
Y_\chi^{2\to2}&= \frac{1}{512\pi^6}\int_{T_{min}}^{T_{max}}\frac{1}{s(T).H(T)}\int_{4M_\chi^2}^\infty ds d\Omega \left|\vec{p_{12}}\right| \left|\vec{p_{34}}\right|\overline{\left|\mathcal{M}\right|}_{12\to 34}^2 \frac{1}{\sqrt{s}} K_1\left(\frac{\sqrt{s}}{T}\right),
\end{split}
\label{eq:beq-ann-yld2to2}
\eea

where the lower and upper limits of the integration over temperature depend on what epoch we are computing the DM yield. In the before EWSB era non-zero DM mass does not affect the resulting yield. In that case the lower limit of the $s$ integral can be taken to be zero instead of $4M_\chi^2$. This is true for any $n\to m$ process with $n,m\in2,3,4$.

\subsection{BEQ for $3\to 2$ process}
\label{sec:app-beq3to2}

For a $3\to 2$ process: $123\to 45$, with particle $1$ carrying a four momenta $p_1$ and so on, we can write the BEQ as:

\bea\begin{split}
\dot{n_\chi}+3 H n_\chi &= \int d\Pi_1 d\Pi_2 d\Pi_3 d\Pi_4 d\Pi_5 \overline{\left|\mathcal{M}\right|}^2_{123\to45} \left(2\pi\right)^4\delta^4\left(p_4+p_5-p_1-p_2-p_3\right)\prod_{i=1}^3f_i\\&=\int d\text{LIPS}_3 d\Pi_4 d\Pi_5 \overline{\left|\mathcal{M}\right|}^2_{123\to45}f_1 f_2 f_3,    
    \end{split}\label{eq:ann-beq-32}
\eea

where $d\text{LIPS}_3=d\Pi_1 d\Pi_2 d\Pi_3\left(2\pi\right)^4\delta^4\left(p_4+p_5-p_1-p_2-p_3\right)$ is the differential Lorentz invariant 3-body phase space. Since we are considering $n\to m$ (with $n,m>2$) processes only before EWSB, hence all SM particles are massless. Also, as the DM mass does not affect the yield much, we compute the yield in the zero DM mass limit. In order to simplify Eq.~\eqref{eq:ann-beq-32} we will first deal with the 2-body phase spaces, following~\cite{Gondolo:1990dk,Edsjo:1997bg}:

\bea
d^3p_4d^3p_5 = \left(4\pi\left|\vec{p_4}\right|\right)dE_4\left(4\pi\left|\vec{p_5}\right|\right)dE_5\frac{1}{2}d\cos\theta.
\eea

We then make a change of the variables as in~\cite{Gondolo:1990dk,Edsjo:1997bg}: $E_+=E_4+E_5,E_-=E_4-E_5,s=2M_\chi^2+2E_4 E_5-2\left|\vec{p_4}\right|\left|\vec{p_5}\right|\cos\theta$. The volume element therefore can be written as:

\bea
\int d\Pi_4d\Pi_5=\int\frac{1}{\left(2\pi\right)^4}\frac{\sqrt{E_+^2-s}}{4}\sqrt{1-\frac{4M_\chi^2}{s}} dE_+ ds,
\eea

where the limits on different variables are: $\left|E_-\right|\leq \sqrt{1-\frac{4M_\chi^2}{s}}\sqrt{E_+^2-s}$, $E_+\geq\sqrt{s}$ and $\sqrt{s}\geq 4M_\chi^2$ . With this the BEQ in Eq.~\eqref{eq:ann-beq-32} reduces to:

\bea\begin{split}
\dot{n_\chi}+3 H n_\chi &=\int_0^\infty ds\int_{\sqrt{s}}^\infty dE_+  e^{-\left(E_1+E_2+E_3\right)/T}\frac{1}{\left(2\pi\right)^4}\frac{\sqrt{E_+^2-s}}{4}\sqrt{1-\frac{4M_\chi^2}{s}}\overline{\left|\mathcal{M}\right|}_{123\to 45} d\text{LIPS}_3\\&=\frac{T}{\left(2\pi\right)^4}\int_{4M_\chi^2}^\infty ds\frac{\sqrt{s}}{4}\sqrt{1-\frac{4M_\chi^2}{s}}\overline{\left|\mathcal{M}\right|}^2_{123\to 45}K_1\left(\frac{\sqrt{s}}{T}\right)d\text{LIPS}_3,
    \end{split}\label{eq:ann-beq-32a}
\eea

where in the second line we applied the conservation of energy: $E_1+E_2+E_3=E_4+E_5$. The integrated 3-body phase space can be expressed in terms of 2-body phase space for massless initial state particles as:

\bea\begin{split}
\int d\text{LIPS}_3 &=\int\frac{ds_{23}}{2\pi}\frac{d\cos\theta_1}{2}\frac{d\phi_1}{2\pi}\frac{\overline{\beta_1}\left(0,\frac{s_{23}}{s}\right)}{8\pi}\frac{d\cos\theta_{23}}{2}\frac{d\phi_{23}}{2}\frac{\overline{\beta_{23}}\left(0,0\right)}{8\pi}.      
    \end{split}\label{eq:ps3a}
\eea

For an isotropic distribution the overall rotation of the system $\left(\cos\theta_1,\phi_1\right)$ can be dropped. The polar angle $\theta_{23}$ is defined relative to the direction of $-\vec{p_1}$. The azimuthal angle $\phi_{23}$ corresponds to the overall rotation and hence it is also trivial. In our case $m_{1,2,3}=0$ as the 3-body phase space consists of the massless SM particles. In the simple case of the massless limit {\footnote{Considering non-zero mass for the DM before EWSB changes the cross-section in percentage level.}}:

\bea\begin{split}
& \overline{\beta_1} = \left(1-\frac{s_{23}}{s}\right), 
\\&\overline{\beta_{23}} =  \sqrt{1-\frac{1}{s_{23}}\left(m_2^2+m_3^2\right)+\frac{1}{s_{23}^2}\left(m_2^2-m_3^2\right)^2}\to 1     
    \end{split}
\eea

Performing all the integrals for overall rotations we have:

\bea\begin{split}
\int d\text{LIPS}_3 &= \int\frac{ds_{23}}{2\pi}\frac{1}{8\pi}\Bigg(1-\frac{s_{23}}{s}\Bigg)\frac{d\cos\theta_{23}}{2}\frac{1}{8\pi}.     
    \end{split}\label{eq:ps3b}
\eea

The variables $s_{23}$ and $\cos\theta_{23}$ can be recasted in terms of the energy fraction $x_i$:

\bea\begin{split}
&x_1 =1-\frac{s_{23}}{s}\\&
x_2 = \frac{1}{2}\left(2-x_1+x_1\cos\theta_{23}\right),
    \end{split}\label{eq:x1x2}
\eea

also the energies of the incoming particles can be written as \footnote{More traditional variables are the Dalitz variables namely: $m_{12}^2=s_{12}$ and $m_{23}^2=s_{23}$~\cite{PhysRevD.98.030001}.}:

\bea\begin{split}
& E_1 = x_1\frac{\sqrt{s}}{2},~E_2 = x_2\frac{\sqrt{s}}{2},~E_3= \frac{\sqrt{s}}{2}\left(2-x_1-x_2\right).     
    \end{split}\label{eq:ps3energy}
\eea

Following Eq.~\eqref{eq:ann-beq-32a} then the yield for $3\to2$ process can be expressed as:

\bea
\begin{split}
\frac{dY_\chi^{3\to2}}{dT} &\simeq-\frac{1}{s(T).H(T)}\frac{1}{64\left(2\pi\right)^7}\int_{4M_\chi^2}^\infty dss^{3/2}\overline{\left|\mathcal{M}\right|}^2_{123\to 45}\sqrt{1-\frac{4M_\chi^2}{s}}\\& K_1\left(\frac{\sqrt{s}}{T}\right)\int_0^1 dx_1 \int_{1-x_1}^1 dx_2. 
\end{split}
\label{eq:beq-ann-yld3to2}
\eea

Since the inclusion of the DM mass changes the yield only in the percentage level, for simplicity, we can ignore the DM mass as well.

\subsection{BEQ for $4\to 2$ process}
\label{sec:app-beq4to2}

For a $4\to 2$ process of the form $1234\to 56$ we can write the BEQ as:

\bea\begin{split}
\dot{n_\chi}+3Hn_\chi &=\int d\Pi_1 d\Pi_2 d\Pi_3 d\Pi_4 d\Pi_5 d\Pi_6\overline{\left|\mathcal{M}\right|}^2_{1234\to56}\left(2\pi\right)^4\delta^4\left(p_5+p_6-p_1-p_2-p_3-p_4\right)\prod_{i=1}^4f_i\\&=\int d\text{LIPS}_4 d\Pi_5 d\Pi_6 \overline{\left|\mathcal{M}\right|}^2_{1234\to56}f_1 f_2 f_3 f_4,     
    \end{split}\label{eq:ann-beq-42}
\eea

where $d\text{LIPS}_4=d\Pi_1 d\Pi_2 d\Pi_3 d\Pi_4\left(2\pi\right)^4\delta^4\left(p_5+p_6-p_1-p_2-p_3-p_4\right)$ is the 4-body phase space. Proceeding as before we can write the BEQ with the redefined variables as:

\bea\begin{split}
\dot{n_\chi}+3 H n_\chi &=\frac{T}{\left(2\pi\right)^4}\int_{4M_\chi^2}^\infty ds\frac{\sqrt{s}}{4}\sqrt{1-\frac{4M_\chi^2}{s}}\overline{\left|\mathcal{M}\right|}^2_{1234\to 56}K_1\left(\frac{\sqrt{s}}{T}\right)d\text{LIPS}_4,
    \end{split}\label{eq:ann-beq-42a}
\eea

where we have again assumed energy conservation $E_1+E_2+E_3+E_4=E_5+E_6$. Similar to the 3-body case, the full 4-body phase space can be decomposed into three 2-body phase space as:

\bea\begin{split}
\int d\text{LIPS}_4 =  \int \frac{ds_{12}}{2\pi}\frac{ds_{34}}{2\pi} d\text{LIPS}_2\left(q_{12},q_{34}\right)  d\text{LIPS}_2\left(\hat{p_1},\hat{p_2}\right)  d\text{LIPS}_2\left(\hat{p_3},\hat{p_4}\right),   
    \end{split}\label{eq:ps4a}
\eea

with

\bea\begin{split}
\int d\text{LIPS}_2\left(q_{12},q_{34}\right) = \frac{\overline{\beta}}{8\pi}\int\frac{d\cos\theta}{2}\frac{d\phi}{2\pi}     
    \end{split}
\eea

and

\bea\begin{split}
\int d\text{LIPS}_2\left(\hat{p_{1(3)}},\hat{p_{2(4)}}\right) = \frac{\overline{\beta_{12(34)}}}{8\pi}\int\frac{d\cos\theta_{12(34)}}{2}\frac{d\phi_{12(34)}}{2\pi},     
    \end{split}
\eea

where $q_{12}=p_1+p_2$, $q_{34}=p_3+p_4$ are the sum of four momenta of the initial particles. The ``hatted'' variables labelled 1 and 2 are in the rest frame of $q_{12}$, and those labelled as 3 and 4 are in the rest frame of $q_{34}$. Gathering all of these together, one can write Eq.~\eqref{eq:ps4a} as:

\bea\begin{split}
\int d\text{LIPS}_4 =  \int \frac{ds_{12}}{2\pi}\frac{ds_{34}}{2\pi} \frac{\overline{\beta}}{8\pi}\int\frac{d\cos\theta}{2}\frac{d\phi}{2\pi}\frac{\overline{\beta_{12}}}{8\pi}\int\frac{d\cos\theta_{12}}{2}\frac{d\phi_{12}}{2\pi}\frac{\overline{\beta_{34}}}{8\pi}\int\frac{d\cos\theta_{34}}{2}\frac{d\phi_{34}}{2\pi}.   
    \end{split}\label{eq:ps4b}
\eea

Again for massless initial states: $\overline{\beta}_{12}=\overline{\beta}_{34}\to 1$, while $\overline{\beta}=\sqrt{1-\frac{2}{s}\left(s_{12}+s_{34}\right)-\frac{1}{s^2}\left(s_{12}-s_{34}\right)^2}$. Then Eq.~\eqref{eq:ps4a} becomes:

\bea\begin{split}
\int d\text{LIPS}_4 &=\frac{1}{4\pi^2(8\pi)^3}\int_0^{\sqrt{s}} ds_{12}\\&\int_0^{(\sqrt{s}-\sqrt{s_{12}})^2} ds_{34} \sqrt{1+\frac{s_{12}^2}{s^2}-\frac{2 s_{12} s_{34}}{s^2}+\frac{s_{34}^2}{s^2}-\frac{2 s_{12}}{s}-\frac{2 s_{34}}{s}}\int\frac{d\cos\theta_{12}}{2} \int \frac{d\cos\theta_{34}}{2}.     
    \end{split}
\eea

Finally the BEQ for $4\to 2$ process can be written using Eq.~\eqref{eq:ann-beq-42a} and Eq.~\eqref{eq:ps4b} as:

\bea
\begin{split}
\frac{dY_\chi^{4\to2}}{dT} &\simeq -\frac{1}{s(T).H(T)}\frac{1}{64\left(2\pi\right)^9}\int_{4M_\chi^2}^\infty ds \sqrt{s}\sqrt{1-\frac{4M_\chi^2}{s}} \overline{\left|\mathcal{M}\right|}^2_{1234\to 56} K_1\left(\frac{\sqrt{s}}{T}\right)\\&\int_0^{\sqrt{s}} ds_{12}\int_0^{(\sqrt{s}-\sqrt{s_{12}})^2} ds_{34} \sqrt{1+\frac{s_{12}^2}{s^2}-\frac{2 s_{12} s_{34}}{s^2}+\frac{s_{34}^2}{s^2}-\frac{2 s_{12}}{s}-\frac{2 s_{34}}{s}}\int\frac{d\cos\theta_{12}}{2} \int \frac{d\cos\theta_{34}}{2},
\end{split}
\label{eq:beq-ann-yld4to2}
\eea

which is in terms of the yield $Y_\chi$. Again, ignoring the DM mass does not change the outcome. 

\section{Computation of squared amplitudes}
\label{sec:app-sq-amp}

Several different processes arise before and after EWSB. We therefore compute the squared amplitudes at two different era considering relevant interactions. Before EWSB there are no decay processes, hence all we have are scatterings. For interactions at a particular dimension we calculate the squared amplitudes for processes with minimum number of initial state particles as yield with large number of initial states is suppressed as we have shown in Appendix.~\ref{sec:app-beq}. In presence of several multiparticle initial states we consider processes upto $3\to 2~(2\to3)$ for dim.5 and dim.7 operators, while for dim.8 operator there is $4\to 2~(2\to 4)$ process. In dim.7 there are $2\to 2$ and  $3\to 2~(2\to 3)$ processes. There are also $4\to 2~(2\to4)$ processes which we do not consider as they will have a sub-dominant contribution. After EWSB, on top of $2\to2$, $3\to2$ and $4\to2$ annihilation there is also $1\to2$ decay both in dim.5 and dim.7 level. We however stick to processes upto $2\to2$ after EWSB. Also, after EWSB all SM particles are massive, together with the DM. For all processes before EWSB the spin averaged squared amplitudes are tabulated in Tab.~\ref{tab:bewsb-amp}, while in Tab.~\ref{tab:aewsb-amp} we tabulate all processes appearing after EWSB.

\FloatBarrier
\begin{table}[htb!]
\begin{center}
\begin{tabular}{|c|c|c|c|c|c|c|c|c|c|c|}
\hline
Operator dim. & Amplitude squared \\[0.5ex] 
\hline\hline
5 & $\overline{\left|\mathcal{M}\right|}^2_{\phi\phi\to\chi\chi}=2\times\frac{2s}{\Lambda^2}\left(1-\frac{4M_\chi^2}{s}\right)$ \\
\hline
7 & $\overline{\left|\mathcal{M}\right|}^2_{XX\to\chi\chi}=12\times\frac{s^3}{9\Lambda^6}\left(1-\frac{4M_\chi^2}{s}\right)$ \\
& $\overline{\left|\mathcal{M}\right|}^2_{\phi\phi\to\chi\chi}=2\times\frac{s^3}{2\Lambda^6}\left(1-\frac{4M_\chi^2}{s}\right)$  \\
\cline{2-2}
&  $\overline{\left|\mathcal{M}\right|}^2_{XX\to\ \chi\chi X}=\frac{2g_2^2(g_s^2)s^2}{3\Lambda^6}\left(1-\frac{4M_\chi^2}{s}\right)$ \\
& $\overline{\left|\mathcal{M}\right|}^2_{\phi\phi\to B\chi\chi}=2\times\frac{g_1^2s^2}{3\Lambda^6}\left(1-\frac{4M_\chi^2}{s}\right)x_1x_2\left(1-\cos\theta_{12}\right)$ \\
& $\overline{\left|\mathcal{M}\right|}^2_{B\phi\to\phi\chi\chi}=2\times\frac{g_1^2s^2}{3\Lambda^6}\left(1-\frac{4M_\chi^2}{s}\right)x_1\left(2-x_1-x_2\right)\left(1-\cos\theta_{13}\right)$ \\
& $\overline{\left|\mathcal{M}\right|}^2_{\phi\phi\to W^i\chi\chi}=6\times\frac{g_2^2s^2}{3\Lambda^6}\left(1-\frac{4M_\chi^2}{s}\right)x_1x_2\left(1-\cos\theta_{12}\right)$ \\
& $\overline{\left|\mathcal{M}\right|}^2_{W^i\phi\to\phi\chi\chi}=6\times\frac{g_2^2s^2}{3\Lambda^6}\left(1-\frac{4M_\chi^2}{s}\right)x_1\left(2-x_1-x_2\right)\left(1-\cos\theta_{13}\right)$ \\
& $\overline{\left|\mathcal{M}\right|}^2_{ff\to\phi\chi\chi}=4\times\frac{N_cs^2}{4\Lambda^6}\left(1-\frac{4M_\chi^2}{s}\right)x_1 x_2\left(1-\cos\theta_{12}\right)$ \\
& $\overline{\left|\mathcal{M}\right|}^2_{f\phi\to f\chi\chi}=4\times\frac{N_cs^2}{2\Lambda^6}\left(1-\frac{4M_\chi^2}{s}\right)x_1 \left(2-x_1-x_2\right)\left(1-\cos\theta_{13}\right)$ \\
& $\sum_i\overline{\left|\mathcal{M}\right|}^2_{ff\to W^i\chi\chi}=3\times\frac{g_2^2N_Cs^2}{2\Lambda^6}\left(1-\frac{4M_\chi^2}{s}\right)x_1x_2\left(1-\cos\theta_{12}\right)$ \\
& $\sum_i\overline{\left|\mathcal{M}\right|}^2_{fW^i\to f\chi\chi}=3\times\frac{g_2^2N_Cs^2}{3\Lambda^6}\left(1-\frac{4M_\chi^2}{s}\right)x_1 \left(2-x_1-x_2\right)\left(1-\cos\theta_{13}\right)$ \\
& $\overline{\left|\mathcal{M}\right|}^2_{ff\to B\chi\chi}=\frac{g_1^2N_Cs^2}{16\Lambda^6}\left(Y_T^2+Y_D^2\right)\left(1-\frac{4M_\chi^2}{s}\right)x_1x_2\left(1-\cos\theta_{12}\right)$\\ 
& $\overline{\left|\mathcal{M}\right|}^2_{fB\to f\chi\chi}=\frac{g_1^2N_Cs^2}{3\Lambda^6}\left(1-\frac{4M_\chi^2}{s}\right)\left(Y_T^2+Y_D^2\right)x_1 \left(2-x_1-x_2\right)\left(1-\cos\theta_{13}\right)$ \\
\cline{2-2}
& $\overline{\left|\mathcal{M}\right|}^2_{XXX\to\chi\chi}=\frac{2g_2^2(g_s^2)s^2}{9\Lambda^6}\left(1-\frac{4M_\chi^2}{s}\right)$ \\
& $\overline{\left|\mathcal{M}\right|}^2_{B\phi\phi\to\chi\chi}=2\times\frac{g_1^2s^2}{12\Lambda^6}\left(1-\frac{4M_\chi^2}{s}\right)x_2\left(2-x_1-x_2\right)\left(1-\cos\theta_{23}\right)$ \\
& $\sum_i\overline{\left|\mathcal{M}\right|}^2_{W^i\phi\phi\to\chi\chi}=6\times\frac{g_2^2s^2}{12\Lambda^6}\left(1-\frac{4M_\chi^2}{s}\right)x_2\left(2-x_1-x_2\right)\left(1-\cos\theta_{23}\right)$ \\
& $\overline{\left|\mathcal{M}\right|}^2_{ff\phi\to\chi\chi}=4\times\frac{N_cs^2}{4\Lambda^6}\left(1-\frac{4M_\chi^2}{s}\right)x_1x_2\left(1-\cos\theta_{12}\right)$ \\
& $\sum_i\overline{\left|\mathcal{M}\right|}^2_{ffW^i\to\chi\chi}=3\times\frac{g_2^2N_Cs^2}{6\Lambda^6}\left(1-\frac{4M_\chi^2}{s}\right)x_1x_2\left(1-\cos\theta_{12}\right)$ \\
& $\overline{\left|\mathcal{M}\right|}^2_{ffB\to\chi\chi}=\frac{g_1^2N_Cs^2}{48\Lambda^6}\left(Y_T^2+Y_D^2\right)\left(1-\frac{4M_\chi^2}{s}\right)x_1x_2\left(1-\cos\theta_{12}\right)$\\ 
\hline
8 & $\overline{\left|\mathcal{M}\right|}^2_{\ell\ell\phi\phi\to\chi\chi}=4\times\frac{s^2}{8\Lambda^8}\left(1-\frac{4M_\chi^2}{s}\right)\left(1-\cos\theta_{12}\right)$ \\
&$\overline{\left|\mathcal{M}\right|}^2_{\ell\ell\to\phi\phi\chi\chi}=4\times\frac{s^2}{8\Lambda^8}\left(1-\frac{4M_\chi^2}{s}\right)\left(1-\cos\theta_{12}\right)$\\
&$\overline{\left|\mathcal{M}\right|}^2_{\phi\phi\to\ell\ell\chi\chi}=4\times\frac{s^2}{2\Lambda^8}\left(1-\frac{4M_\chi^2}{s}\right)\left(1-\cos\theta_{34}\right)$\\
 &$\overline{\left|\mathcal{M}\right|}^2_{\ell\phi\to\ell\chi\chi}=4\times\frac{s^2}{4\Lambda^8}\left(1-\frac{4M_\chi^2}{s}\right)\left(1-\cos\theta_{13}\right)$\\
\hline\hline\end{tabular}
\end{center}
\caption{Table for amplitude squared before EWSB. From top to bottom we have the $2\to2$, $2\to3$, $3\to2$ processes for dim.5 and dim.7, while $4\to2$ and $2\to4$ processes for dim.8 interactions. Here $X\in W^i_\mu,B_\mu,G^a_\mu$ are the SM gauge bosons corresponding to $SU(2)_L$, $U(1)_Y$ and $SU(3)_c$ with $i=1,2,3$ and $a=1...8$. All SM states are massless, while the DM is massive in this epoch.}
\label{tab:bewsb-amp}
\end{table}
\FloatBarrier

\FloatBarrier
\begin{table}[htb!]
\begin{center}
\begin{tabular}{|c|c|c|c|c|c|c|c|c|c|c|}
\hline
Operator dim. & Amplitude squared \\[0.5ex] 
\hline\hline
5 & $\overline{\left|\mathcal{M}\right|}^2_{h\to\chi\chi}=\frac{8m_h^2v_h^2}{\Lambda^2}\left(1-\frac{4M_\chi^2}{m_h^2}\right)$ \\
& $\overline{\left|\mathcal{M}\right|}^2_{hh\to\chi\chi}=\frac{s}{2\Lambda^2}\left(1-\frac{4M_\chi^2}{s}\right)$  \\
& $\overline{\left|\mathcal{M}\right|}^2_{VV\xrightarrow{h}\chi\chi}=\frac{2s^3}{3\Lambda^2}\left(1-\frac{4M_\chi^2}{s}\right)\frac{1}{\left(s-m_h^2\right)^2+\Gamma_h^2 m_h^2}\left(1-\frac{4m_v^2}{s}+\frac{12m_v^4}{s^2}\right)$  \\
& $\overline{\left|\mathcal{M}\right|}^2_{ff\xrightarrow{h}\chi\chi}=\frac{N_cs^2 m_f^2}{\Lambda^2}\left(1-\frac{4M_\chi^2}{s}\right)\frac{1}{\left(s-m_h^2\right)^2+\Gamma_h^2 m_h^2}\left(1-\frac{4m_f^2}{s}\right)$  \\
& $\overline{\left|\mathcal{M}\right|}^2_{gg\xrightarrow{h}\chi\chi}=\frac{g_s^4 s m_h^4}{4608\pi^4\Lambda^2}\frac{\mathcal{F}\left(x\right)^2}{\left(s-m_h^2\right)^2+\Gamma_h^2 m_h^2}\left(1-\frac{4M_\chi^2}{s}\right)$  \\
\hline
7 & $\overline{\left|\mathcal{M}\right|}^2_{h\to\chi\chi}=\frac{2 m_h^2 v_h^6}{\Lambda^6}\left(1-\frac{4M_\chi^2}{s}\right)$ \\
& $\overline{\left|\mathcal{M}\right|}^2_{hh\to\chi\chi}=\frac{9sv_h^4}{2\Lambda^6}\left(1-\frac{4M_\chi^2}{s}\right)$ \\
& $\overline{\left|\mathcal{M}\right|}^2_{hh\to\chi\chi}=\frac{s^3}{8\Lambda^6}\left(1-\frac{2M_\chi^2}{s}\right)\left(1-\frac{4m_h^2}{s}\right)^2$ \\
\cline{2-2}
& $\overline{\left|\mathcal{M}\right|}^2_{\gamma\gamma\to\chi\chi}=\frac{64s_w^4s^3}{9\Lambda^6}\left(1-\frac{4M_\chi^2}{s}\right)$ \\
& $\overline{\left|\mathcal{M}\right|}^2_{WW\to\chi\chi}=\frac{8s^5}{9 m_W^4 \Lambda^6}\left(1-\frac{2m_W^2}{s}\right)^2\left(1-\frac{4M_\chi^2}{s}\right)\left(1-\frac{4m_W^2}{s}+\frac{12m_W^4}{s^2}\right)$ \\
& $\overline{\left|\mathcal{M}\right|}^2_{ZZ\to\chi\chi}=\frac{8c_w^4s^5}{9m_Z^4 \Lambda^6}\left(1-\frac{2m_Z^2}{s}\right)^2\left(1-\frac{4M_\chi^2}{s}\right)\left(1-\frac{4m_Z^2}{s}+\frac{12m_Z^4}{s^2}\right)$ \\
& $\overline{\left|\mathcal{M}\right|}^2_{\gamma Z\to\chi\chi}=\frac{8s_w^2c_w^2s^5}{9m_Z^4 \Lambda^6}\left(1-\frac{2m_Z^2}{s}\right)^2\left(1-\frac{4M_\chi^2}{s}\right)\left(1-\frac{2m_Z^2}{s}+\frac{9m_Z^4}{s^2}\right)$ \\
& $\overline{\left|\mathcal{M}\right|}^2_{gg\to\chi\chi}=\frac{64s^3}{9\Lambda^6}\left(1-\frac{4M_\chi^2}{s}\right)$ \\\cline{2-2}
& $\overline{\left|\mathcal{M}\right|}^2_{WW\to\chi\chi}=\frac{g_2^4v_h^4s^3}{288m_V^4\Lambda^6}\left(1-\frac{4M_\chi^2}{s}\right)\left(1-\frac{4m_W^2}{s}+\frac{12m_W^4}{s^2}\right)$ \\
& $\overline{\left|\mathcal{M}\right|}^2_{ZZ\to\chi\chi}=\frac{g_2^4v_h^4s^3}{1152c_w^4m_V^4\Lambda^6}\left(1-\frac{4M_\chi^2}{s}\right)\left(1-\frac{4m_Z^2}{s}+\frac{12m_Z^4}{s^2}\right)$ \\
& $\overline{\left|\mathcal{M}\right|}^2_{WW\xrightarrow{h}\chi\chi}=\frac{2v_h^4s^3}{3\Lambda^6}\left(1-\frac{4M_\chi^2}{s}\right)\frac{1}{\left(s-m_h^2\right)^2+\Gamma_h^2 m_h^2}\left(1-\frac{4m_W^2}{s}+\frac{12m_W^4}{s^2}\right)$  \\
& $\overline{\left|\mathcal{M}\right|}^2_{ZZ\xrightarrow{h}\chi\chi}=\frac{2v_h^4s^3}{3\Lambda^6}\left(1-\frac{4M_\chi^2}{s}\right)\frac{1}{\left(s-m_h^2\right)^2+\Gamma_h^2 m_h^2}\left(1-\frac{4m_Z^2}{s}+\frac{12m_Z^4}{s^2}\right)$  \\
& $\overline{\left|\mathcal{M}\right|}^2_{ff\xrightarrow{h}\chi\chi}=\frac{N_cs^2v_h^4m_f^2}{\Lambda^6}\left(1-\frac{4M_\chi^2}{s}\right)\frac{1}{\left(s-m_h^2\right)^2+\Gamma_h^2 m_h^2}\left(1-\frac{4m_f^2}{s}\right)$  \\
& $\overline{\left|\mathcal{M}\right|}^2_{gg\xrightarrow{h}\chi\chi}=\frac{g_s^4 s m_h^4 v_h^4}{4608\pi^4\Lambda^6}\frac{\mathcal{F}\left(x\right)^2}{\left(s-m_h^2\right)^2+\Gamma_h^2 m_h^2}\left(1-\frac{4M_\chi^2}{s}\right)$ \\
\hline
8 & $\overline{\left|\mathcal{M}\right|}^2_{\nu\nu\to\chi\chi}=\frac{v_h^4s^2}{4\Lambda^8}\left(1-\frac{4M_\chi^2}{s}\right)\left(1-\frac{4m_\nu^2}{s}\right)$ \\
\hline\hline\end{tabular}
\end{center}
\caption{Table for amplitude squared after EWSB where $s(c)_w$ is the (co)sine of the weak mixing angle. All SM states are massive along with the DM.}
\label{tab:aewsb-amp}
\end{table}
\FloatBarrier

\section{Thermally averaged cross-section}
\label{sec:app-therm-avg}

Here we would like to furnish the derivation of thermally averaged cross-section for $2\to2$, $3\to2$ and $4\to2$ process, which is going to be utilized for determining the thermalization condition for the DM. Since we are interested in determining the rate of interaction at high temperature, hence we stick to the before EWSB scenario where all SM particles are massless.

\subsection{Thermally averaged $2\to2$ cross-section}
\label{sec:therm2to2}

Let us first determine the thermally averaged $2\to2$ cross-section for a process $12\to 34$:

\bea\begin{split}
\langle\sigma v\rangle_{2\to2} &=\frac{\int\sigma v_{\text{rel}}~\exp(-E_1/T)~\exp(-E_2/T)d^3p_1 d^3p_2}{\int \exp(-E_1/T)~\exp(-E_2/T)d^3p_1 d^3p_2},     
    \end{split}\label{eq:sigv22}
\eea

where we have used the definition of cross section as:

\bea
d\sigma=\frac{\overline{\left|\mathcal{M}\right|^2}_{2\to2}}{\mathcal{F}}d\text{LIPS}_2,
\eea

where $\mathcal{F}=4E_1 E_2 \left|\vec{v}_{\text{rel}}\right|$ is the flux factor and $d\text{LIPS}_2$ is the Lorentz invariant 2-body differential phase space. The momentum-space volume element can be written in terms of the redefined variables as (similar to Appendix.~\ref{sec:app-beq3to2}):

\bea
d^3p_1 d^3p_2 = 2\pi^2 E_1 E_2 dE_+ dE_- ds.
\eea

With this we can write the numerator of Eq.~\eqref{eq:sigv22} as~\cite{Gondolo:1990dk,Edsjo:1997bg}: 

\bea\begin{split}
&\int\sigma v_{\text{rel}}~\exp(-E_1/T)~\exp(-E_2/T)d^3p_1 d^3p_2=2\pi^2 T \int ds\sigma(s-4M_\chi^2)\sqrt{s}K_1\left(\frac{\sqrt{s}}{T}\right),     
    \end{split}
\eea
where again we can ignore the DM mass as that is not going to affect our results. The denominator is derived in the massless limit of the SM particles as:

\bea\begin{split}
&\int \exp(-E_1/T)~\exp(-E_2/T)d^3p_1 d^3p_2= 64\pi^2 T^6.
    \end{split}\label{eq:neq2}
\eea

Combining the numerator and denominator we find Eq.~\eqref{eq:sigv22} takes the form:

\bea
\langle\sigma v\rangle_{2\to2}=\frac{1}{32T^5}\int_{4M_\chi^2}^\infty ds\sigma(s-4M_\chi^2)\sqrt
{s}K_1\left(\sqrt{s}/T\right),
\label{eq:sigv22a}
\eea

where we can ignore the DM mass and the lower limit of the integral then turns out to be zero.

\subsection{Thermally averaged $3\to2$ \& $2\to3$ cross-section}
\label{sec:therm3to2}

The thermally averaged $3\to2$ cross-section can be expressed as~\cite{Cline:2017tka,Pierre:2019vao,Bhattacharya:2019mmy} follows, where we will ignore the DM mass, as taking that into account makes no substantial change in the results:

\bea\begin{split}
\langle\sigma v^2\rangle_{3\to2} &=\frac{1}{\prod_{i=1}^3 n_i^{eq}}\int d\Pi_1...d\Pi_5 \left(2\pi\right)^4 \delta^{(4)}\left(p_4+p_5-p_1-p_2-p_3\right)\overline{\left|\mathcal{M}\right|}^2_{3\to2}\prod_{i=1}^{3} f_i\\&=\frac{1}{\prod_{i=1}^3 n_i^{eq}} \frac{T}{(2\pi)^4}\int_0^\infty ds\frac{\sqrt{s}}{4} K_1\left(\frac{\sqrt{s}}{T}\right)  \overline{\left|\mathcal{M}\right|}^2_{3\to2} d\text{LIPS}_3,    
    \end{split}\label{eq:sigv32a}
\eea

where again we have performed a change of variables in the second line and considered $d\text{LIPS}_3=d\Pi_1d\Pi_2d\Pi_3\left(2\pi\right)^4 \delta^{(4)}\left(p_4+p_5-p_1-p_2-p_3\right)$. Now, in the massless limit the denominator reads:

\bea\begin{split}
\prod_{i=1}^3 n_i^{eq} &= \frac{g_1g_2g_3}{\left(2\pi\right)^9}\int \exp(-E_1/T)~\exp(-E_2/T)\exp(-E_3/T)d^3p_1 d^3p_2d^3p_3= g_1g_2g_3\frac{T^9}{\pi^6}.   
    \end{split}
\eea

One can write the expression for $d\text{LIPS}_3$ following Eq.~\eqref{eq:ps3b}. Together, the final expression can be read from Eq.~\eqref{eq:sigv32a}:

\bea\begin{split}
\langle\sigma v^2\rangle_{3\to2} &=\frac{\pi^6}{g_1g_2g_3T^8}\frac{1}{\left(2\pi\right)^4}\int_0^\infty ds\frac{\sqrt{s}}{4} K_1\left(\frac{\sqrt{s}}{T}\right)  \overline{\left|\mathcal{M}\right|}^2_{3\to2} d\text{LIPS}_3\\&=\frac{1}{g_1g_2g_3}\frac{1}{8192\pi T^8}\int_0^\infty dss^{3/2} K_1\left(\frac{\sqrt{s}}{T}\right)  \overline{\left|\mathcal{M}\right|}^2_{3\to2}\int_0^1 dx_1\int_{1-x_1}^1 dx_2.     
    \end{split}
\eea

Similarly,

\bea\begin{split}
\langle\sigma v\rangle_{2\to3} &=\frac{1}{\prod_{i=1}^2 n_i^{eq}} \frac{T}{(2\pi)^4}\int_0^\infty ds\frac{\sqrt{s}}{4} K_1\left(\frac{\sqrt{s}}{T}\right)  \overline{\left|\mathcal{M}\right|}^2_{2\to3} d\text{LIPS}_3,
    \end{split}\label{eq:sigv2to3}
\eea

where $d\text{LIPS}_3=\left(2\pi\right)^4 \delta^{(4)}\left(p_3+p_4+p_5-p_1-p_2\right)d\Pi_3d\Pi_4d\Pi_5$. The denominator of Eq.~\eqref{eq:sigv2to3} can again be obtained as:

\bea\begin{split}
\prod_{i=1}^2 n_i^{eq} &=  \frac{g_1g_2}{\left(2\pi\right)^6}\int \exp(-E_1/T)\exp(-E_2/T)d^3p_1 d^3p_2=g_1g_2\frac{T^6}{\pi^4}.\label{eq:n1n2eq}  
    \end{split}
\eea

Then, on simplification one obtains:

\bea\begin{split}
\langle\sigma v\rangle_{2\to3} &= \frac{1}{g_1g_2}\frac{1}{16 T^5}\int_0^\infty ds\frac{\sqrt{s}}{4} K_1\left(\frac{\sqrt{s}}{T}\right) \overline{\left|\mathcal{M}\right|}^2_{2\to3} d\text{LIPS}_3,   
    \end{split}\label{eq:sigv2to3a}
\eea

\subsection{Thermally averaged $4\to2$ \& $2\to4$ cross-section}
\label{sec:therm4to2}

Proceeding as before we can write the $4\to2$ thermally averaged cross-section for a process $1234\to56$ as:

\bea\begin{split}
\langle\sigma v^3\rangle_{4\to2} &= \frac{1}{\prod_{i=1}^4 n_i^{eq}}\int d\Pi_1...d\Pi_6 \left(2\pi\right)^4 \delta^{(4)}\left(p_5+p_6-p_1-p_2-p_3-p_4\right)\overline{\left|\mathcal{M}\right|}^2_{4\to2}\prod_{i=1}^{4} f_i\\&=\frac{1}{\prod_{i=1}^4 n_i^{eq}} \frac{T}{\left(2\pi\right)^4}\int_0^\infty ds\frac{\sqrt{s}}{4}\overline{\left|\mathcal{M}\right|}^2_{1234\to 56}K_1\left(\frac{\sqrt{s}}{T}\right)d\text{LIPS}_4,     
    \end{split}
\eea

with $d\text{LIPS}_4=\left(2\pi\right)^4 \delta^{(4)}\left(p_5+p_6-p_1-p_2-p_3-p_4\right)d\Pi_1...d\Pi_4$. Again the denominator in the massless limit:

\bea\begin{split}
\prod_{i=1}^4 n_i^{eq} &=g_1g_2g_3g_4\frac{\left(4\pi\right)^4}{\left(2\pi\right)^{12}}\int\exp(-E_1/T)~\exp(-E_2/T)\exp(-E_3/T) \exp(-E_4/T)\\&E_1^2 E_2^2 E_3^2 E_4^2 dE_1 dE_2 dE_3 dE_4=g_1g_2g_3g_4\frac{T^{12}}{\pi^8}.     
    \end{split}
\eea

Therefore, the final expression for $4\to2$ process:

\bea\begin{split}
 \langle\sigma v^3\rangle_{4\to2} &= \frac{1}{g_1g_2g_3g_4}  \frac{\pi^8}{T^{11}}\frac{1}{\left(2\pi\right)^4}\int_0^\infty ds\frac{\sqrt{s}}{4}\overline{\left|\mathcal{M}\right|}^2_{1234\to 56}K_1\left(\frac{\sqrt{s}}{T}\right)d\text{LIPS}_4\\&=\frac{1}{g_1g_2g_3g_4}\frac{\pi^2}{64\left(8\pi\right)^3 T^{11}}\int_0^\infty ds\frac{\sqrt{s}}{4}\overline{\left|\mathcal{M}\right|}^2_{1234\to 56}K_1\left(\frac{\sqrt{s}}{T}\right)\\&\int_0^{\sqrt{s}} ds_{12}\int_0^{(\sqrt{s}-s_{12})^2} ds_{34} \sqrt{1+\frac{s_{12}^2}{s^2}-\frac{2 s_{12} s_{34}}{s^2}+\frac{s_{34}^2}{s^2}-\frac{2 s_{12}}{s}-\frac{2 s_{34}}{s}}\int\frac{d\cos\theta_{12}}{2} \int \frac{d\cos\theta_{34}}{2}    \end{split}
\eea

where we have exploited Eq.~\eqref{eq:ps4b} for obtaining $d\text{LIPS}_4$. One can similarly write the thermally averaged cross-section for a $2\to4$ process using Eq.~\eqref{eq:n1n2eq}:

\bea\begin{split}
\langle\sigma v\rangle_{2\to4}&=\frac{1}{\prod_{i=1}^2 n_i^{eq}} \frac{T}{(2\pi)^4}\int_0^\infty ds\frac{\sqrt{s}}{4} K_1\left(\frac{\sqrt{s}}{T}\right)  \overline{\left|\mathcal{M}\right|}^2_{2\to4} d\text{LIPS}_4\\&=\frac{1}{g_1g_2}\frac{1}{16T^5}\int_0^\infty ds\frac{\sqrt{s}}{4} K_1\left(\frac{\sqrt{s}}{T}\right)  \overline{\left|\mathcal{M}\right|}^2_{2\to4} d\text{LIPS}_4.    
    \end{split}
\eea

\section{Condition for thermalization}
\label{sec:app-therm-cond}

The condition whether the DM is in thermal equilibrium with the SM bath is determined by the ratio $\mathcal{R}=\frac{\Gamma_{n\to m}}{H}$, which quantifies if the rate of some $n\to m$ (with $n,m\in 1,2,3,4$) reaction $\Gamma_{n\to m}$ is larger, equal or less than the rate of expansion or the Hubble $H$ depending on which the DM can be out of equilibrium (non-thermal) or in equilibrium (thermal) with the SM bath. Now, the reaction rate is given by: 

\bea
\Gamma_{n\to m}=\begin{cases}
n_{\text{SM}}^{n-1}\langle\sigma v\rangle_{n\to m}& \text{if}~n,m>1\\
\Gamma_{\text{decay}} & \text{if}~n=1,
               \end{cases}
\eea

where $\Gamma_{\text{decay}}$ is the decay width for a process and $n_{\text{SM}}$ is the number density of the SM bath that can be determined following Eq.\ref{eq:num-density}, and is given by:

\bea
n_{\text{SM}} =
    \begin{cases}
      g_{\text{SM}}\left(T\right)\frac{3}{4}\frac{\zeta(3)}{\pi^2}  T^3 & \text{if}~T>>m_\text{SM}\\
      g_{\text{SM}}\left(T\right) \left(\frac{m_\text{SM} T}{2\pi}\right)^{3/2}e^{-m_\text{SM}/T} & \text{if}~T\leq m_\text{SM},
    \end{cases} 
\eea

with $g_{\text{SM}}\left(T\right)$ being the DOF of the SM particles. Now, the thermal averaged cross-section for $n\to m$ process is given by~\cite{Cline:2017tka,Pierre:2019vao,Bhattacharya:2019mmy}:

\bea\begin{split}
\langle\sigma v^{n-1}\rangle_{n\to m} &=\frac{1}{S_f}\frac{1}{\prod_{i=2}^{n} n_i^{\text{eq}}}\int\prod_{i=2}^{n+m}d\Pi_i\left(2\pi\right)^4\delta^{(4)}\left(\sum_{i=2}^n p_i-p_f\right)\overline{\left|\mathcal{M}\right|}^2_{n\to m}\prod_{i=2}^n f_i,     
    \end{split}\label{eq:sig-avg}
\eea

where $g_i$ is the DOF and $f_i$ is the phase space distribution for the species $i$. $S_f$ is the symmetry factor: $S_f=\prod_{i=2}^n n_i!$, where $n_i$ is the is the number of identical particles of species $i$ in the final state. The Hubble rate, on the other hand, is given by:

\bea
H\left(T\right) = 1.66 \sqrt{g_{\star\rho}\left(T\right)}\frac{T^2}{M_{\text{pl}}},
\eea

where $g_{\star\rho}$ is the DOF for the SM bath and $M_{\text{pl}}$ is the reduced Planck mass. Therefore, we need to calculate $\mathcal{R}=\frac{n_{\text{SM}}^{n-1}\langle\sigma v^{n-1}\rangle_{n\to m}}{H\left(T\right)}$ for $2\to2$, $3\to2$, $2\to3$, $4\to2$ and $2\to4$ processes before EWSB, while after EWSB apart from annihilation there is also decay that itself determines the rate. Following Eq.~\eqref{eq:sig-avg} we can calculate thermally averaged cross-section for $n\to m$ processes with $n,m\in 2,3,4$. One should note, for $2\to n$ process with $n\in 2,3,4$ the thermally averaged cross-section reads $\langle\sigma v\rangle$ and has the unit of $\text{GeV}^{-2}$, for $3\to2$ process the thermally averaged cross-section goes $\langle\sigma v^2\rangle\sim\text{GeV}^{-5}$ and for $4\to2$ it is $\langle\sigma v^3\rangle\sim\text{GeV}^{-8}$.


\section{Computation of the 1-loop integral}
\label{sec:app-loop}

With a proper choice of momenta directions, the contribution due to the loop shown in Fig.~\ref{fig:inflt-dec} can be written as:

\bea\begin{split}
&\frac{\mu_h}{\Lambda}\int\frac{d^4k}{\left(2\pi\right)^4}\frac{i}{\left(k+p\right)^2-m^2+i\epsilon}\frac{i}{k^2-m^2+i\epsilon}\\&=\frac{\mu_h}{\Lambda}\int_0^1 dx\int\frac{d^4\kappa}{\left(2\pi\right)^4}\frac{1}{\left(\kappa^2-M^2+i\epsilon\right)^2},
    \end{split}
\eea

where $k$ is the loop momenta and $\kappa=k+x.p$ with $M^2=m^2-x\left(1-x\right)p^2$, also $p^2=M_I^2$ is the inflaton mass (on-shell). One can then perform the Wick rotation and write the integral in terms of the Euclidean momenta as:

\bea\begin{split}
L&=\frac{\mu_h}{16\pi^2\Lambda}\int_0^1 dx\int_0^{\Lambda^2} dK_E^2\frac{K_E^2}{\left(K_E^2+M^2\right)^2}=\frac{\mu_h}{16\pi^2\Lambda}\int_0^1 dx \Biggl[-1+2\ln\Bigl(\frac{\Lambda}{M}\Bigr)\Biggr],
    \end{split}
\eea

where a cut-off $\Lambda$ is imposed over the loop momenta. The integral over $x$ can be performed by substituting the expression for $M$. However, we ignore the mass of the particle going in the loop. 

\bibliography{ref_fimp}
\end{document}